\documentclass[fleqn,usenatbib]{mnras}

\usepackage{newtxtext,newtxmath}
\usepackage{xspace}
\usepackage[T1]{fontenc}
\usepackage{ae,aecompl}
\usepackage{url}
\usepackage{appendix}

\usepackage{graphicx}	% Including figure files
\usepackage{amsmath}	% Advanced maths commands
\usepackage{float}

\usepackage[utf8]{inputenc}

\newcommand{\nup}{$\nu_{\rm peak}^S$}

\title[The spectra of IceCube neutrinos]{The spectra of IceCube neutrino candidate sources - I. Optical spectroscopy of blazars\\}

\author[S. Paiano et al.]{Simona Paiano$^{1,2,3}$\thanks{E-mail:
simona.paiano@inaf.it}, Renato Falomo$^{4}$, Aldo Treves$^{5,6}$, Paolo Padovani$^{7,8}$,   \newauthor Paolo Giommi$^{9,10,11}$, Riccardo Scarpa$^{12,13}$ \\
$^{1}$INAF - Osservatorio Astronomico di Roma, via Frascati 33, I-00040, Monteporzio Catone, Italy \\
$^{2}$INAF - IASF Milano, via Corti 12, I-20133, Milano, Italy \\
$^{3}$INAF - IASF Palermo, via Ugo La Malfa, 153, I-90146, Palermo, Italy \\ 
$^{4}$INAF - Osservatorio Astronomico di Padova, vicolo dell'Osservatorio 5, I-35122, Padova, Italy\\
$^{5}$Universita' dell'Insubria, via Valeggio, 22100, Como, Italy\\
$^{6}$INAF - Osservatorio Astronomico di Brera, via Bianchi 46, I-23807, Merate (Lecco), Italy\\
$^{7}$European Southern Observatory, Karl-Schwarzschild-Str. 2, D-85748 Garching bei M\"unchen, Germany\\
$^{8}$Associated to INAF - 
Osservatorio di Astrofisica e Scienza dello Spazio, Via Piero Gobetti 93/3, I-40129 Bologna, Italy\\
$^{9}$Institute for Advanced Study, Technische Universit{\"a}t M{\"u}nchen, Lichtenbergstrasse 2a, D-85748 Garching bei M\"unchen, Germany\\
$^{10}$Associated to Agenzia Spaziale Italiana, ASI, via del Politecnico s.n.c., I-00133 Roma Italy \\
$^{11}$ICRANet, Piazzale della Repubblica 10, I-65122, Pescara, Italy\\
$^{12}$Instituto de Astrofisica de Canarias, C/O Via Lactea, s/n E38205 - La Laguna (Tenerife) - SPAIN\\
$^{13}$Universidad de La Laguna, Dpto. Astrofsica, s/n E-38206 La Laguna (Tenerife) - SPAIN\\
}
\date{Received:~\today; Accepted:~\today }

\begin{document}
\label{firstpage}
\pagerange{\pageref{firstpage}--\pageref{lastpage}}
\maketitle

\begin{abstract}
There is mounting evidence that ultra-energetic neutrinos of astrophysical origin
may be associated with blazars. Here 
we investigate a unique sample of 47 blazars, $\sim 20$ of which could be new neutrino sources. 
In particular, we focus on 17 objects of yet unknown redshift, for which we present optical spectroscopy secured at the Gran Telescopio Canarias and the ESO Very Large Telescope.
We find all sources but one (a quasar) to be BL Lac objects. 
For nine targets we are able to determine the redshift (0.09~$<$~z~$<$~1.6), while for the others we set a lower limit on it, based on 
 either the robust detection of intervening absorption systems or on an estimation derived from the absence of spectral signatures of the host galaxy. 
In some spectra we detect forbidden and semi-forbidden emission lines with luminosities in the range $10^{40} - 10^{41}$ erg s$^{-1}$. We also report on the spectroscopy of seven blazars possibly associated  with energetic neutrinos that partially meet the criteria of our sample and are discussed in the Appendix. These results represent the starting point of our investigation into the real nature of these objects and their likelihood of being neutrino emitters.
\end{abstract}


\section{Introduction}  \label{sec:intro} 

The IceCube Neutrino Observatory at the South Pole\footnote{\url{http://icecube.wisc.edu}} has detected tens of high-energy neutrinos of likely astrophysical and extragalactic origin reaching beyond the PeV ($10^{15}$ eV) range  \citep[e.g.][and references therein]{schneider2019,stettner2019,Aartsen2020}.
These neutrinos are thought to be generated when very high-energy (VHE) cosmic rays (CRs) interact with matter or radiation creating charged and neutral mesons, which then decay into neutrinos, $\gamma$-rays, and other particles. At variance with CRs, neutrinos and $\gamma$-rays are the ``messengers'' that can travel cosmological distances undeflected. However, $\gamma$-ray spectra are softened by pair-production interactions with the extragalactic background light (EBL) at $E \gtrsim 100$ GeV in an energy- and redshift-dependent fashion \citep[see, e.g.,][and references therein]{biteau2020,CTA2020}. This means that the extragalactic photon sky is almost completely dark at the energies sampled by IceCube ($\gtrsim 60$ TeV). Therefore, only neutrinos can provide information on the VHE physical processes that generated them. 

So far, only one astronomical object has been significantly associated (at the $3 - 3.5\sigma$ level) in space and time with IceCube neutrinos, i.e., the powerful blazar TXS\,0506+056 at $z=0.3365$ \citep{icfermi,iconly,padovani2018,paiano2018}. 
Other possible associations include 3HSP~J095507.9+355101, an extreme blazar with synchrotron peak frequency\footnote{ From a spectral energy distribution (SED) point of view blazars are divided based on the rest-frame frequency of the low-energy (synchrotron) hump (\nup) into low- (LBL/LSP: \nup~$<10^{14}$~Hz [$<$ 0.41 eV]), intermediate- (IBL/ISP: $10^{14}$~Hz$ ~<$ \nup~$< 10^{15}$~Hz [0.41 -- 4.1 eV)], and high-energy (HBL/HSP: \nup~$> 10^{15}$~Hz [$>$ 4.1 eV]) peaked sources respectively \citep{padgio95,Abdo_2010}. Extreme blazars are characterized by 
\nup ~$> 2.4 \times 10^{17}$~Hz [$>$ 1 keV] \citep{biteau2020}.} \nup
$\gtrsim 5\times 10^{17}$ Hz, $z=0.557$, and well within the neutrino event error region of IceCube-200107A, that exhibited a very hard X-ray flare at the time of the IceCube event \citep{paiano2020,giommi2020b}. 
The case for some blazars being neutrino sources is mounting \citep[see e.g.][]{righi2019}.

Blazars are a rare type of Active Galactic Nuclei \citep[AGN; see][for reviews]{falomo2014, Padovani_2017} characterised by strong and highly variable radiation across the entire electromagnetic spectrum. This is produced by energetic charged particles moving in a magnetic field within a relativistic jet that is seen at a small angle with respect to the line of sight \citep{UP95,Padovani_2017}.

 From an optical spectroscopy point of view, blazars have always been divided in two 
classes, i.e., flat-spectrum radio quasars (FSRQs) and BL Lac objects (henceforth, BLL). FSRQs 
display strong, quasar-like broad emission lines while BLL sometimes exhibit only 
absorption features, in some cases show at most weak (a few \AA) emission lines, and often 
are totally featureless \citep{UP95}. Note that most FSRQs are LBLs, with a  
small number of them being IBLs. 

Several studies reported hints of a correlation between blazars and the arrival direction of astrophysical neutrinos (e.g. \citealt{Padovani_2016,Lucarelli_2019} and references therein). 
Very recently some of the authors of this paper have extended the detailed dissection of the region around the IceCube-170922A event related to TXS\,0506+056 carried out by \cite{padovani2018} to all the 70 public IceCube high-energy neutrinos that are well reconstructed (so-called tracks) \citep{giommi2020a}. This provided a $3.2\,\sigma$ (post-trial\footnote{A trial
correction for the ``{\it Look Elsewhere Effect}'' is needed 
\citep[e.g.,][]{Patrignani_2016}. 
This stems from a simple fact: in the ideal case of 20 completely
independent tests, for example, a result more
significant than $\sim 2\sigma$ will be observed simply by chance. The final p-value can in this case be trial-corrected by multiplying it by the number of tests carried out.}) correlation excess of $\gamma$-ray detected IBLs and HBLs with IceCube neutrinos. 
No excess was found for LBLs. 
Given that TXS\,0506+056, as 3HSP J095507.9+355101, is also a blazar of the IBL/HBL type \citep{Padovani_2019} this result, together with previous findings, consistently points to a growing evidence for a connection between some IceCube neutrinos and IBL and HBL blazars. Moreover, this means that out of the 47 IBLs and HBLs in Table 5 of \cite{giommi2020a} $\approx 20$ could be new neutrino sources waiting to be identified 
(see their Table 6). 

To make further progress at this point we need optical 
spectra, which are required to determine the redshift, and hence the luminosity involved, measure the main properties of the spectral lines, and  possibly estimate the mass of the central black hole.

In this paper we present the spectroscopy of a fraction of the 47 objects selected by \cite{giommi2020a}, which together with results taken from the literature cover $\sim 80$ per cent of the sample. 
The paper is the first of a series  of publications arising from the project ``The spectra of IceCube Neutrino (SIN) candidate sources'' whose purpose is to: (1) determine the nature of the sources, e.g.  are they 
EHBLs like 3HSP J0955 or are they 
``masquerading BL Lacs''? \cite{Padovani_2019}, in fact, have shown that, despite appearances, TXS\,0506+056 is {\it not} a blazar of the BL Lac type but instead a masquerading BL Lac, i.e., intrinsically an FSRQ with hidden broad lines. In other words, its extremely weak emission lines are due to a very bright, Doppler-boosted jet continuum, which is washing out the lines. Are they extreme blazars as 3HSP J095507.9+355101? (2) model their SEDs 
 using all available multi-wavelength data, including the spectral slopes and the nucleus-to-host ratios (when measured) derived in this paper, starting with the objects with available redshift 
to determine the physical parameters, which affect the efficiency of neutrino production, and subsequently the expected neutrino emission from each blazar; 
(3) determine the likelihood of a physical connection between the neutrino and the blazar, as some of us did in \cite{Petropoulou_2015} (see also, e.g., \citealt{Petropoulou_2020} for 3HSP J095507.9+355101), and derive important insight on the physical parameters of the blazar jet, such as its proton content. All of the above will provide fundamental information to study the neutrino -- blazar connection.

The present paper is also a continuation of the ongoing systematic study of the properties
of {\it Fermi} BL Lacs, which some of the present authors have carried out in the last decade 
\citep[e.g][]{paiano2017tev, paiano20173fgl, paiano2017ufo1, landoni2018, paiano2019ufo2, 
paiano2020b}. This study has provided  measurements of the redshifts and 
line properties (if detected) of $\gamma$-ray blazars, which are useful to address
several related research questions, namely their evolution, their connection with 
ultra-high-energy CRs, and the EBL and Galactic and extragalactic magnetic fields.

This paper is structured as follows. Section 2 describes the sample we used, while Section 3 deals with the observations and data reduction. Section 4 discusses our methods and results, while Section 5 provides notes on individual sources. 
Finally, Section 6 summarizes our conclusions. 

We use a $\Lambda$CDM cosmology with Hubble constant $H_0 = 70$ km
s$^{-1}$ Mpc$^{-1}$, matter density $\Omega_{\rm m,0} = 0.3$, and dark energy density $\Omega_{\Lambda,0} = 0.7$. 
%Spectral indices are defined by $S_{\nu} \propto \nu^{+\alpha}$ where $S_{\nu}$ is the flux at frequency $\nu$.

%%%%%%%%%%%%%%%%%%%%%%%%%%%%%%%%%%%%%%%%%%%%%%%%%%%%%%%%

\section{The sample} 
\label{sec:sample} 
Our sample includes all 47 IBLs and HBLs in Table 5 of \cite{giommi2020a} plus M87. %E TABELLA 1
These are all $\gamma$-ray sources and have by selection $|b_{\rm II}|>10^{\circ}$ and  \nup~$>10^{14}$~Hz.
These sources are matched to 30 neutrino events out of 70 with  reasonably well-defined IceCube positions (i.e., with an area of the error ellipse smaller than that of a circle with radius r~$= 3^{\circ}$). In various cases more than one blazar is within the IceCube error ellipse; the source list is 
reported in Table~1.

A literature scan shows that for 19 of them a redshift is known (see Table~1) and in three cases it is rather uncertain. 
Of the remaining 28 objects, we performed spectroscopy and present the spectra for 17. 
Combining these observations with the literature results, one has a coverage of 36 out of the 47 objects in \cite{giommi2020a}, which corresponds to $\sim$~80 per cent. 
%\red{PP: Excluding M87 I get 36 [19+17]  sources out of 47, or 77 per cent, with redshift. That's because one source, 3HSPJ140449.6+65543, was missing from Tab. 1.}
The 11 objects left are targets of current ongoing observations by our group  and will be discussed in a future publication.
A journal of the observations of the 17 sources is given in Table~2. 
Additional spectroscopic observations of observed sources of still unknown redshift are planned with different instrument configuration. 
%Note that 7 have been observed with the Gran Telescopio Canarias (GTC) and 10 with the Very Large Telescope (VLT).

We also observed some targets included in a preliminary version of the \cite{giommi2020a}'s list. These are still $\gamma$-ray emitting blazars without a redshift determination, which turned out not to fulfil all the final criteria adopted by those authors, especially in regards to the size of the IceCube error ellipse and the \nup~cut. These seven additional targets are discussed in Appendix~A.

\section{Observations and data reduction} 
\label{sec:obsdata} 

The optical spectra were collected using two different instruments. 
For seven sources in the northern hemisphere we used the 10.4~m Gran Telescopio Canarias (GTC) at the Roque de Los Muchachos (La Palma) with the spectrograph OSIRIS \citep{cepa2003}. For the other 10 targets the 8~m Very Large Telescope (VLT) at Paranal equipped with FORS2 \citep{appenzeller1998} was utilized. 
In the former case the instrument was configured with the grism R1000B (slit width of 1.2") that covers the spectral range 4100~-~7750~$\textrm{\AA}$ at a  resolution R~$\sim$~600. 
In the latter case, grism GRIS\_300V+10 was used (slit width of 1.3") covering 4700~-~8600~$\textrm{\AA}$ at R~$\sim$~200.

Data reduction of the GTC observations was performed adopting standard IRAF procedures \citep{tody1986, tody1993} for long slit spectroscopy following the same scheme given in \citet{paiano2017tev}. 
For FORS2 data we adopted the reduction pipeline provided by the EsoReflex environment \citep{freudling2013} 
including extraction of 1D spectra.

In both cases the accuracy of the wavelength calibration  based on the scatter of the polinomial fit (pixel vs wavelength)  is $\sim$0.1~$\textrm{\AA}$ over the whole observed spectral range. 

In order to perform an optimal correction of CRs and other artifacts, the observation of each source was divided in at least three individual exposures. This procedure allows us also to check for possible spurious features. 
The final spectrum results from the combination of all individual exposures.

Spectro-photometric standard stars were secured for each night in order to perform the relative flux calibration.  
In addition we assessed the absolute flux calibration from the measured magnitude of the source (see Table~3) as derived from the acquisition images obtained during the observation.  For each acquisition image we measure the observed magnitude of a number of stars with known magnitude from SDSS and/or Pan-STARRS catalogues and derive the calibration constant of the frame. In all cases the accuracy is better than 0.1 magnitude. 
We note that for about $\sim$50 per cent of the observed sources we found a difference of $\sim$0.5~-~1 mag between the present observations and previous epochs (see Section 5 and Table~1 and 3).

Finally all spectra were dereddened applying the extinction law of \citet{cardelli1989} and assuming the value of Galactic extinction E(B-V) derived from the NASA/IPAC Infrared Science 
Archive\footnote{http://irsa.ipac.caltech.edu/applications/DUST/}
\citep{schlafly2011}. 

%%%%%%%%%%%%%%%%%%%%%%%%%%%%%%%%%%%%%%%%%%%%%%%%%%%%%%%%%%%%%%%%%%%

\section{Results} 
\label{sec:results}

All flux calibrated and dereddened spectra of the 17 neutrino blazar candidates are displayed in Fig.~\ref{fig:spectra}. 
They are also available in the online database ZBLLAC\footnote{http://web.oapd.inaf.it/zbllac/} \citep{landoni2020}. 
For each spectrum, we evaluate the signal-to-noise ratio (S/N) in a number of spectral regions and the averaged value is given in Table~3. 

From the spectroscopic point of view, we find that all observed targets but 
one, 4FGL~J0244.7+1316, which has a quasar-like spectrum, can be classified as 
BLL. Their spectra are characterized by a power-law emission and in some {\bf cases} also the signature of the host galaxy stellar population is present.

The spectra were carefully inspected in order to identify emission and/or absorption features. When a feature was found, we checked its reliability verifying that it is present in the individual exposures.
For 9 out of 17 targets, we are able to detect absorption and/or emission lines that allow us to determine a firm redshift, which ranges between 0.09 and 1.6 (see Table~3). 

For 6 objects the optical spectrum is characteristic of BLL where both the non-thermal component and the host galaxy are clearly visible \citep{falomo2014}.  In Table 5 we report the EW of the main absorption features of the host galaxies.
For these targets we have decomposed the observed spectrum into a power law and a template for the host galaxy \citep{mannucci2001}.  The decomposition was obtained from a best fit of the two components with free parameters (nucleus-to-host ratio, spectral slope of the non thermal component).
In all above cases the decomposition is a good representation of the observed spectrum (see Figure~\ref{fig:decomposition}). 
We find that the ratio (evaluated at 6500 \AA)  of these two components, the nucleus-to-host ratio (N/H), ranges from 0.3 to 5 (see also Section~5).

It is of interest to search for narrow emission lines as signature of recent star formation and/or nuclear activity in the BLL \citep{bressan2006, paiano2017tev}. 
We measure the equivalenth width (EW) of these features by integrating the flux of the line above the local linear continuum and estimated the dominant error by assuming slightly different continua (based on the 1$\sigma$ level difference from  the assumed continuum). 
In 5 objects we detected weak and narrow emission lines due to [O~II] (EW~=~0.6 - 1.0~$\textrm{\AA}$), [O~III] (EW~=~0.4~-~1.7~$\textrm{\AA}$), [N~II] (EW$\sim$1.3~$\textrm{\AA}$) and [S~II] and [Ne~V].
For 3 sources of known redshift no emission lines of [O~II] and [O~III] are detected and we set an upper limit of the line luminosity based on the estimated minimum Equivalent Width (EW) at the expected wavelength of the line.
Three sources exhibit broad emission lines due to Mg~II, C~III] and C~II]. 
The properties of these emission lines are summarized in Table~4. For
4FGL~J2227.9+0036 only absorption systems due to an intervening medium are observed and therefore only a lower limit to the redshift can be derived.

Finally for the remaining seven sources the optical spectrum appears featureless and well described by a power law continuum. For these objects, we evaluate the minimum detectable EW from which a lower limit of the redshift (0.25~-~0.7, see also Table~3) is derived following the procedure described in \citet{paiano2017tev}. 
Briefly it is assumed that the optical spectrum is the superposition of a non-thermal component described by a power law (F$_\lambda \sim \lambda^\alpha$) and the starlight component of an elliptical host galaxy of M(R)=-22.9 \citep{sbarufatti2005imaging}. 
Under the above assumption, the visibility of the stellar features depend on the nucleus-to-host ratio and therefore from the observed magnitude of the source and the minimum EW of the spectrum a lower limit to the redshift can be estimated.
Since the distribution of the absolute magnitude of the host galaxies of BL Lacs has a width of $\sim$ 1 magnitude 
(see \cite{sbarufatti2005imaging}) the redshift lower limits given in Table represent the most probable value. If the host galaxy of a source is 0.5 mag lower or higher than the assumed average value also the redshift limit will be greater or smaller by $\sim$ 0.05-0.1 (depending on the redshift) respectively. 

%%%%%%%%%%%%%%%%%%%%%%%%%%%%%%%%%%%%%%%%%%%%%%%%%%%%%%%%%%%%%%%%%%%

\section{Notes on individual sources} 
\label{sec:notes}

\begin{itemize}
\item[] \textbf{4FGL~J0224.2+1616}: % ---------------------------------------------
We obtained a high quality S/N ($\sim$70) spectrum of the object (g~=~19.8). 
The spectrum is characterized by a featureless continuum that is described by a power law ($\alpha\sim$-0.80) taking into account a significant reddening (E(B-V)=0.20). 
On the basis of the minimum EW ($\sim$0.4~$\textrm{\AA}$) a lower limit of the redshift z~$>$~0.5 can be set due to the lack of absorption by the starlight of the host galaxy.
Under the above condition, in the spectral range from 5500~-~6500~$\textrm{\AA}$ we can set an upper limit to the flux for possible narrow emission lines  $\lesssim$~5$\times$10$^{-17}$~erg~cm$^{-2}$~s$^{-1}$.

\item[] \textbf{4FGL~J0239.5+1326}: % ---------------------------------------------
 We secured a spectrum (S/N$\sim$45) of the object (with a magnitude g~=~20.6), characterized by a featureless continuum and described by a power law ($\alpha \sim$-0.55) considering the reddening with E(B-V)=0.10. 
The upper limit on the EW for the lines is 0.5~$\textrm{\AA}$ and from the non detection of absorption lines from the host galaxy, we set a lower limit of the redshift of z~$>$~0.7. 
This condition implies a line flux $\lesssim$~2.0$\times$10$^{-17}$~erg~cm$^{-2}$~s$^{-1}$ for possible narrow emission lines at the spectral range 5500~-~6500~$\textrm{\AA}$.

\item[] \textbf{4FGL~J0244.7+1316}: % ---------------------------------------------
Our optical spectrum exhibits a prominent (EW$\sim$124~$\textrm{\AA}$, FWHM$\sim$74~$\textrm{\AA}$) emission line at 5557~$\textrm{\AA}$ identified as Mg~II~2800 at z~=~0.9846. 
Two additional narrow emission features attributed to [Ne~V] and [O~II] are clearly detected (see Table~4 and Fig.~\ref{fig:closeup}) at the same redshift.
The object has a clear quasar-like appearance.

\item[] \textbf{4FGL~J0344.4+3432}: % ---------------------------------------------
Our  spectrum of the target (g~=~19.7) is featureless and described by a power law ($\alpha \sim$ -0.90) with a minimum detectable EW$\sim$0.9~$\textrm{\AA}$. This allows us to set a lower limit to the redshft  z~$>$~0.25 considering an elliptical host galaxy for the BLL.
Under the above assumption, we set an upper limit line flux of  $\lesssim$~1.3$\times$10$^{-16}$~erg~cm$^{-2}$~s$^{-1}$ for possible narrow emission lines in the spectral range 5500~-~6500~$\textrm{\AA}$.

\item[] \textbf{4FGL~J0525.6-2008}: % ---------------------------------------------
The optical spectrum shows the typical absorption features (G-band, H$_{\beta}$, Mg~I, Ca+Fe, Na~I) of the old stellar population characteristic of elliptical galaxies at z~=~0.0913. 
Indeed from the images (see e.g. the Pan-STARRS images), the target appears as an elliptical galaxy of g~=~17.1  (from aperture photometry of radius 2.7 arcsec).  
In addition to the absorption features, we also detect weak emission lines of [O~III]~5007 (see Fig.~\ref{fig:closeup}), [N~II]~6548,6584, and [S~II]~6716-6731 (see Table~4) at the same redshift. 
From the spectral decomposition (see Fig.~\ref{fig:decomposition}), we derive a nucleus-to-host ratio N/H$\sim$4.5 and $\alpha$~=~-1.0 for the nuclear power law component.

\item[] \textbf{4FGL~J0649.5-3139}: % ---------------------------------------------
From the optical spectrum obtained by \citet{pena2017} (and reported in the ZBLLAC database), an absorption feature at $\sim$4400~$\textrm{\AA}$ was found. The identification with Mg~II from intervening absorption sets a lower limit to the redshift z$\geq$0.563. 
We obtain a better quality (S/N$\sim$210) optical spectrum to search for possible emission lines in order to determine a firm redshift. Our spectrum is featureless and exhibits a continuum power law ($\alpha$ $\sim$-1.15). 
No emission lines are present with EW$>$0.20~$\textrm{\AA}$ corresponding to an upper limit of line flux of $\lesssim$~9$\times$10$^{-18}$~erg~cm$^{-2}$~s$^{-1}$ for possible narrow emission lines in the spectral range 5500~-~6500~$\textrm{\AA}$. We cannot confirm the absorption feature at $\sim$4400~$\textrm{\AA}$ because it is outside our spectral range.

\item[] \textbf{4FGL~J0854.0+2753}: % ---------------------------------------------
A noisy SDSS spectrum suggests a redshift z~=~0.4937. 
We obtained a moderate S/N spectrum for which we clearly detect the absorption features of Ca~II (at 5873, 5925~$\textrm{\AA}$) and G-band (at 6426~$\textrm{\AA}$), obtaining z~=~0.4930.

\item[] \textbf{4FGL~J1040.5+0617}: %   ---------------------------------------------
The tentative redshift z~=~0.735 was proposed by \citet{maselli2015} on the basis of a single weak emission line from a SDSS spectrum.
We obtained a high quality S/N$\sim$120 spectrum that allows us to clearly detect a broad emission line (EW~=~4.69$\pm$0.5 $\textrm{\AA}$) attributed to Mg~II~2800 at 4874~$\textrm{\AA}$ and a narrow line due to [O~II]~3727 at 6468~$\textrm{\AA}$ (see the close up in Fig.~\ref{fig:closeup}). 
The redshift is z~=~0.740.

\item[] \textbf{4FGL~J1043.6+0654}: % ---------------------------------------------
We obtain a high quality S/N ($\sim$150) of the source (g~=~19.4). 
The spectrum is characterized by a featureless continuum and well described by a power law with $\alpha$~=~-1.15. 
No emission or absorption lines are detected down to EW$\sim$0.20$ \textrm{\AA}$ that yields a lower limit of the redshift of z$>$0.7 and a line flux $\lesssim$~1.2$\times$10$^{-17}$~erg~cm$^{-2}$~s$^{-1}$ for possible narrow emission lines at the spectral range 5500~-~6500~$\textrm{\AA}$.

\item[] \textbf{4FGL~J1258.7-0452}: % ------------------------------------------------
In our spectrum the Ca~II absorption doublet is apparent at 5576, 5627~$\textrm{\AA}$ (see Fig. \ref{fig:closeup}) and, together with the G-band absorption line at 6103~$\textrm{\AA}$, yields z~=~0.4179. 
No emission lines are present with EW$>$0.3~$\textrm{\AA}$. 
This corresponds to an [O~II] and an [O~III] line luminosity  ~$\lesssim$~3.0$\times$10$^{40}$~erg~s$^{-1}$ and ~$\lesssim$~2$\times$10$^{40}$~erg~s$^{-1}$ 
respectively. 
The decomposition of the spectrum (see Fig.~\ref{fig:decomposition}) provides a N/H$\sim$5, and $\alpha$~=~-1.6 for the nuclear power law component.

\item[] \textbf{4FGL~J1300.0+1753}: % % -------------------------------------------
A noisy SDSS spectrum does not display any significant features.
Our spectrum (S/N$\sim$40) is dominated by a power law continuum emission with $\alpha$ $\sim$-1.30. 
No features are detected down to EW~=~0.7~$\textrm{\AA}$, corresponding to a lower limit of z~$>$~0.6 based on the non-detection of the absorption features from the starlight of the elliptical host galaxy.
We can set an upper limit line flux of $\sim$2.5$\times$10$^{-17}$~erg~cm$^{-2}$ for narrow emission lines in the spectral range 5500~-~6500~$\textrm{\AA}$.

\item[] \textbf{4FGL~J1359.1-1152}: % ---------------------------------------------
Our spectrum is characterized by the typical absorption lines (Ca~II, G-band, H$_{\beta}$ Mg~I, Ca+Fe, Na~I and H$_{\alpha}$) due to old stellar population of the host galaxy at the redshift z~=~0.242. 
No emission line are detected down to EW~=~0.6~$\textrm{\AA}$. This corresponds to an [OIII] line luminosity $\lesssim$~2.1$\times$10$^{40}$~erg~s$^{-1}$.
From the decomposition into template galaxy and a non-thermal emission (see Fig.~\ref{fig:decomposition}), we find a N/H$\sim$0.5, and $\alpha$~=~-0.8 for the nuclear power law component.

\item[] \textbf{4FGL~J1440.0-2343}: % ---------------------------------------------
The spectrum is characterized by a non-thermal continuum with the signature of Ca~II doublet and G-band absorption features (see Fig.~\ref{fig:closeup}) from the host galaxy at z~=~0.309. 
In addition we detect emission lines at 4878~$\textrm{\AA}$ (EW~=~1.0~$\textrm{\AA}$) and 6553~$\textrm{\AA}$ (EW~=~0.4~$\textrm{\AA}$), that we identify with [O~II] and [O~III] at the same redshift.
The host galaxy is well detected in the decomposition (see Fig.~\ref{fig:decomposition}) where N/H~=~2.5 and the power law index $\alpha$ $\sim$-1 of the nuclear component are estimated.

\item[] \textbf{4FGL~J1447.0-2657}: % ---------------------------------------------
The spectrum shows the absorption lines due to Ca~II, G-band, and H$_{\beta}$ attributed to the stellar population of the host galaxy at z~=~0.3315. 
No emission lines are detected in the spectrum. 
The upper limit on the EW is $\sim$0.5~$\textrm{\AA}$ that corresponds to an [OII] and [OIII] line flux of ~$\lesssim$~7.8 and 6.5$\times$10$^{-17}$~erg~cm$^{-2}$~s$^{-1}$ and a line luminosity upper limit ~$\lesssim$~2.8 and 2.5$\times$10$^{40}$~erg~s$^{-1}$.
From the decomposition (see Fig.~\ref{fig:decomposition}) the spectrum is characterized by a non-thermal power law continuum with $\alpha$ $\sim$ -1.3 and we find a N/H~=~1.5. 

\item[] \textbf{4FGL~J2223.3+0102}: % ---------------------------------------------
There are two spectra secured by the SDSS survey. In both cases no convincing absorption or emission features are apparent.
We obtained a higher quality spectrum of the target (g$\sim$19.3). The continuum is well described by a power law index of $\alpha$~=~-0.80.
In spite of the much higher S/N ($\sim$150) no spectral features are detected with EW$>$0.15~$\textrm{\AA}$ and this implies an upper limit to the line flux ~$\lesssim$~1.5$\times$10$^{-17}$~erg~cm$^{-2}$~s$^{-1}$ for possible narrow emission lines in the spectral range 5500~-~6500~$\textrm{\AA}$.
From the non detection of the absorption lines of the host galaxy, we set a redshift lower limit of z$>$0.7. 

\item[] \textbf{4FGL~J2227.9+0036}: % ---------------------------------------------
The spectrum is dominated by a non-thermal continuum well described by a power law ($\alpha$~=~-0.70).  
We detect an absorption doublet at 5853, 5868~$\textrm{\AA}$ and EW~=~1.30, 0.80~$\textrm{\AA}$ (see Fig.~\ref{fig:spectra} and Fig.~\ref{fig:closeup}) that corresponds to a Mg~II intervening absorption system at z$=$1.0935. 
This feature is also present in a medium-quality SDSS spectrum, but it was not correctly identified 
because such features are not searched for by the SDSS pipeline.
Two addition absorption lines are found at 4988~$\textrm{\AA}$ (EW~=~0.70~$\textrm{\AA}$) and 5443~$\textrm{\AA}$ (EW~=~0.30~$\textrm{\AA}$) corresponding to Fe~II (2382,2600) at the same redshift.
Therefore the target is at z$>$1.0935.
No emission lines are detected in the spectrum. Taking into account the redshift lower limit, the [O~II] emission lines can be located at $\gtrsim$~7800~$\textrm{\AA}$ where the minimum EW is 0.5~$\textrm{\AA}$. This corresponds to a line flux  ~$\lesssim$~3.5$\times$10$^{-17}$~erg~cm$^{-2}$~s$^{-1}$.

\item[] \textbf{4FGL~J2326.2+0113}: % ---------------------------------------------
The SDSS spectrum shows some emission lines, which suggest a redshift z$=$1.59. 
We obtained a better S/N ($\sim$120) spectrum that is characterized by a power law continuum ($\alpha$ $\sim$~-0.25). 
We clearly detect emission lines of C~III], C~II], and Mg~II at z=1.597 (see Fig.~\ref{fig:spectra} and Tab.~4).
In addition we also detect an Mg~II intervening absorption system at z$=$1.084.

\end{itemize}

%%%%%%%%%%%%%%%%%%%%%%%%%%%%%%%%%%%%%%%%%%%%%%%%%%%%%%%%%%%%%%%%%%%%%%

\section{Summary and future perspectives} 
\label{sec:summary}

We presented optical spectroscopy of 17 extragalactic sources that are candidate for being the astronomical counterparts of neutrino events. Apart from one target, which has a quasar-like spectrum (4FGL~J0244.7+1316), all of the remaining  objects have a spectral shape that is well consistent with a BLL classification. 
We clearly detected absorption and/or emission features in the spectra of 10 targets.
For 9 of them we are able to determine a firm redshift (0.09~$<$~z~$<$~1.6) from absorption lines due to the old stellar population of their host galaxies and in 5 cases also from the emission lines.
For one source (4FGL~J2227.9+0036) we set a spectroscopic lower limit to the redshift, based on the detection of intervening absorption systems attributed to Mg~II and Fe~II.  
Five  sources show in their optical spectra the clear modulation due to the flux from the host galaxy (see Figure~\ref{fig:decomposition}). 
The closest object (z~=~0.0913) in our sample is 4FGLJ0525.6-2008 and its optical spectrum is fully dominated by the host galaxy. Despite the good quality of the spectra in seven objects we do not detect any spectral feature as is expected for this class of sources. Nevertheless for these we can estimate a lower limit to the redshift from the absence of the absorption lines of their host galaxy. 
Only for five sources the spectrum exhibits emission features (see Table~4). These are forbidden lines attributed to [O~II], [O~III] and [N~II] of luminosity (1~$-$~4)$\times$10$^{41}$, $\sim$3$\times$10$^{40}$, and $\sim$2$\times$10$^{41}$ erg s$^{-1}$ respectively. 
Moreover, semi-forbidden lines of C~III] and C~II] are revealed in the high redshift (z~=~1.595) BLL 4FGL~J2326.2+0113 and a 
broad emission line of  Mg~II 2800 \AA~is found in the spectra of two targets.
Overall the spectral properties of the objects in this study are similar to those of a sample of 55 hard $\gamma$-ray BLL investigated by \citet{paiano2020b} both as regards the continuum shape and redshift range of the objects.
In both samples, when weak [O~III]~5007 emission line is detected, their luminosity is comparable (in the range \textit{L}$_{[\textit{OIII}]}$ = ~(2~-~8)~$\times$10$^{40}$~erg s$^{-1}$).
 These issues will be fully discussed in the second paper of this series.

We will be using the results of this paper to carry out a detailed investigation of the nature of these blazars. 
In particular, following \cite{Padovani_2019} and based on the continuum and line powers, and the overall SEDs, we will be assessing if any of these blazars can be classified as ``masquerading'', as is the case for TXS~0506+056, or extreme BLL as 3HSP~J095507.9+355101. 
Moreover to study in depth the neutrino -- blazar connection with a statistically meaningful sample, a model of their photon and neutrino SEDs using a lepto-hadronic code \citep[e.g.][]{Petropoulou_2020} will be explored.

%%%%%%%%%%%%%%%%%%%%%%%%  TABLES  %%%%%%%%%%%%%%%%%%%%%%%%%%%%%%%

\newpage
\setcounter{table}{0}
\begin{table*}
\begin{center}
\caption{The sample of 47 neutrino candidate blazars. In boldface are the targets considered in this paper.} \label{tab:table1}
\begin{tabular}{llllllllll}
\hline 
Object   &    Counterpart     &  IC event & Sep. & RA    &  DEC  &  Mag.  &  \nup  & z & z  \\   
  Name         &                    &           & (degrees) & (J2000)  &(J2000) &  g         &     & & Ref. \\
\hline
%\hline
4FGL~J0103.5+1526 & 5BZG~J0103+1526 & IC-160331A & 0.32 & 01:03:26 & +15:26:24 & 18.4 & 15.0 & 0.2461 & SDSS\\
\hline
4FGL~J0158.8+0101 & 5BZU~J0158+0101 & IC-090813A & 0.28 & 01:58:52 & +01:01:32 & 21.4 & 14.1 & 0.4537 & ZBLLAC \\
\hline
\bf{4FGL~J0224.2+1616}  & VOU~J022411+161500 & IC-111216A & 2.89 & 02:24:12 &  +16:15:00  & 19.6 & 14.5 & ? & *\\ %VOUJ36.04921+16.250
%\hline
4FGL~J0232.8+2018  & 3HSP~J023248.6+201717 & IC-111216A & 1.91 & 02:32:48 & +20:17:17 &  17.2 & 18.5 & 0.1390 & NED\\ %Firm z is 1ES0229+200
\hline
\bf{4FGL~J0239.5+1326}  & 3HSP~J023927.2+13273 & IC-161103A & 1.27 & 02:39:27 &  +13:27:39 & 21.2 & 15.0 & ? & * \\ %20.2
%\hline
\bf{4FGL~J0244.7+1316}  & CRATES~J024445+132002 & IC-161103A &  0.80 & 02:44:46 &  +13:20:07 & 18.9 & 14.5 & ? & * \\
\hline
4FGL~J0339.2-1736  & 3HSP~J033913.7-17360 & IC-141109A & 1.24 & 03:39:14 & $-$17:36:01 & 16.8 & 15.6 & 0.07 & 6dFGS \\ %PKS0336-177
\hline
\bf{4FGL~J0344.4+3432}  & 3HSP~J034424.9+34301 & IC-150831A & 1.20 & 03:44:25 &  +34:30:17 & 18.9 & 15.7 & ? & * \\
\hline
4FGL~J0509.4+0542      & TXS~0506+056         & IC-170922A & 0.07 & 05:09:26 & +05:42:21 & 15.4 & 14.5 & 0.3365 & ZBLLAC \\
\hline
\bf{4FGL~J0525.6-2008}  & CRATES~J052526-201054 & IC-150428A & 0.86 & 05:25:28 & $-$20:10:48 & 17.6 & 14.5 & ? & * \\
\hline
3FGL~J0627.9-1517       & 3HSP~J062753.3-151957 & IC-170321A & 1.30 & 06:27:53 & $-$15:19:57 & 19.4 & 17.3 & 0.3102 & ZBLLAC \\
\hline
\bf{4FGL~J0649.5-3139}  & 3HSP~J064933.6-31392 & IC-140721A & 0.78 & 06:49:34 & $-$31:39:20 & 19.3 & 17.0 & $>$ 0.563 & ZBLLAC \\ %3FHL~J0649-3138
\hline
\bf{4FGL~J0854.0+2753}  & 3HSP~J085410.1+27542 & IC-150904A & 0.44 & 08:54:10 &  +27:54:22  & 20.1 & 16.1 & 0.4937? & SDSS\\
\hline
4FGL~J0946.2+0104   & 3HSP~J094620.2+010451 & IC-190819A & 2.24 & 09:46:20 & +01:04:52 & 20.0 & $>$18.0 & 0.576 & SDSS \\ %1RXS J094620.5+010459
%\hline
4FGL~J1003.4+0205    & 3HSP~J100326.6+02045 & IC-190819A & 2.35 & 10:03:27 & +02:04:56 & 19.7 & 15.8 & ? & SDSS\\
\hline
\bf{4FGL~J1040.5+0617}  & GB6~J1040+0617       & IC-141209A & 0.37 & 10:40:32 &  +06:17:22  & 20.3 & 14.5 & 0.735? & SDSS \\
%\hline
\bf{4FGL~J1043.6+0654}  & 5BZB~J1043+0653      & IC-141209A & 0.94 & 10:43:24 &  +06:53:10  & 19.8 & 14.5 & ? & * \\
\hline
4FGL~J1055.7-1807  & VOU~J105603-180929 & IC-171015A & 2.6 & 10:56:03 & $-$18:09:30 & 20.5 & 14.1 & ? & * \\ 
\hline
4FGL~J1117.0+2013 & 3HSP~J111706.2+201407 & IC-130408A & 2.02 & 11:17:06 & +20:14:07 & 16.8 & 16.5 & 0.138 & SDSS \\ %5BZB J1117+2014
4FGL~J1124.0+2045 & 3HSP~J112405.3+204553 & IC-130408A & 3.61 & 11:24:05 & +20:45:53 & 18.8 & 15.3 & ? & SDSS \\ %5BZB J1124+2045
4FGL~J1124.9+2143   & 3HSPJ112503.6+21430 & IC-130408A & 3.48 & 11:25:03 & +21:43:00 & 18.5 & 15.8 & ? & SDSS \\ %SDSSJ112503.64+214300
\hline
4FGL~J1230.8+1223   & M87                 & IC-141126A & 0.93  & 12:30:49 & +12:23:30 & 11.6 & - & 0.004 & NED \\
4FGL~J1231.5+1421  & 3HSP~J123123.1+14212 & IC-141126A & 1.06 & 12:31:24 & +14:21:24 & 17.7 & 16.0 & 0.2558 & SDSS \\
\hline
3FGL~J1258.4+2123  & 3HSP~J125821.5+21235 & IC-151017A & 3.2 & 12:58:21 & +21:23:51 &  20.5 & 16.7 & 0.6265 & ZBLLAC\\
\hline
\bf{4FGL~J1258.7-0452}  & 3HSP~J125848.0-04474 & IC-150926A & 0.54 & 12:58:48 & $-$04:47:45 & 19.1 & 17.0 & ? &  *\\  
\hline
\bf{4FGL~J1300.0+1753}  & 3HSP~J130008.5+17553 & IC-151017A & 3.45 & 13:00:09 &  +17:55:38  & 20.1 & 14.5 & ? & SDSS \\
4FGL~J1314.7+2348     & 5BZB~J1314+2348      & IC-151017A & 3.68 & 13:14:44 & +23:48:26 & 17.1 & $\geq$14 & 0.15~? & M14 \\
\hline
4FGL~J1321.9+3219 & 5BZB~J1322+3216      & IC-120515A  & 1.54 & 13:22:47 & +32:16:09  & 19.3 & 14.5 & ? & 5BZCAT \\
\hline
\bf{4FGL~J1359.1-1152}  & VOU~J135921-115043 & IC-120123A & 1.75 & 13:59:21 & $-$11:50:44 & 19.8  & 14.0 & ? & * \\
\hline
4FGL~J1404.8+6554  & 3HSP~J140449.6+65543 & IC-140216A & 1.66 & 14:04:50 &  +65:54:32 &  {19.1} & {16.0}  & {0.362} & {SDSS}\\
\hline
4FGL~J1439.5-2525 & VOU~J143934-252458 & IC-170506A & 1.81 & 14:39:35 & $-$25:24:58 & 19.0 &  14.0 & 0.16 & NED \\
\bf{4FGL~J1440.0-2343}	 & 3HSP~J143959.4-23414 & IC-170506A & 2.83 & 14:39:59 & $-$23:41:40 & 18.9 & 16.2 & ? & * \\
\bf{4FGL~J1447.0-2657}  & 3HSP~J144656.8-26565 & IC-170506A & 0.95 & 14:46:57 & $-$26:56:58 & 19.4 & 17.6 & ? & * \\ 
\hline
4FGL~J1507.3-3710 & VOU~J150720-370902 & IC-181014A & 2.0 & 15:07:21 & $-$37:09:03 & 17.5 & 14.5 & ? & * \\
\hline
4FGL~J1528.4+2004  & 3HSP~J1528+2004      & IC-110521A & 3.23  & 15:28:36 &  +20:04:20 & 20.2 & 16.2  & ? & *  \\
4FGL~J1533.2+1855  & 3HSP~J153311.2+18542 & IC-110521A & 2.47 & 15:33:11 & +18:54:29 & 18.6 & 17.0 & 0.307 & SDSS \\
4FGL~J1554.2+2008  & 3HSP~J155424.1+20112 & IC-110521A & 2.82 & 15:54:24 & +20:11:25 & 18.1 & 17.3 & 0.2223 & SDSS \\ %UFO in 4FGL
\hline
4FGL~J1808.2+3500 & CRATES~J180812+350104    & IC-110610A &  0.55 & 18:08:11 & +35:01:19  &  15.4 &  14.5  & ? & deMe20  \\ %Menezes 0.269:
4FGL~J1808.8+3522 & 3HSP~J180849.7+35204 & IC-110610A & 0.21 & 18:08:50 & +35:20:43 & 15.8 & {15.0} & 0.141 & ZBLLAC\\
\hline
\end{tabular}
\end{center}
\raggedright
\footnotesize \textit{Notes.} Column~1: {\it Fermi} name of the target in the 4FGL (or 3FGL in two cases) catalogue; 
Column~~2: Counterpart name of the 
%(sources in boldface are considered in this work); Column~~2: Counterpart name of the 
$\gamma$-ray source; Column~~3: IceCube track; Column~~4: Angular separation (degrees) between the target and the centroid of the IceCube track; Column~~5~-~6: Right ascension and declination of the optical counterpart; Column~7: g magnitude from SDSS survey and PANSTARRs; Column~8:  Frequency of the synchrotron peak ($\nu_{\textit{peak}}^{\textit{syn}}$); Column~9: Redshift from the literature; Column~10: Reference for the redshift (SDSS: SDSS Data Release 16 (DR16): 
\cite{Ahumada2020}; ZBLLAC: Database of BL Lac objects spectra - http://www.oapd.inaf.it/zbllac/; NED: NASA/IPAC Extragalactic Database; 6dFGS: 6dF Galaxy Survey Database - Final Data Release (DR3): \cite{Jones2009}; M14: \citealt{massaro2014}; deMe20: \citealt{deMenezes2020}).  VOU: \cite{giommi2020a}; 5BZCAT: \cite{BZCAT}; an asterisk  means that no redshift or optical spectra are found in the literature.
\\
%\tablenotetext{}{
%\raggedright
% } 
\end{table*}

\setcounter{table}{0}
\begin{table*}
\begin{center}
\caption{The sample of 47 neutrino candidate blazars - \textit{continued} } %\label{tab:table1}
\begin{tabular}{llllllllll}
\hline 
Object   &    Counterpart     &  IC event & Sep. & RA    &  DEC  &  Mag.  &  \nup  & z & z  \\   
  Name         &                    &           & (degrees) & (J2000)  &(J2000) &  g         &     & & Ref. \\
\hline
4FGL~J2030.5+2235 & 3HSP~J203031.6+22343 & IC-100710A & 1.31 & 20:30:32 & +22:34:39 & 20.3 & 16.2 & ? & * \\ %UFO in 4FGL
4FGL~J2030.9+1935 & 3HSP~J203057.1+19361 & IC-100710A & 1.82 & 20:30:57 &  +19:36:12 & 18.6 & 15.8 & ? & *\\
\hline
4FGL~J2133.1+2529c & 3HSPJ213314.3+25285 & IC-150714A & 2.18 & 21:33:14 & +25:28:59 & 18.8 & 15.2 & 0.294 & 5BZCAT \\
\hline
\bf{4FGL~J2223.3+0102}  & 3HSP~J222329.5+01022 & IC-140114A & 1.33 & 22:23:30 &  +01:02:26  & 18.6 & 15.5 & ? & SDSS \\ 
\bf{4FGL~J2227.9+0036}  & 5BZB~J2227+0037      & IC-140114A & 0.64 & 22:27:58 &  +00:37:06  & 18.4 & 14.5 & ? &  SDSS \\
\hline
\bf{4FGL~J2326.2+0113}  & CRATES~J232625+011147 & IC-160510A & 1.31 & 23:26:26 &  +01:12:09  & 20.5 & 14.0  & ? & SDSS \\
\hline
4FGL~J2350.6-3005 & 3HSPJ235034.3-30060 & IC-190104A & 3.32 & 23:50:34 & $-$30:06:03 & 18.1 & 15.7 & 0.2328 & 6dFGS \\
4FGL~J2351.4-2818  & IC~5362 & IC-190104A & 1.07 & 23:51:36 & $-$28:21:53 & 16.3 & 14.5 & 0.0276 & NED \\
4FGL~J2358.1-2853  & CRATES~J235815-285341 & IC-190104A & 2.47 & 23:58:17 & $-$28:53:34 & 19.5 & 14.0 & ? & * \\
\hline
\end{tabular}
\end{center}
\raggedright
%\footnotesize - Continued -  
%\tablenotetext{}{
%\raggedright
% } 
\end{table*}

\setcounter{table}{1}
\begin{table*}
\begin{center}
\caption{The sample of 17 neutrino candidate blazars and journal of the observations.}\label{tab:table2}
%\centering
\begin{tabular}{llllllll}
\hline 
Object Name  & E(B-V) & Telescope & Instrument & Date & t$_{Exp}$ & seeing  & Air Mass \\   
             &        &            &      & (s)       &   (")  &  \\
\hline
4FGL~J0224.2+1616  &  0.20 & GTC & OSIRIS & 22 December 2019  & 7200 & 1.7 & 1.30 \\
4FGL~J0239.5+1326  &  0.10 & GTC & OSIRIS &05 October  2019  & 7200 & 1.8  & 1.05 \\
4FGL~J0244.7+1316  &  0.09 & GTC & OSIRIS &09 October  2019  & 4500 & 0.7  & 1.04 \\
4FGL~J0344.4+3432  &  0.27 & GTC & OSIRIS &10 October  2019  & 3000 & 0.7  & 1.47\\
4FGL~J0525.6-2008  &  0.04 & VLT & FORS2 & 23 October  2019  & 2400 & 0.8  & 1.06 \\
4FGL~J0649.5-3139  &  0.10 & VLT & FORS2 &23 October  2019  & 5400 & 1.0  & 1.07\\ 
4FGL~J0854.0+2753  &  0.03 & GTC & OSIRIS &21 October  2019  & 6600 & 1.4  & 1.25\\
4FGL~J1040.5+0617  &  0.02 & GTC & OSIRIS &15 March    2019  & 7200 & 1.6  & 1.22\\ 
4FGL~J1043.6+0654  &  0.03 & VLT & FORS2 &26 December 2019  & 5400 & 0.5  & 1.35 \\
4FGL~J1258.7-0452  &  0.02 & VLT & FORS2 &18 February 2020  & 5400 & 0.5  & 1.15 \\
4FGL~J1300.0+1753  &  0.03 & GTC & OSIRIS &02 April    2019  & 9000 & 1.7  & 1.31\\
4FGL~J1359.1-1152  &  0.08 & VLT & FORS2 &28 February 2020  & 2700 & 2.0  & 1.14\\
4FGL~J1440.0-2343  &  0.10 & VLT & FORS2 &24 March    2020  & 2700 & 1.9  & 1.08\\
4FGL~J1447.0-2657  &  0.11 & VLT & FORS2 &24 March    2020  & 5400 & 1.3  & 1.00\\
4FGL~J2223.3+0102  &  0.06 & VLT & FORS2 &04 October  2019  & 5400 & 0.9  & 1.14 \\
4FGL~J2227.9+0036  &  0.05 & VLT & FORS2 &03 October  2019  & 2700 & 0.8  & 1.35 \\
4FGL~J2326.2+0113  &  0.03 & VLT & FORS2 &03 October  2019  & 5400 & 0.7  & 1.17 \\
\hline
\end{tabular}
\end{center}
\raggedright
\footnotesize \textit{Notes}. Column~1: Name of the target ; Column~2: E(B-V) taken from the NASA/IPAC Infrared Science Archive (https://irsa.ipac.caltech.edu/applications/DUST/);  Column~3: Telescope used for the observation; Column~4: Date of the observation; Column~5: Total integration time (sec); Column~6: Average seeing during the observation (arcsec), Column~6: Air mass during the observation.\\
%\tablenotetext{}{
%\raggedright
% } 
\end{table*}

% TABLE  - start

\setcounter{table}{2}
\begin{table*}
\begin{center}
\caption{Properties of the optical spectra of the 17 neutrino candidate blazars studied in this work.}\label{tab:table3}
%\centering
%\begin{tabular}{llcccccccll}
%{\bf THIS TABLE IS UNDER REVISION BY RF}
\begin{tabular}{llcllll}
\hline
Object Name      &  g  & S/N  &   EW$_{min}$  & z &  Line type & $\alpha$  \\  
                 &     &      &   ($\textrm{\AA}$) &  &  &   \\  
\hline
4FGL~J0224.2+1616  &  19.8 $\pm$ 0.1 & 70  & 0.40  & $>$0.5       & h  & 0.80 $\pm$ 0.2  \\
4FGL~J0239.5+1326  &  20.6 $\pm$ 0.2 & 45  & 0.50  & $>$0.7       & h  & 0.55 $\pm$ 0.15 \\
4FGL~J0244.7+1316  &  18.9 $\pm$ 0.1 & 110 &   -   & 0.9846 $\pm$ 0.0005   & e  &   1.0 $\pm$ 0.2 \\
4FGL~J0344.4+3432  &  19.7 $\pm$ 0.2 & 45  & 0.55  & $>$0.25       & h & 0.90 $\pm$ 0.2   \\
4FGL~J0525.6-2008  &  17.6 $\pm$ 0.1 & 200 &   -   & 0.0913  $\pm$ 0.0005  & e,g & 1.00 $\pm$ 0.2\\ %16.7*
4FGL~J0649.5-3139  &  20.3 $\pm$ 0.2 & 210 & 0.20  & $>$0.7        & h  & 1.15 $\pm$ 0.2  \\ %19.9
4FGL~J0854.0+2753  &  19.9 $\pm$ 0.1 & 15  &   -   & 0.4930 $\pm$ 0.0005   & g & 0.80 $\pm$ 0.2 \\
4FGL~J1040.5+0617  &  19.4 $\pm$ 0.1 & 120 &   -   & 0.740  $\pm$ 0.001   & e  & 0.75 $\pm$ 0.15 \\ 
4FGL~J1043.6+0654  &  19.7 $\pm$ 0.1 & 150 & 0.20  & $>$0.7    & h  & 1.15 $\pm$ 0.2     \\ %19.4
4FGL~J1258.7-0452  &  18.8 $\pm$ 0.1 & 170 &   -   & 0.4179 $\pm$ 0.0006   & g  & 1.6$\pm$ 0.3 \\ %18.4
4FGL~J1300.0+1753  &  20.1 $\pm$ 0.2 & 40  & 0.7   & $>$0.6    & h    & 1.3 $\pm$ 0.3       \\
4FGL~J1359.1-1152  &  19.4 $\pm$ 0.1 & 65  &   -   & 0.242 $\pm$ 0.001    & g  & 0.80 $\pm$ 0.1 \\ %18.1
4FGL~J1440.0-2343  &  18.6 $\pm$ 0.1 & 90  &   -   & 0.309 $\pm$ 0.001   & e,g & 1.0 $\pm$ 0.2        \\ %17.9
4FGL~J1447.0-2657  &  19.4 $\pm$ 0.1 & 80  &   -   & 0.3315 $\pm$ 0.0004   & g  & 1.30 $\pm$ 0.2 \\ %18.6
4FGL~J2223.3+0102  &  19.3 $\pm$ 0.1 & 150 & 0.15  & $>$0.7            & h & 0.8 $\pm$ 0.10  \\ %18.9
4FGL~J2227.9+0036  &  19.3 $\pm$ 0.1 & 105 &   -   & >1.0935   & i   & 0.7 $\pm$ 0.10         \\ %18.9
4FGL~J2326.2+0113  &  19.9 $\pm$ 0.2 & 120 &   -   & 1.595  $\pm$ 0.001   & e,i & 0.25 $\pm$ 0.05 \\ %19.4
\hline
\end{tabular}
\end{center}
\raggedright
\footnotesize \textit{Notes}. Column~1: Name of the target; Column~2: Magnitude (g) measured from the acquisition image; Column~3: Median S/N of the spectrum; Column~4: Minimum equivalent width (EW$_{min}$) derived in the 5500 - 6500 $\textrm{\AA}$ range (the measure is provided only in case of featureless spectrum); Column~5: Redshift.  The error is evaluated as combination of the uncertanity of the centroid of the features with the overall accuracy of the wavelength calibration; Column~6: Type of detected line to estimate the redshift: \textit{e} = emission line, \textit{g} = galaxy absorption line, \textit{i}= intervening absorption assuming Mg~II 2800~$\textrm{\AA}$ identification, \textit{h}= lower limit derived on the lack of detection of host galaxy absorption lines assuming a BLL elliptical host galaxy with M(R) = -22.9. Since the distribution of BLL host galaxies has a dispersion (1$\sigma$) of $\sim$ 0.5 mag, these limits may be change by 0.05-0.1 depending on the redshift limit \citep[see details in][]{paiano2017tev}. Column-7: Spectral index $\alpha$of the continuum described by a power law F$_\lambda \sim \lambda^{-\alpha}$ \\
%\tablenotetext{}{
%\raggedright
% } 
\end{table*}
%------------------------------------------------------------------------

% Table 4 ---------------------------

\setcounter{table}{3}
\begin{table*}
\begin{center}
\caption{Properties of the emission lines.}\label{tab:table4}
%\centering
%\begin{tabular}{llcccccccll}
\begin{tabular}{llllll}
\hline
Object Name  &   z  & $\lambda$  &   EW  &   Line ID &  \textit{L} (line)   \\ 
             &             &  ($\textrm{\AA}$) &     ($\textrm{\AA}$) &    & (erg s$^{-1}$)  \\ 
\hline
4FGL~J0244.7+1316  & 0.9846 & 5557  & 124 $\pm$ 3.0   & Mg~II~2800        & 1.5$\times$10$^{44}$ \\
                   &        & 6795  & 0.65 $\pm$ 0.2   & [Ne~V]~3426       & 3.8$\times$10$^{41}$  \\  %0.9835
                   &        & 7396  & 0.90 $\pm$ 0.2  & [O~II]~3727       & 4.2$\times$10$^{41}$  \\
4FGL~J0525.6-2008  & 0.0913 & 5464  & 1.70 $\pm$ 0.3  & [O III]~5007      & 2.2$\times$10$^{40}$  \\ 
                   %&       & 7146  & 1.30 $\pm$ 0.3  & [N~II]~6548       & 1.9$\times$10$^{40}$  \\
                   &        & 7184  & 1.35 $\pm$ 0.3  & [N~II]~6584       & 1.9$\times$10$^{40}$  \\
                   &        & 7340  & 1.20 $\pm$ 0.2  & [SII]~6716-6731   & 1.6$\times$10$^{40}$  \\
%4FGL~J0854.0+2753 & 0.4930  -  This is too noisy !!!    
4FGL~J1040.5+0617  & 0.740  & 4874  & 4.69 $\pm$ 0.5  & Mg~II~2800        & 1.1$\times$10$^{42}$  \\ 
                   &        & 6468  & 0.58 $\pm$ 0.2  & [O~II]~3727       & 1.1$\times$10$^{41}$  \\ %0.7355
4FGL~J1258.7-0452  & 0.4179 & 5284  & $<$0.3 & [O~II]~3727       & $\lesssim$~3.0$\times$10$^{40}$  \\ 
                   &        & 7099  & $<$0.3 & [O~III]~5007      & $\lesssim$~2.0$\times$10$^{40}$  \\ 
4FGL~J1359.1-1152  & 0.242  & 6218  & $<$0.6 & [O~III]~5007      & $\lesssim$~2.1$\times$10$^{40}$  \\
4FGL~J1440.0-2343  & 0.309  & 4878  & 0.98 $\pm$ 0.3  & [O~II]~3727       & 7.9$\times$10$^{40}$  \\
                   &        & 6553  & 0.36 $\pm$ 0.2  & [O III]~5007      & 2.8$\times$10$^{40}$  \\
4FGL~J1447.0-2657  & 0.3315 & 4962  & $<$0.6 & [O~II]~3727       & $\lesssim$~2.8$\times$10$^{40}$  \\
                   &        & 6667  & $<$0.5 & [O III]~5007      & $\lesssim$~2.5$\times$10$^{40}$  \\
4FGL~J2326.2+0113  & 1.595  & 4953  & 2.67 $\pm$ 0.6  & C~III]~1909       & 2.6$\times$10$^{42}$  \\ 
                   &        & 6041  & 0.82 $\pm$ 0.3  & C~II]~2326        & 7.4$\times$10$^{41}$  \\ %1.597
                   &        & 7272  & 0.81 $\pm$ 0.3  & Mg~II~2800        & 7.1$\times$10$^{41}$  \\ %1.597
\hline
\end{tabular}
\end{center}
\raggedright
\footnotesize \textit{Notes}. Column~1: Name of the target; Column~2: Redshift; Column~3: Barycenter of the detected line; Column~4: Measured equivalent width of the line; Column~5: Line identification; Column~6: Line luminosity. \\
%\tablenotetext{}{
%\raggedright
% } 
\end{table*}

% Table 5 --------------------------------------------------------------------------------

\setcounter{table}{4}
\begin{table*}
\begin{center}
\caption{Equivalent width of the main host galaxy absorption lines.}\label{tab:table5}
%\centering
%\begin{tabular}{llcccccccll}
\begin{tabular}{llllll}
\hline
Object Name  &   Redshift  & Ca II  &   G-band &   Mg I 5175 &   Na I 5892  \\ 
             &  &  ($\textrm{\AA}$) &  ($\textrm{\AA}$) &  ($\textrm{\AA}$)  & ($\textrm{\AA}$)  \\ 
\hline
4FGL J0525.6-2008  & 0.0913 & *  & 8.0 $\pm$ 1   &  25 $\pm$ 5    &  6 $\pm$ 1 \\
4FGL J0854.0+2753 & 0.4930  & 18 $\pm$ 3   & 6 $\pm$ 2   &  *   &  * \\
4FGL J1258.7-045  & 0.4179  & 2.5 $\pm$ 0.5   & 1 $\pm$ 0.3   &  *   &  * \\
4FGL J1359.1-115  & 0.242  & 15 $\pm$ 1   & 6 $\pm$ 0.5   &  9 $\pm$ 1    &  * \\
4FGL J1440.0-234  & 0.309 & 3.5 $\pm$ 0.5   & 1.3 $\pm$ 0.2   &  3 $\pm$ 0.7    &  * \\
4FGL J1447.0-265   & 0.3315 & 6.5 $\pm$ 0.7   & 2.5 $\pm$ 0.8   &  18 $\pm$ 3    &  * \\
\hline
\end{tabular}
\end{center}
\raggedright
\footnotesize \textit{Notes}.  Column~1: Name of the target; Column~2: Redshift; Column~3: EW of the doublet CaII 3934, 3968 \AA; Column~4: EW of G-band 4304 \AA; Column~5: EW of Mg I 5175 \AA~blend; Column~6: EW of Na I 5892 \AA. 
EW measurements and errors follow the procedure outlined in Section 4 for the emission lines. \\ 
%\tablenotetext{}{
%\raggedright
% } 
\end{table*}

%%%%%%%%%%%%%%%%%%%%%%%%%%% FIGURES %%%%%%%%%%%%%%%%%%%%%%%%%%%%

\clearpage
%\newpage
\setcounter{figure}{0}
\begin{figure*}%[htbp]
\includegraphics[width=0.25\textwidth, angle=-90]{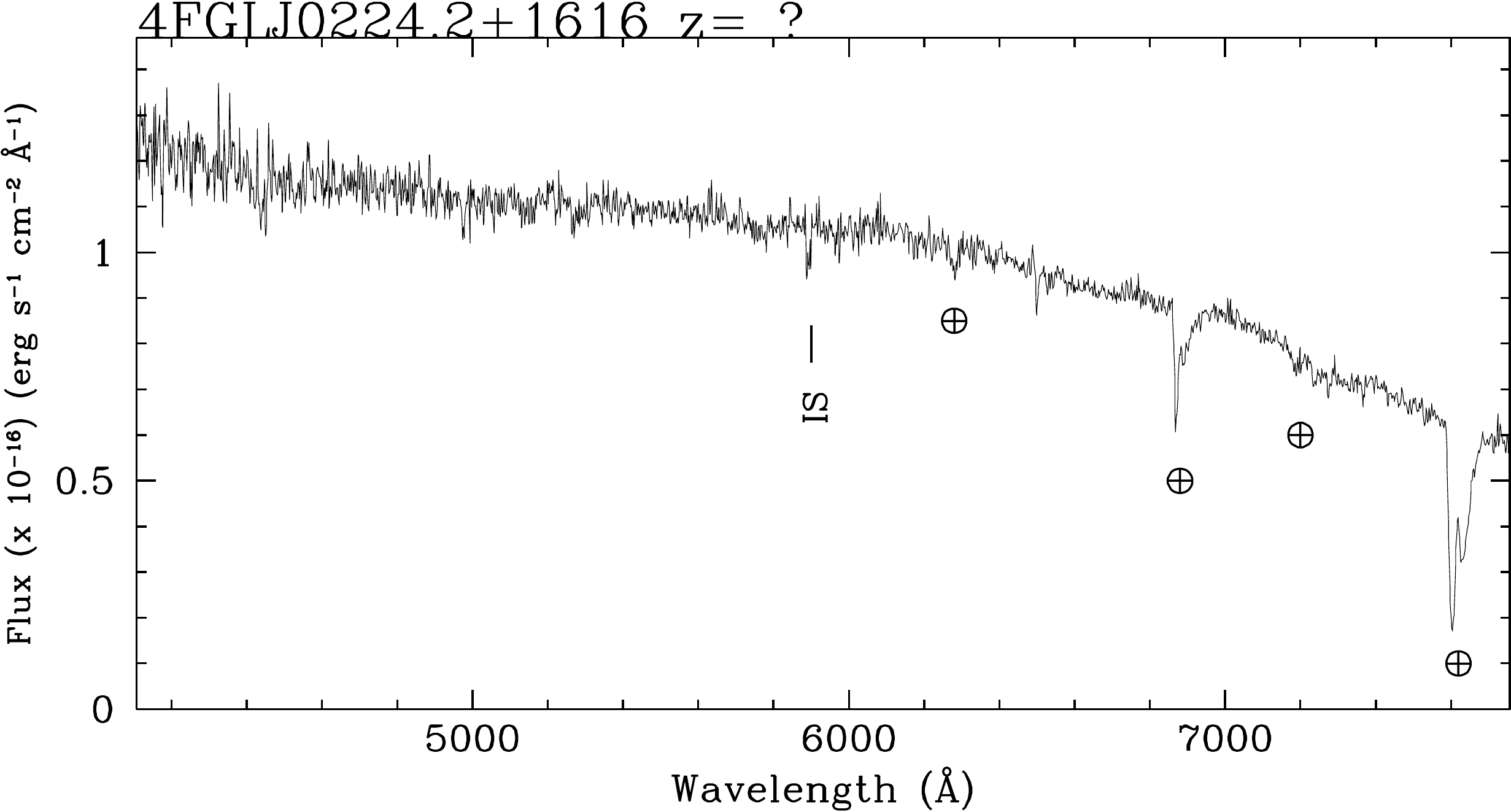}
\includegraphics[width=0.25\textwidth, angle=-90]{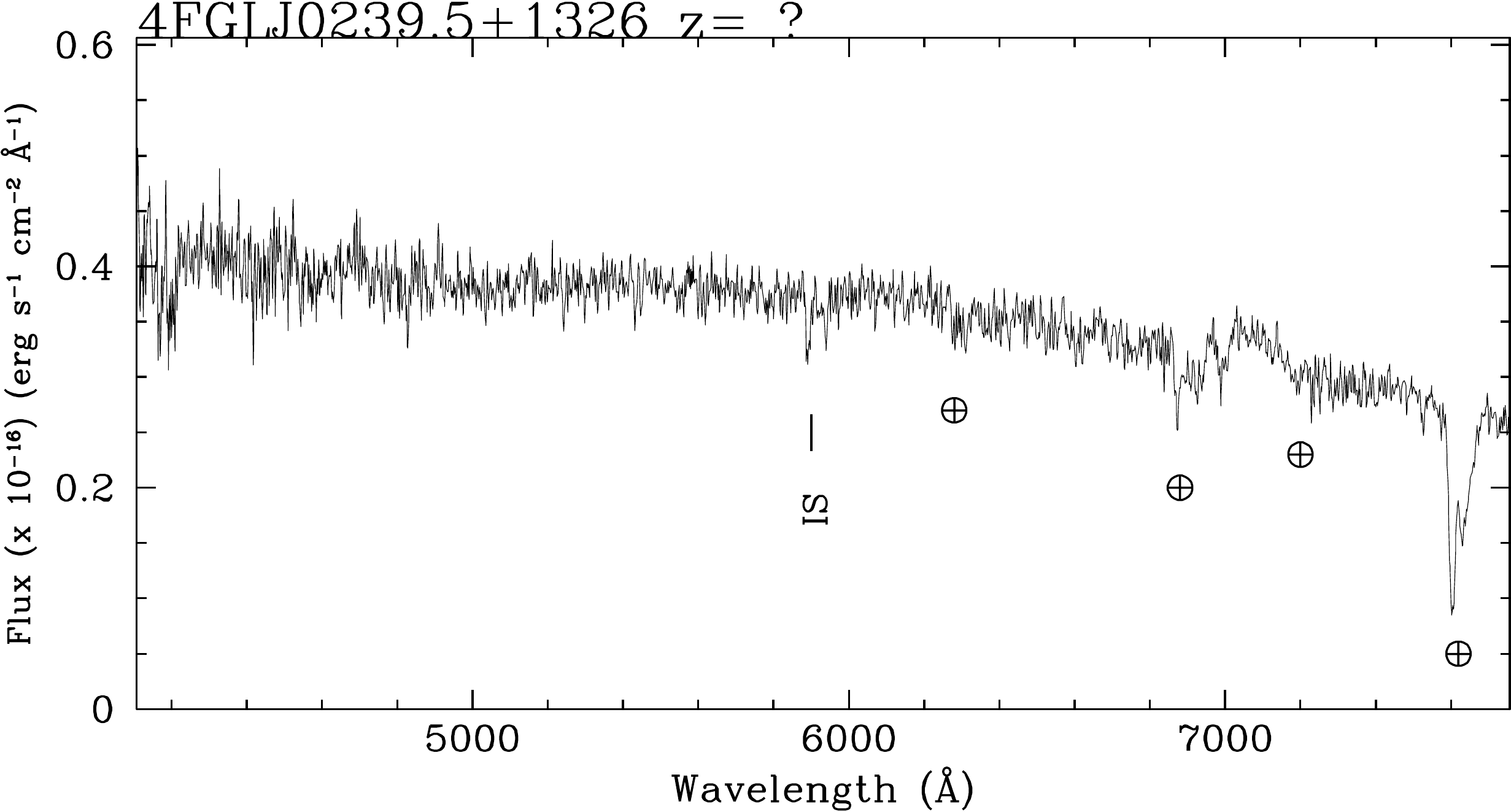}
\includegraphics[width=0.25\textwidth, angle=-90]{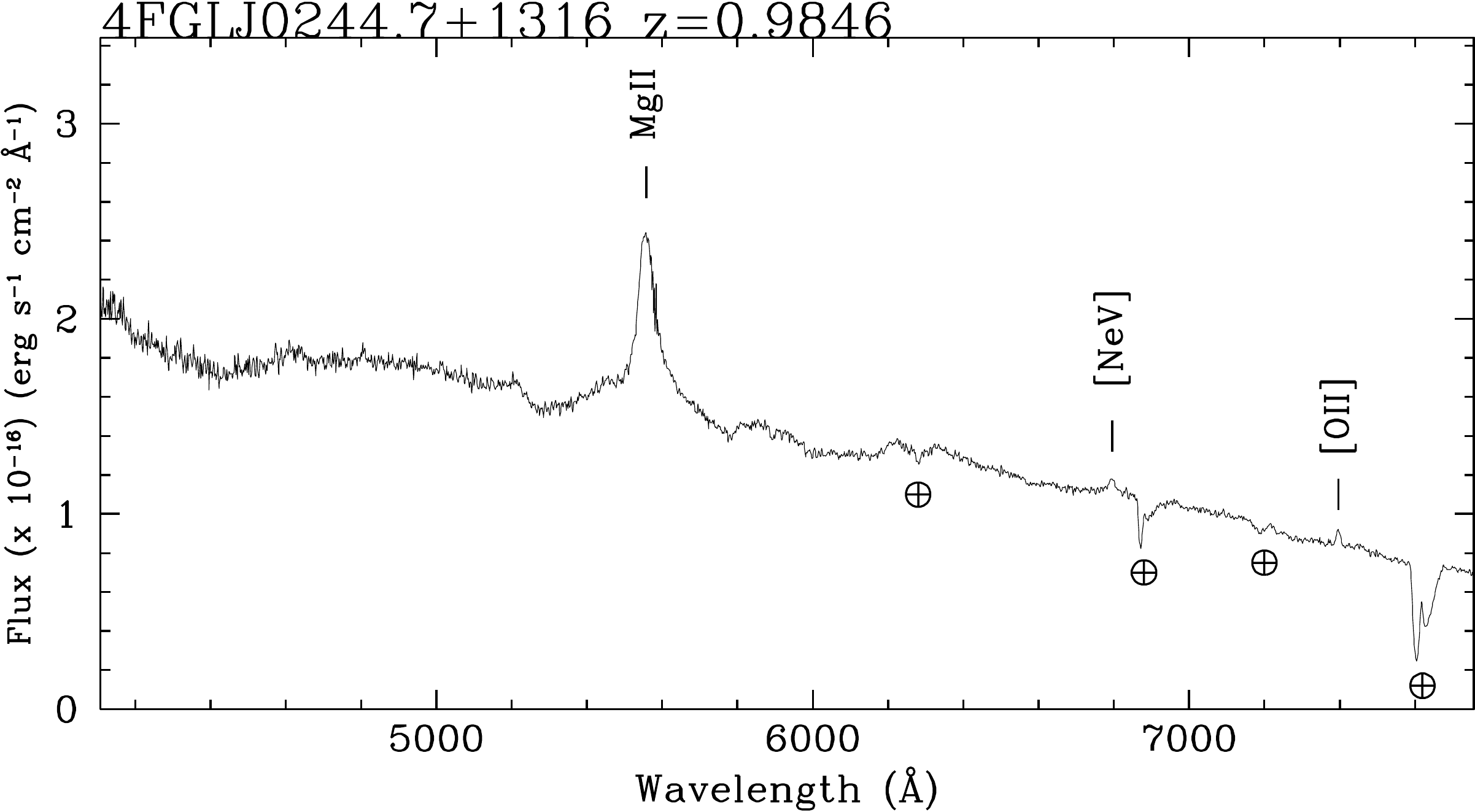}
\includegraphics[width=0.25\textwidth, angle=-90]{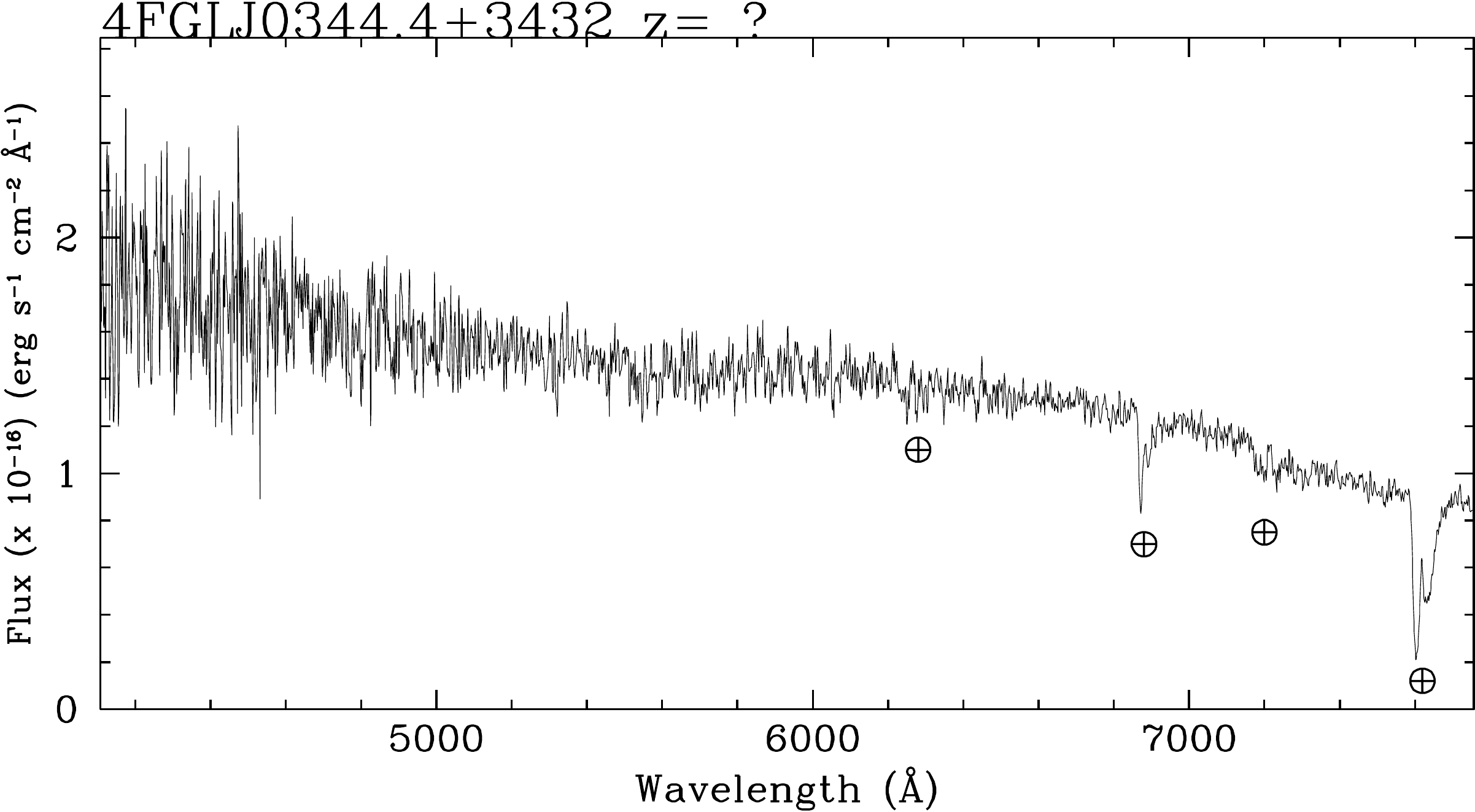}
\includegraphics[width=0.25\textwidth, angle=-90]{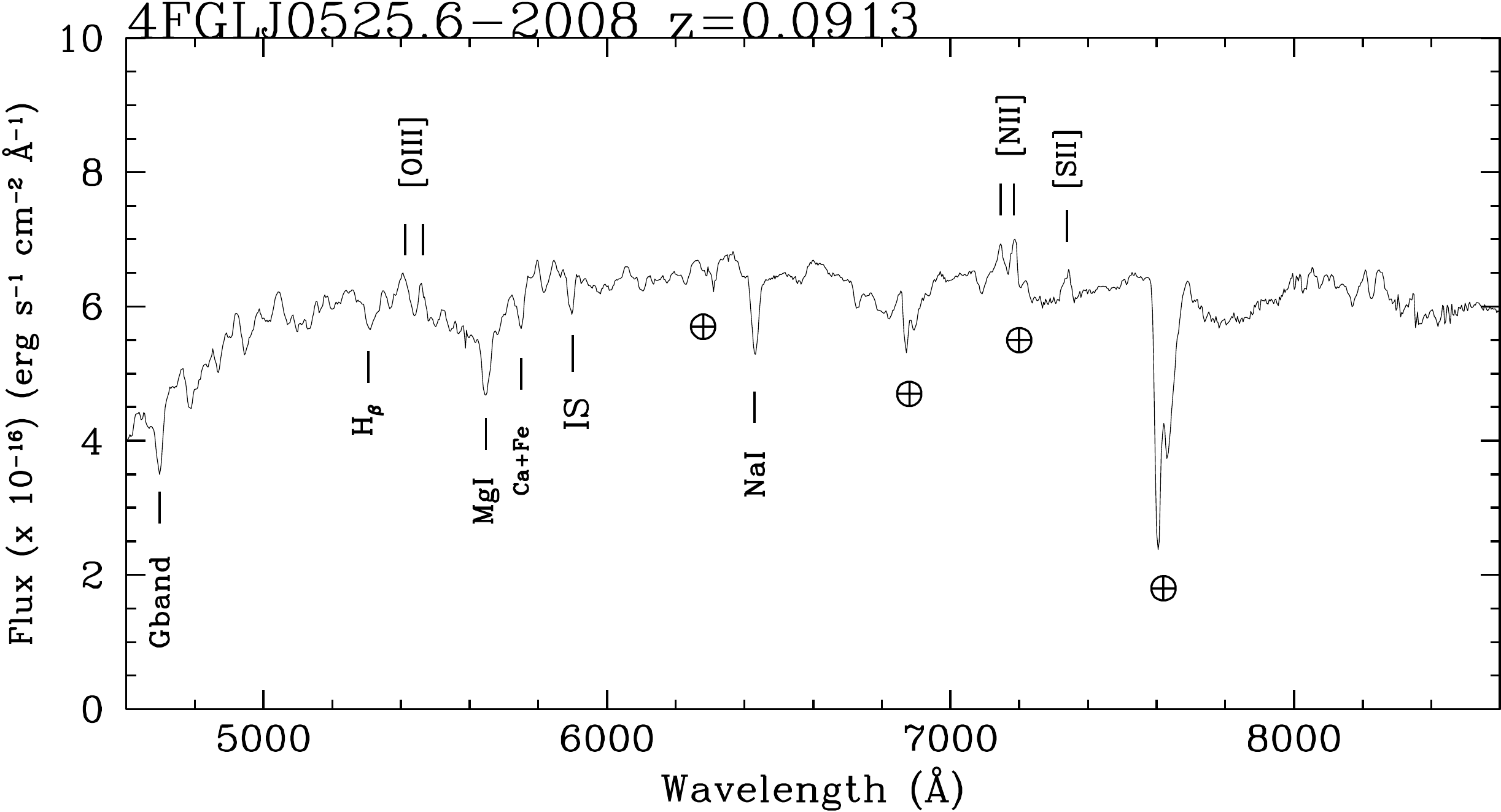}
\includegraphics[width=0.25\textwidth, angle=-90]{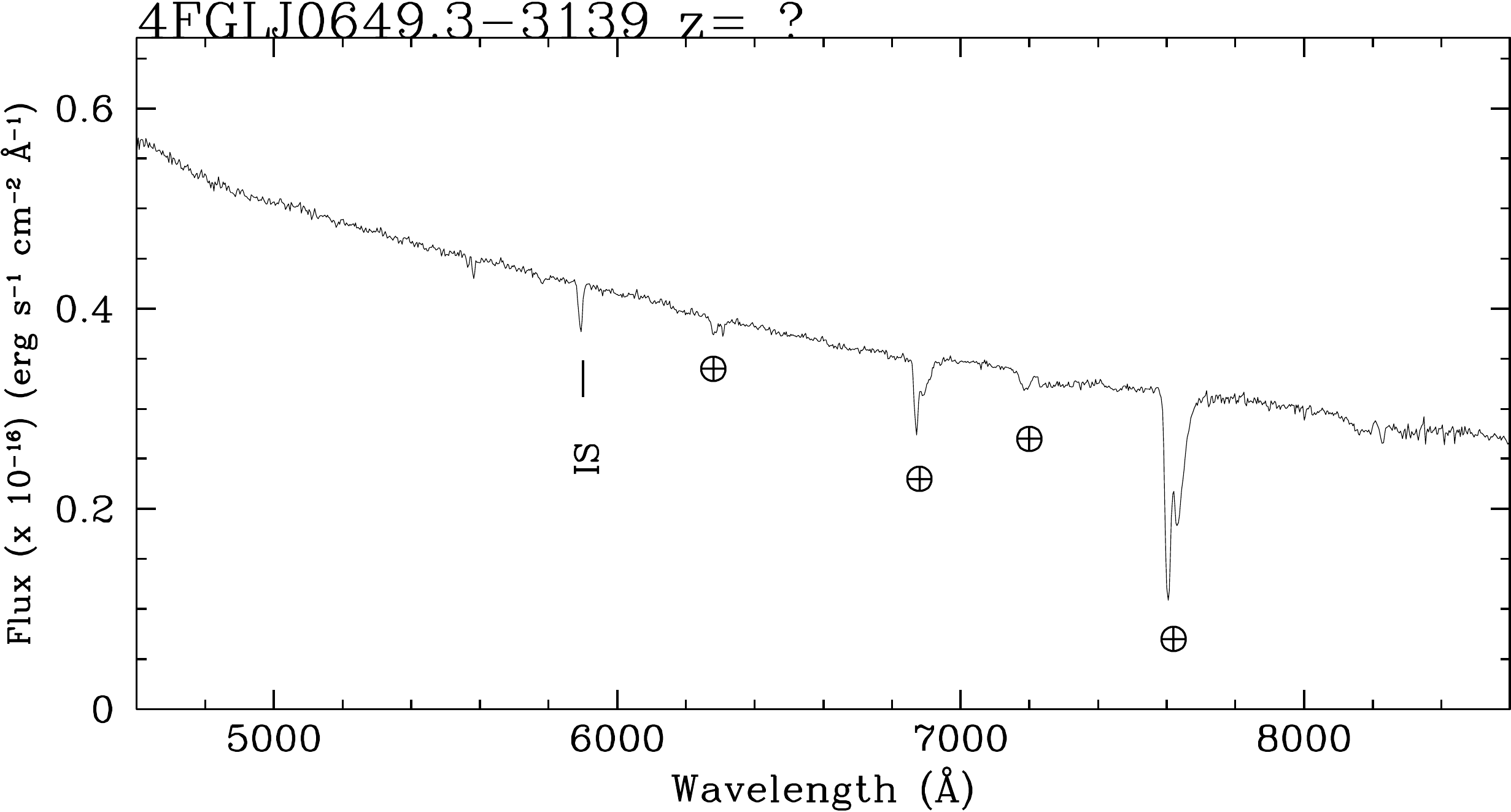}
\includegraphics[width=0.25\textwidth, angle=-90]{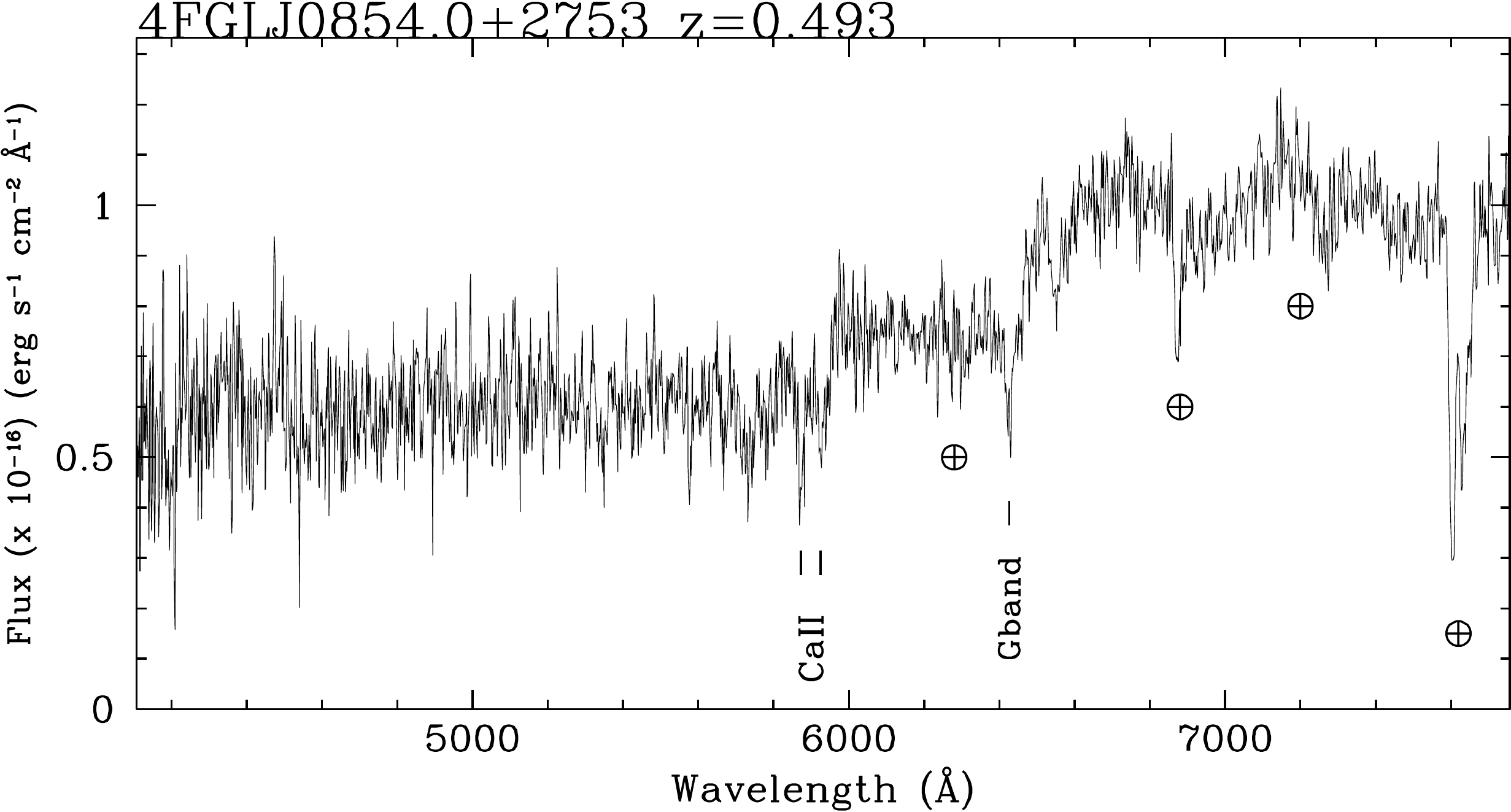}
\includegraphics[width=0.25\textwidth, angle=-90]{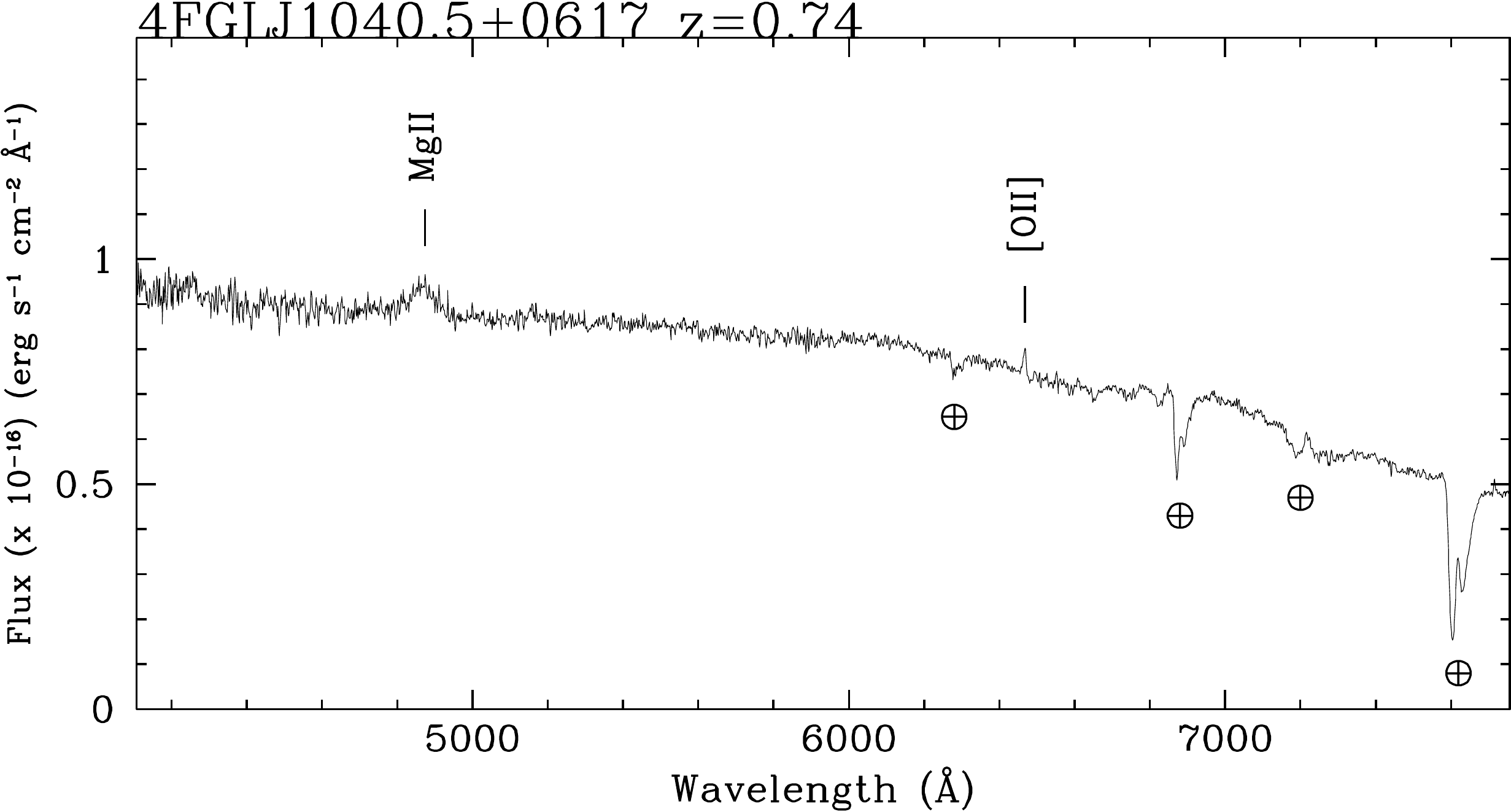}
\includegraphics[width=0.25\textwidth, angle=-90]{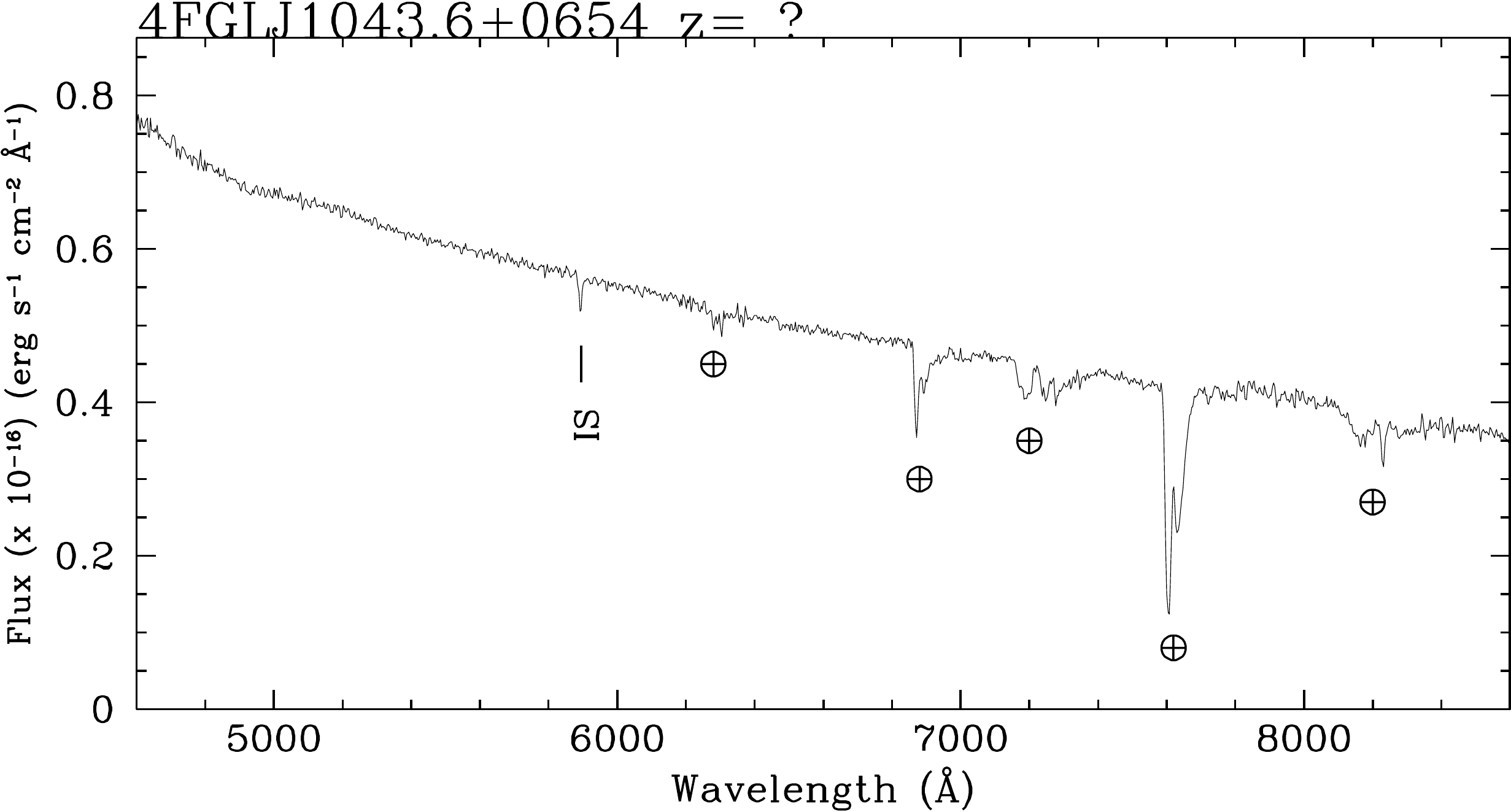}
\includegraphics[width=0.25\textwidth, angle=-90]{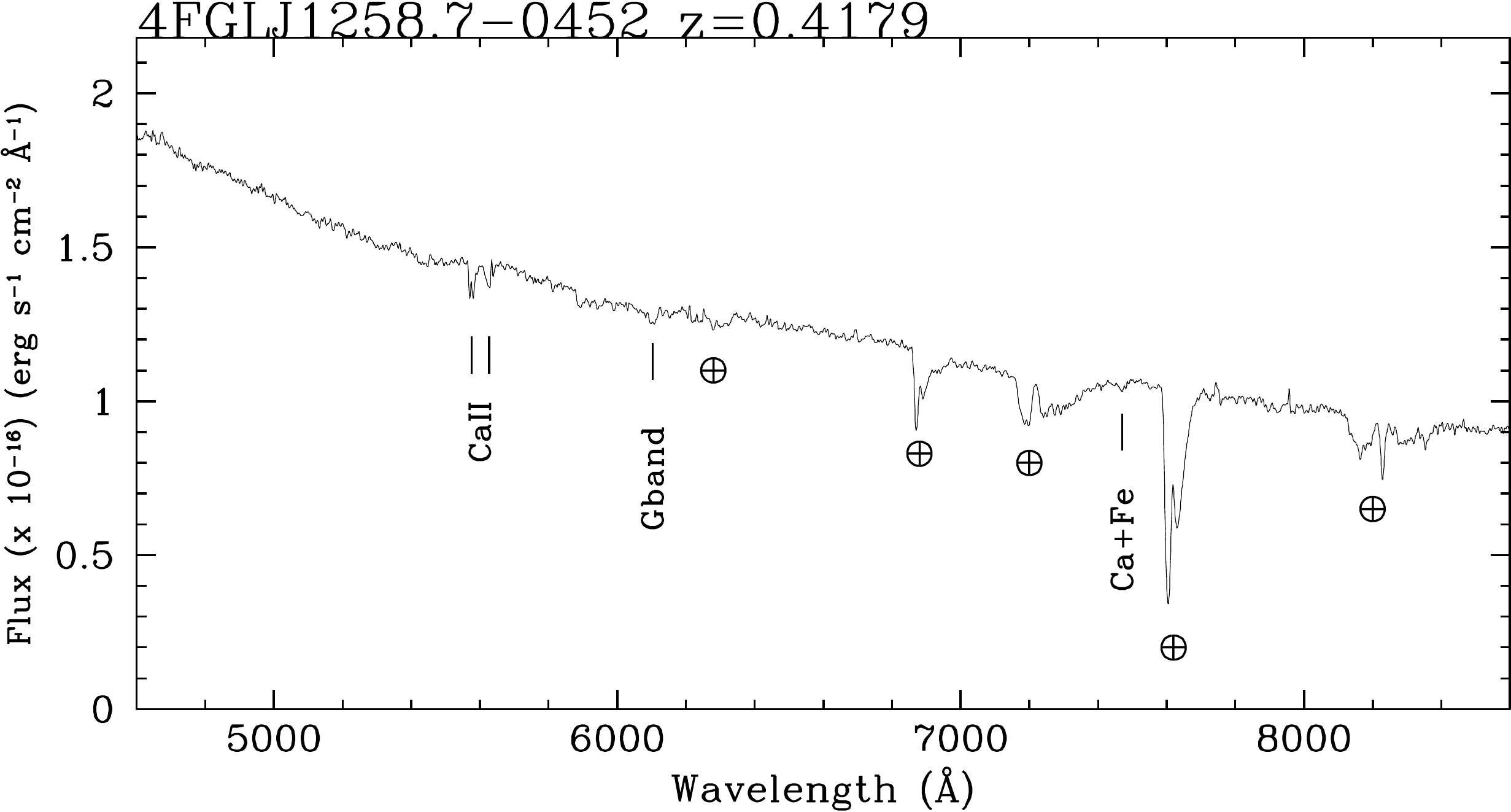}
\caption{Flux calibrated and dereddened spectra of the neutrino candidate blazars obtained at GTC and VLT. The main telluric bands are indicated by $\oplus$, the absorption features from interstellar medium of our galaxies are labelled as IS (Inter-Stellar).}
\label{fig:spectra}
\end{figure*}%[htbp]

\setcounter{figure}{0}
\begin{figure*}%[htbp]
\includegraphics[width=0.25\textwidth, angle=-90]{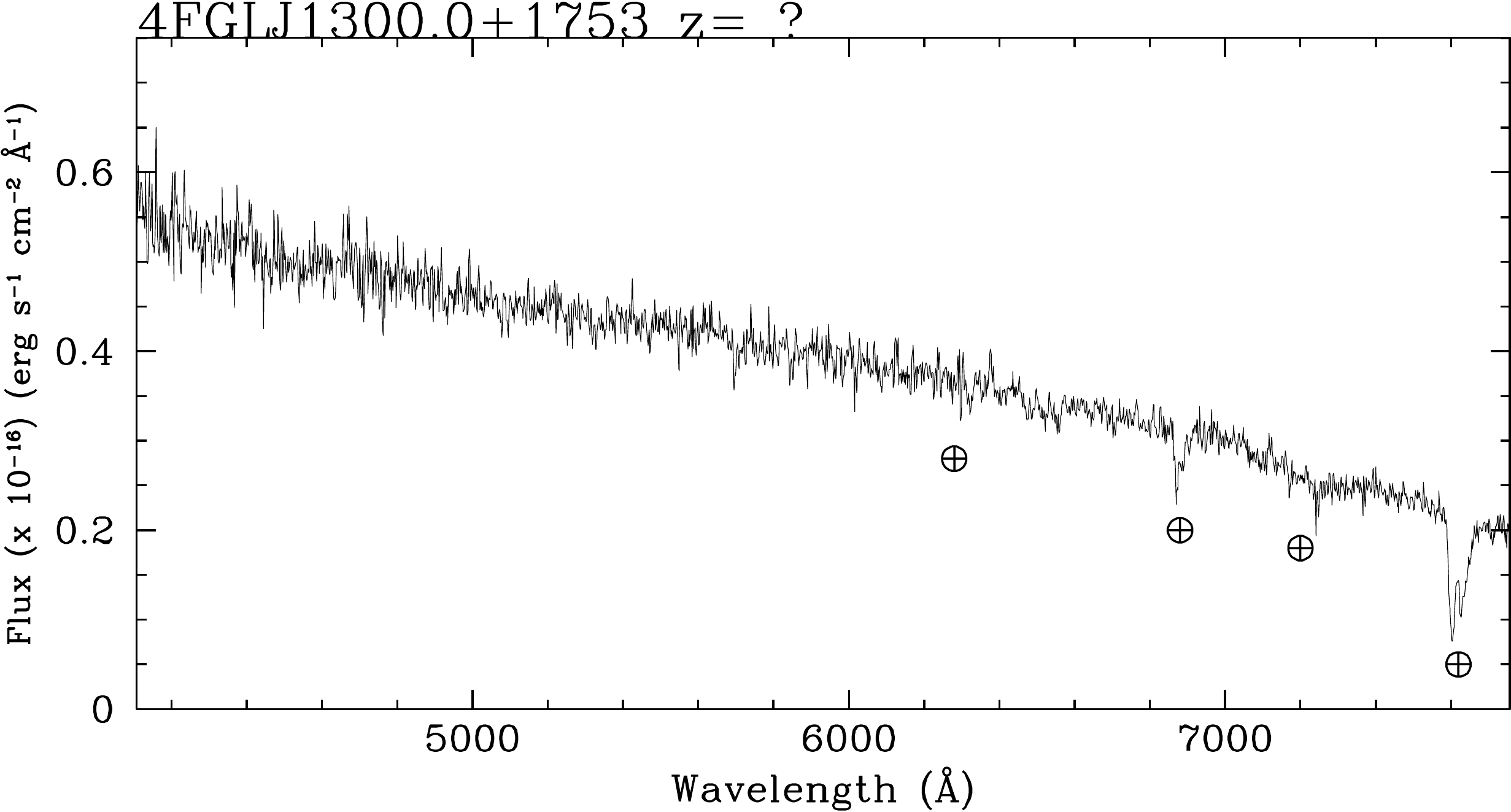}
\includegraphics[width=0.25\textwidth, angle=-90]{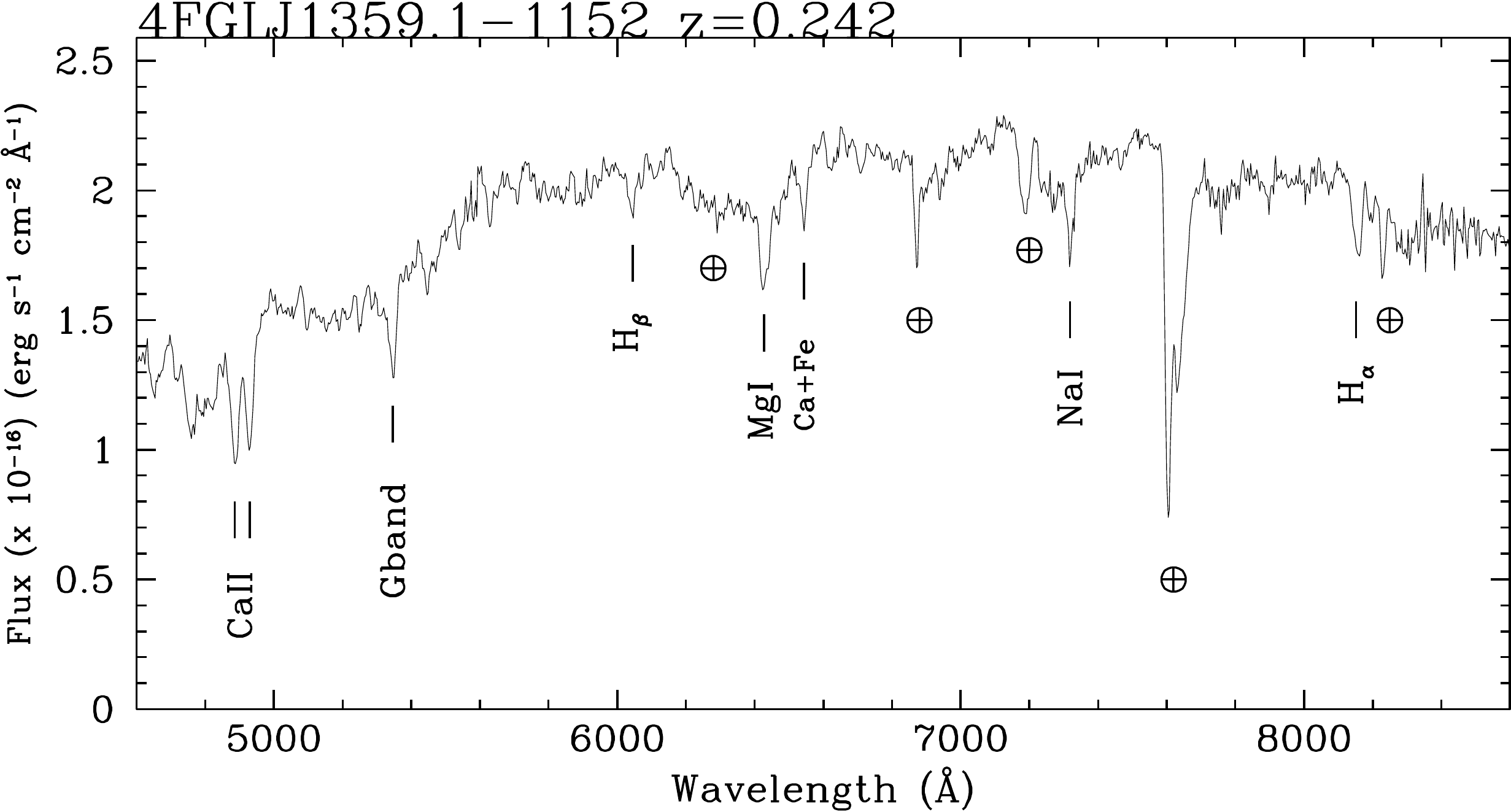}
\includegraphics[width=0.25\textwidth, angle=-90]{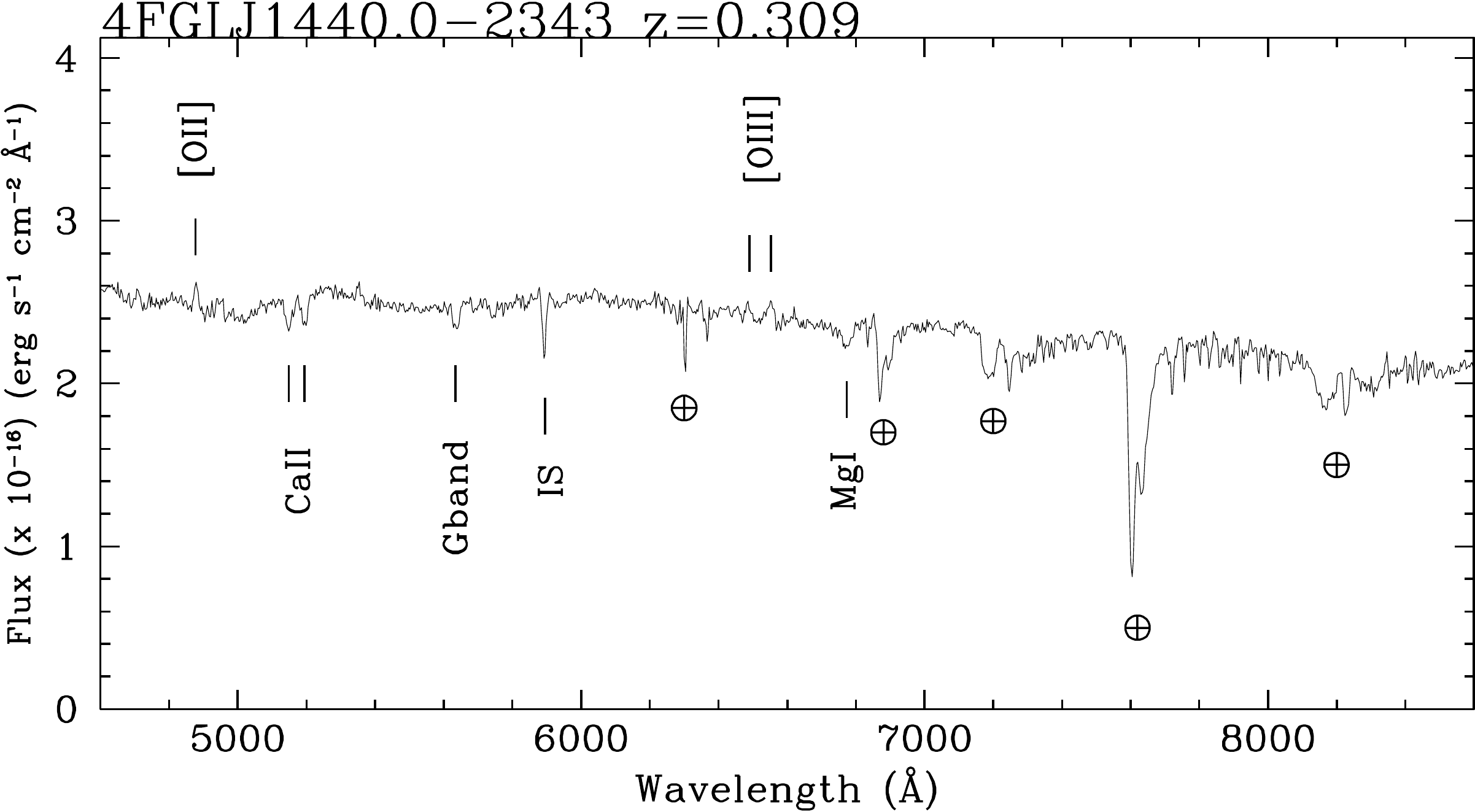}
\includegraphics[width=0.25\textwidth, angle=-90]{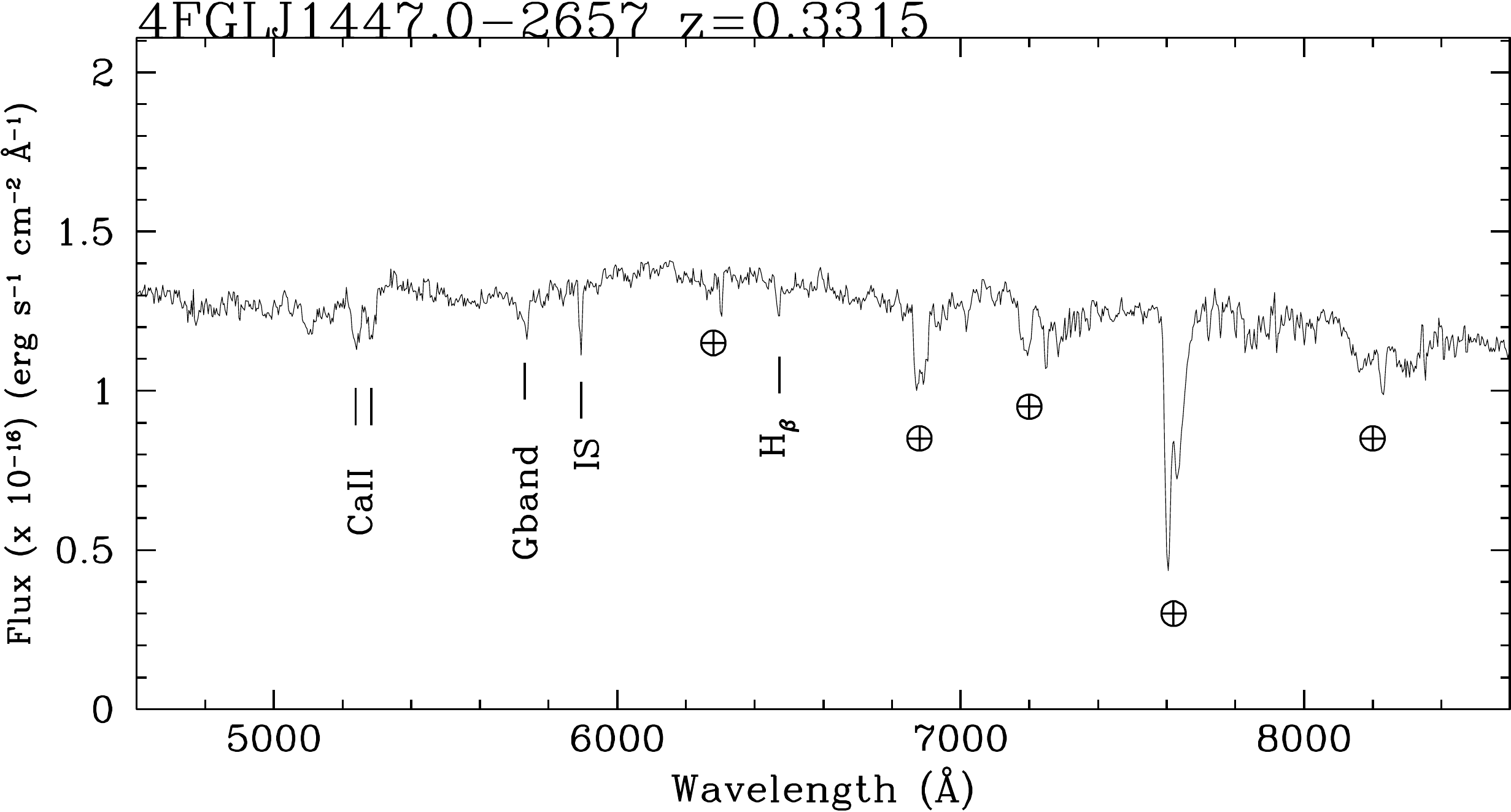}
\includegraphics[width=0.25\textwidth, angle=-90]{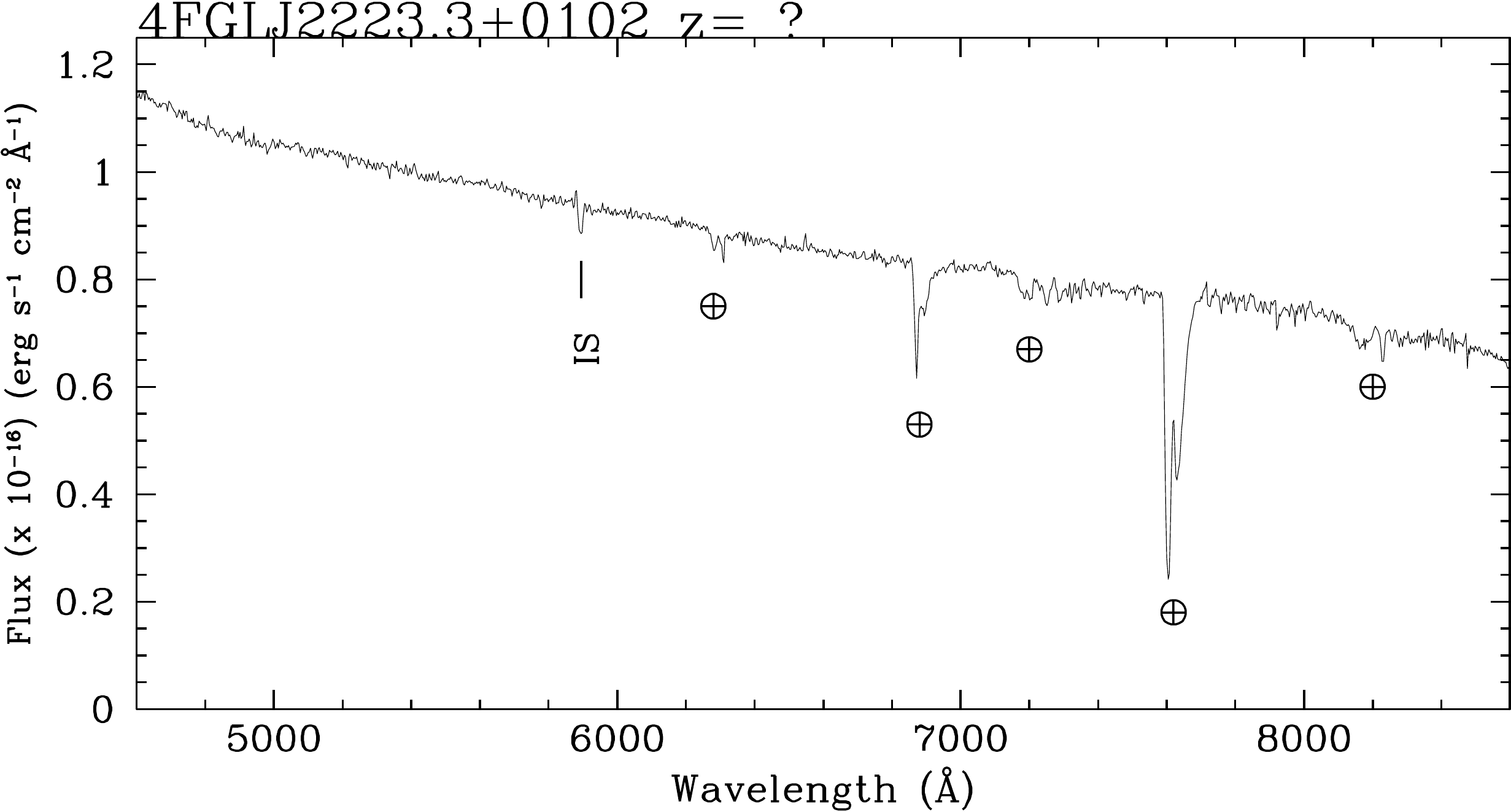}
\includegraphics[width=0.25\textwidth, angle=-90]{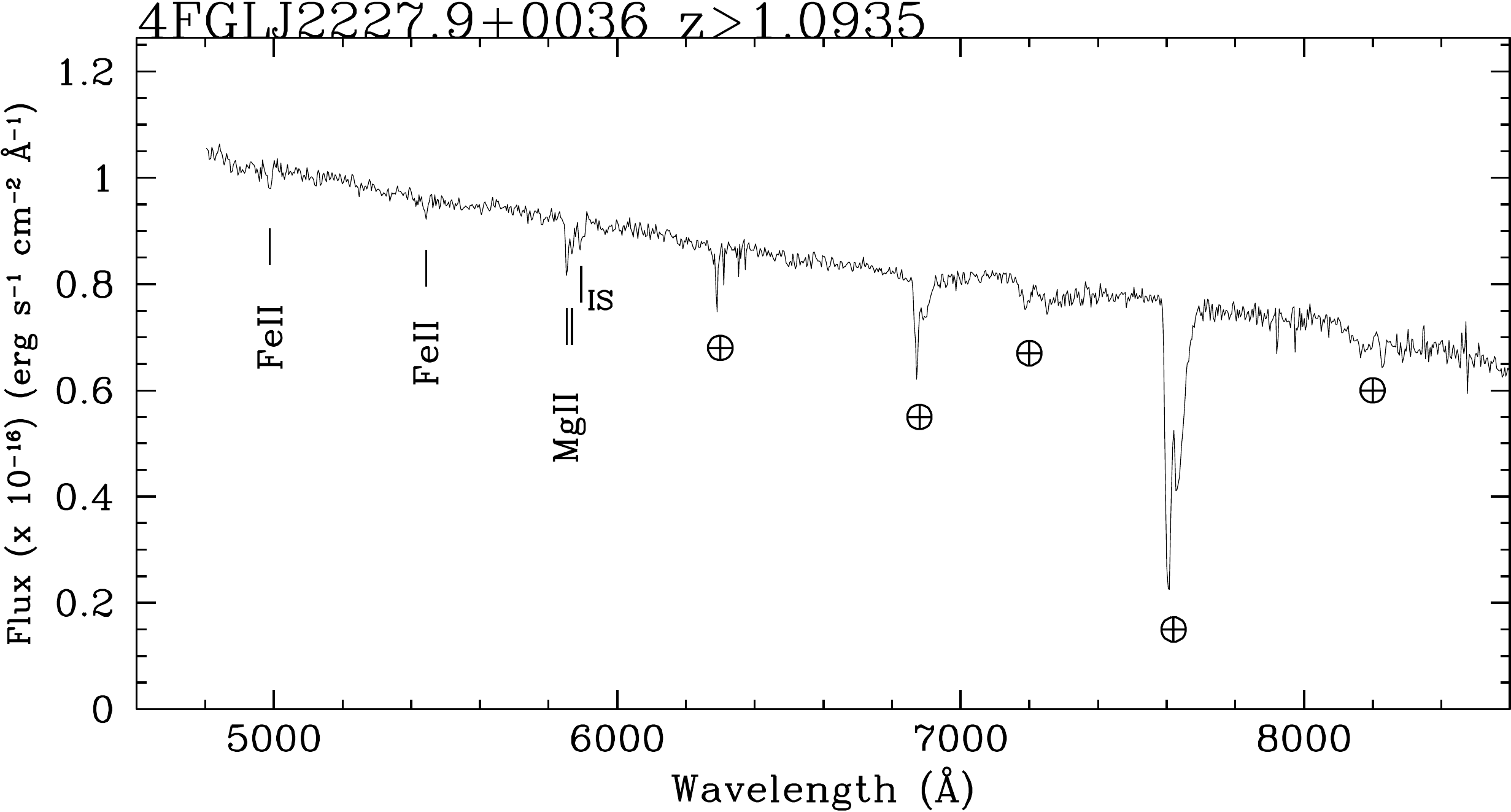}
\includegraphics[width=0.25\textwidth, angle=-90]{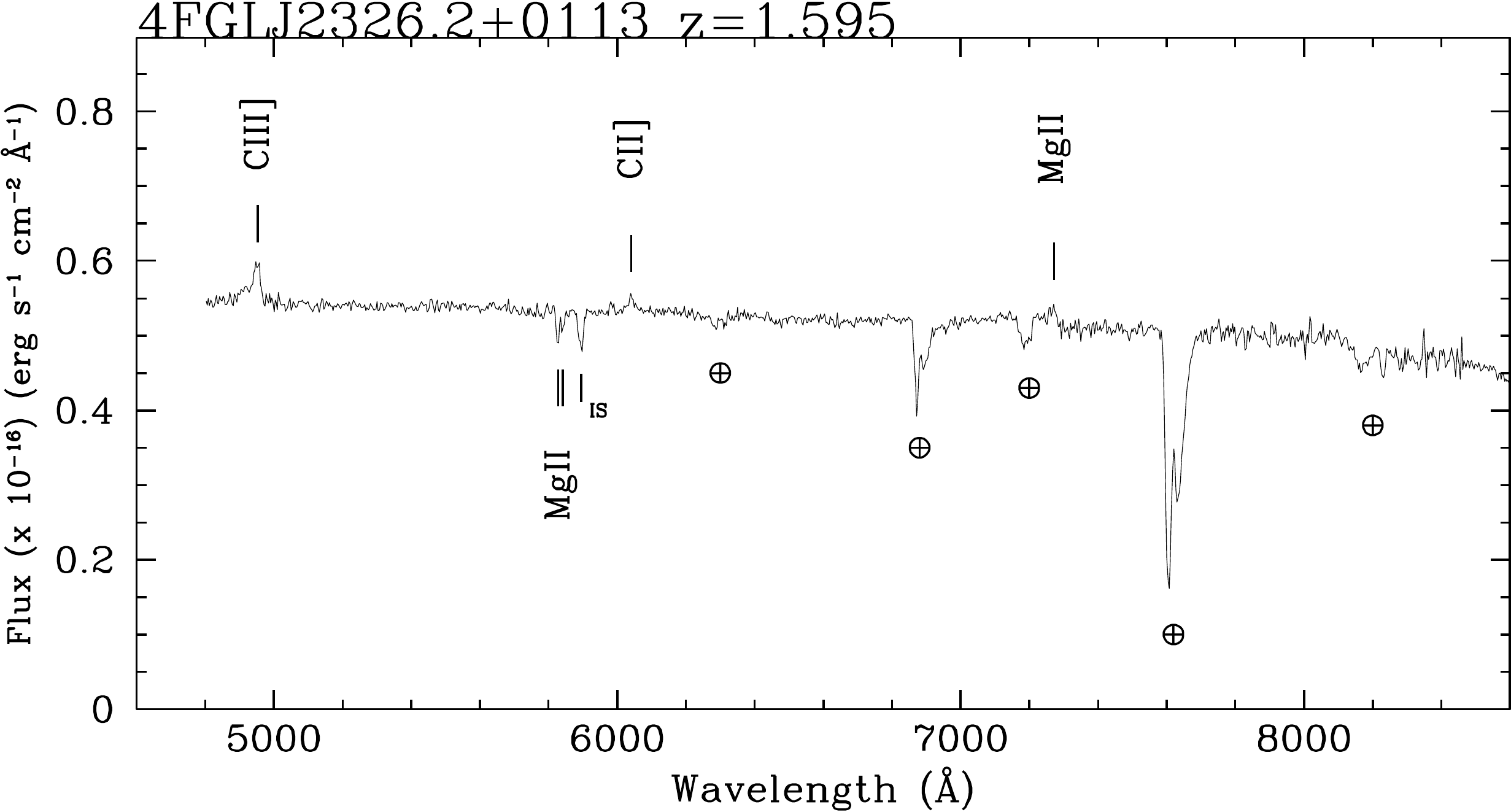}
\caption{- \textit{Continued}}
%\label{fig:spectra}
\end{figure*}%[htbp]

\setcounter{figure}{1}
\begin{figure*}%[htbp]
\includegraphics[width=0.29\textwidth, angle=-90]{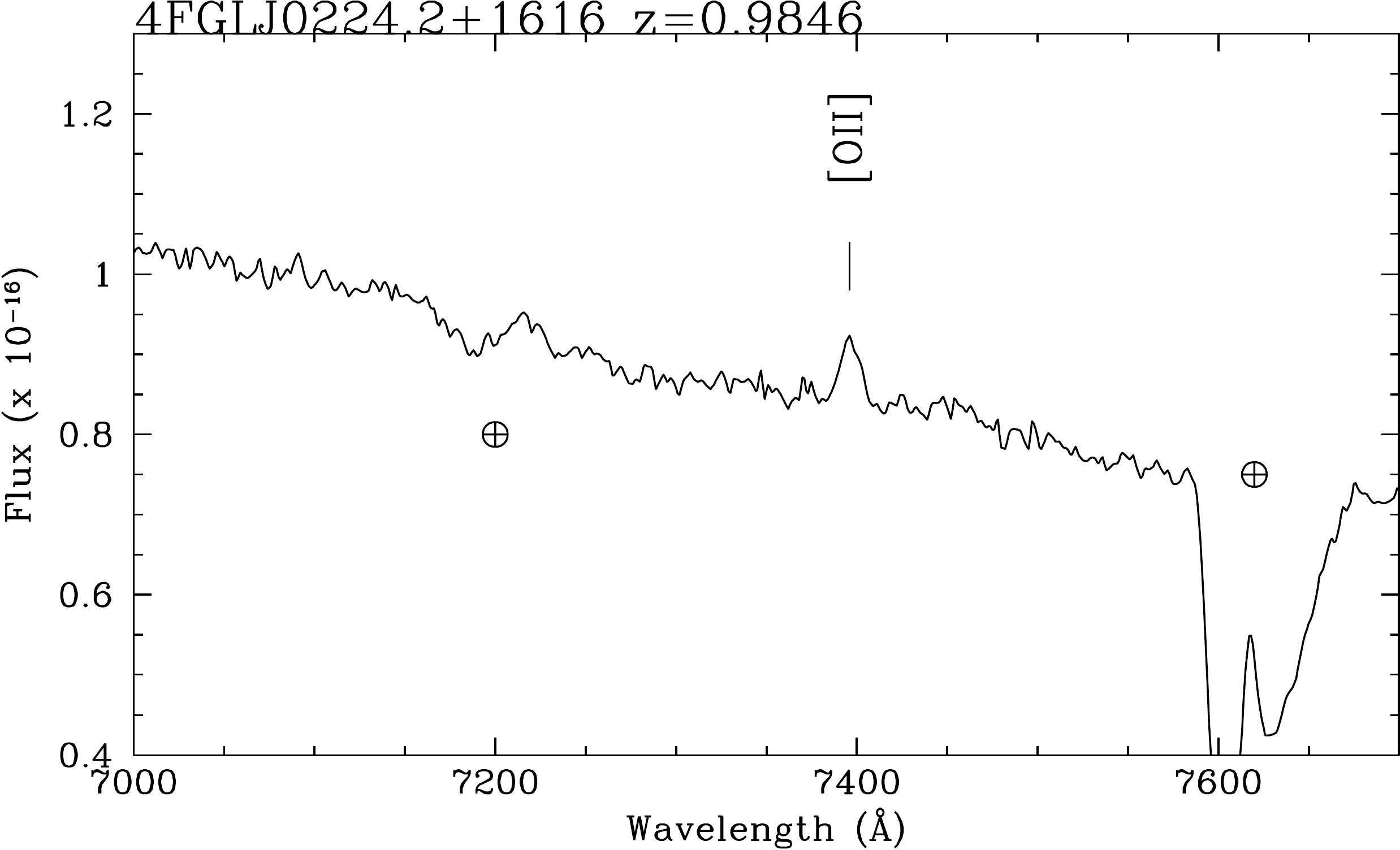}
\includegraphics[width=0.29\textwidth, angle=-90]{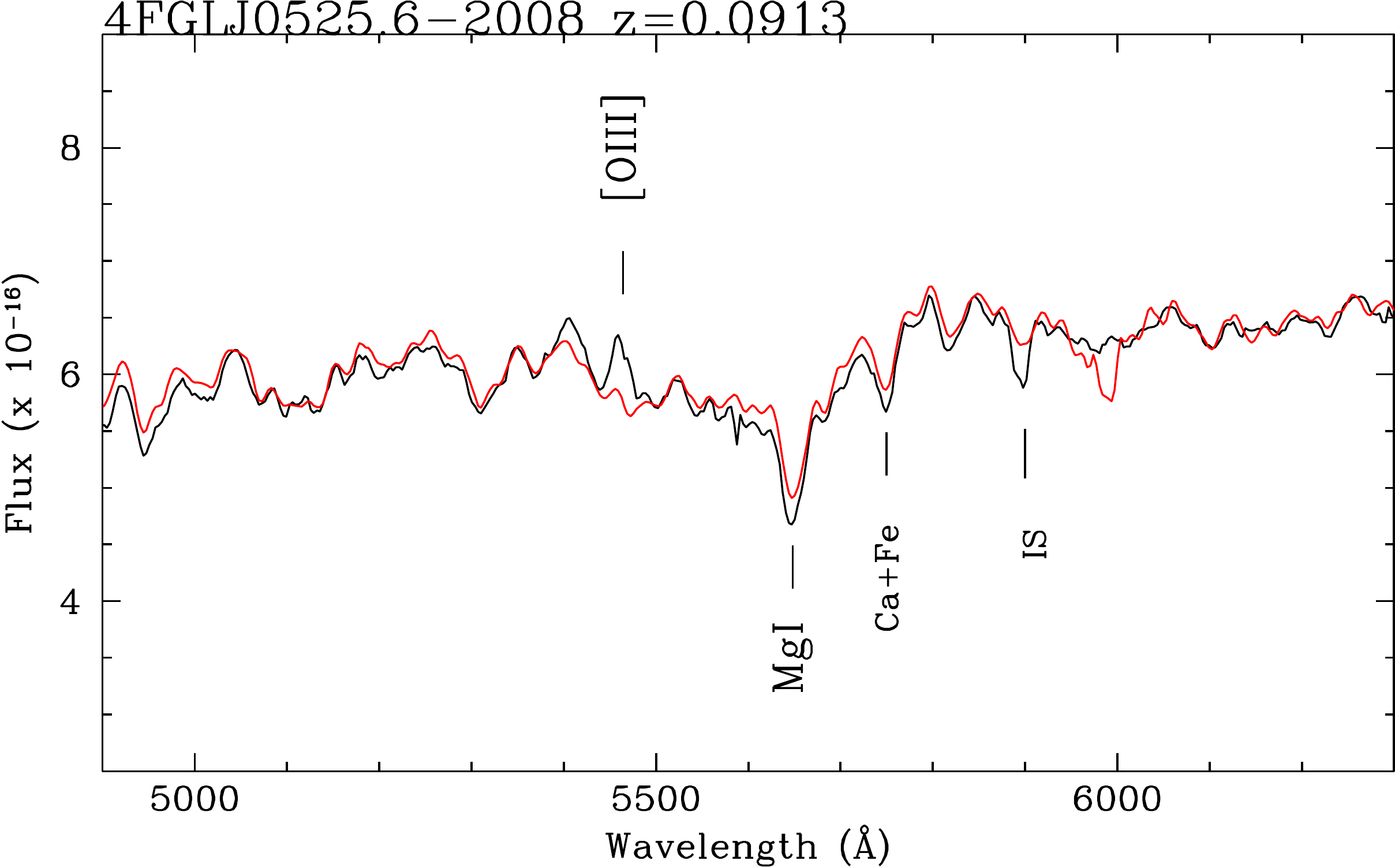}
\includegraphics[width=0.29\textwidth, angle=-90]{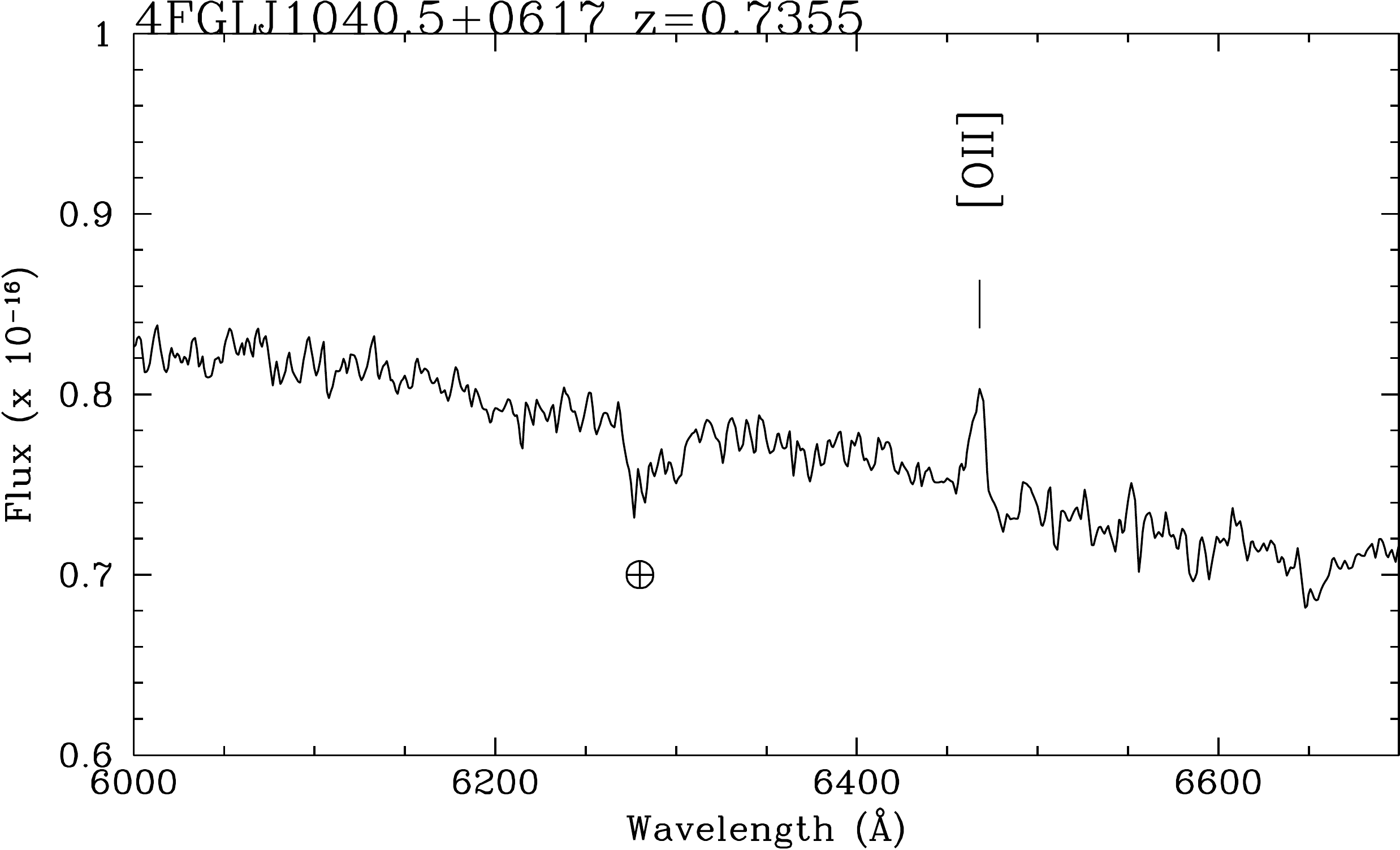}
\includegraphics[width=0.29\textwidth, angle=-90]{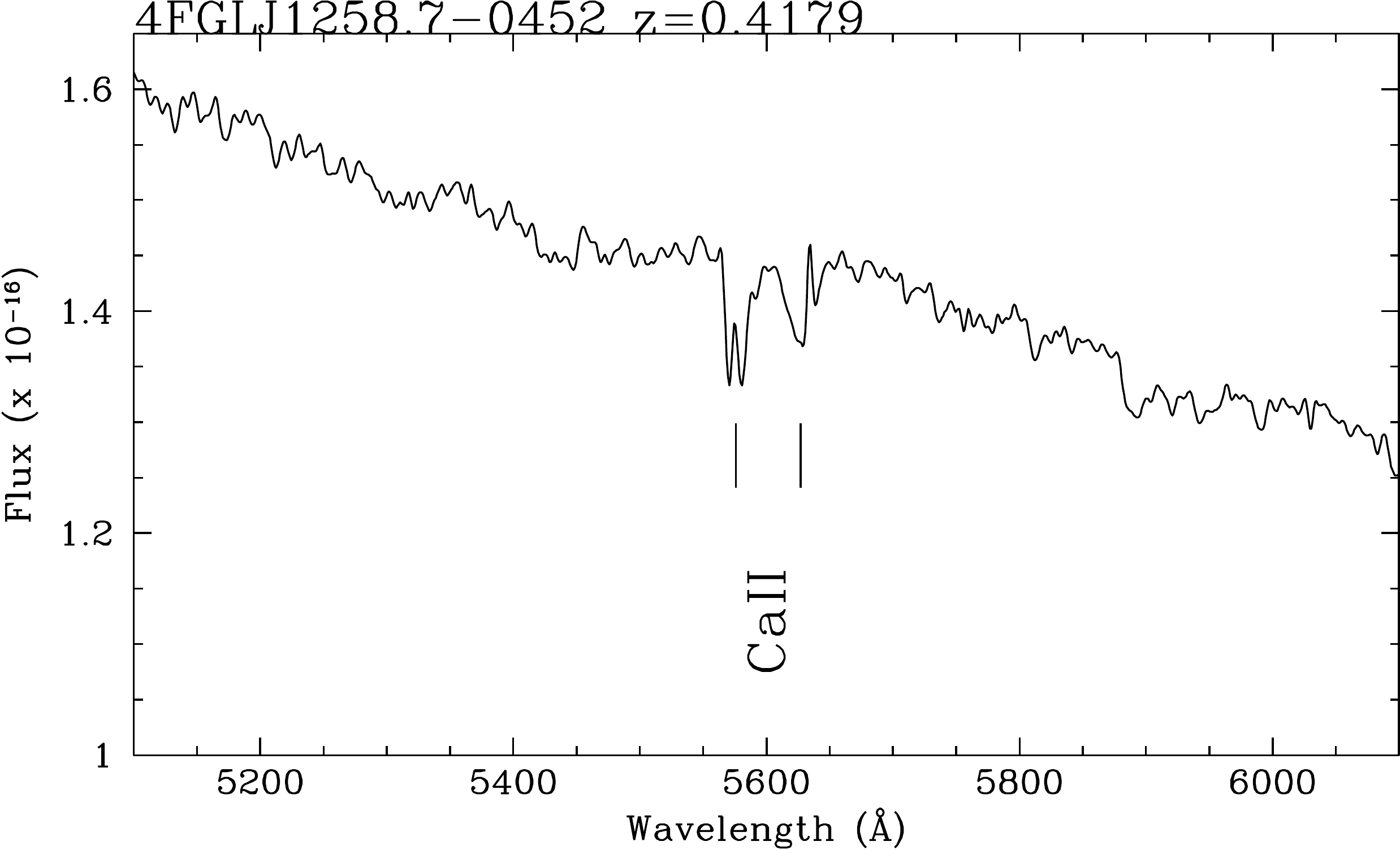}
\includegraphics[width=0.29\textwidth, angle=-90]{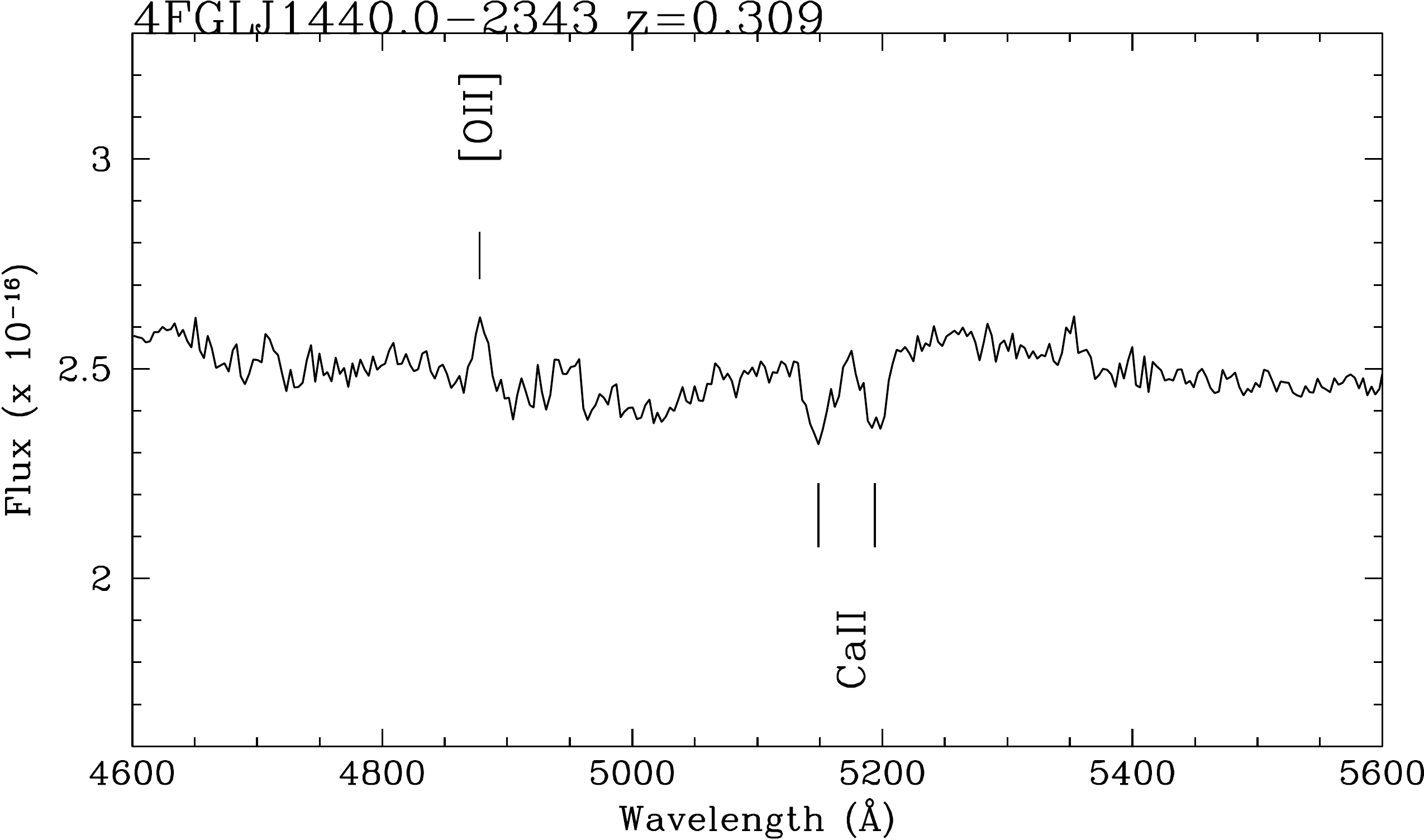}
\includegraphics[width=0.29\textwidth, angle=-90]{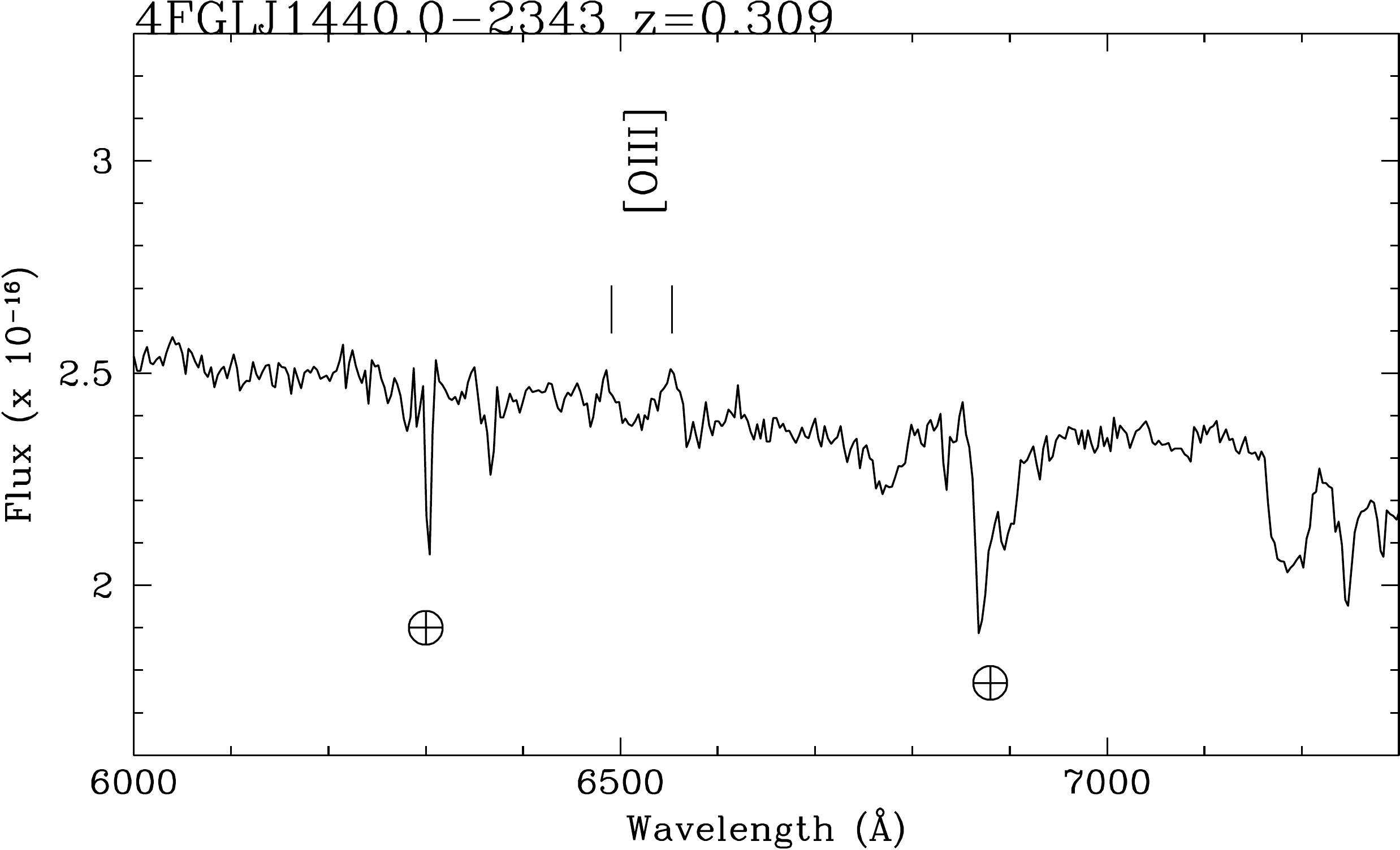}
\includegraphics[width=0.29\textwidth, angle=-90]{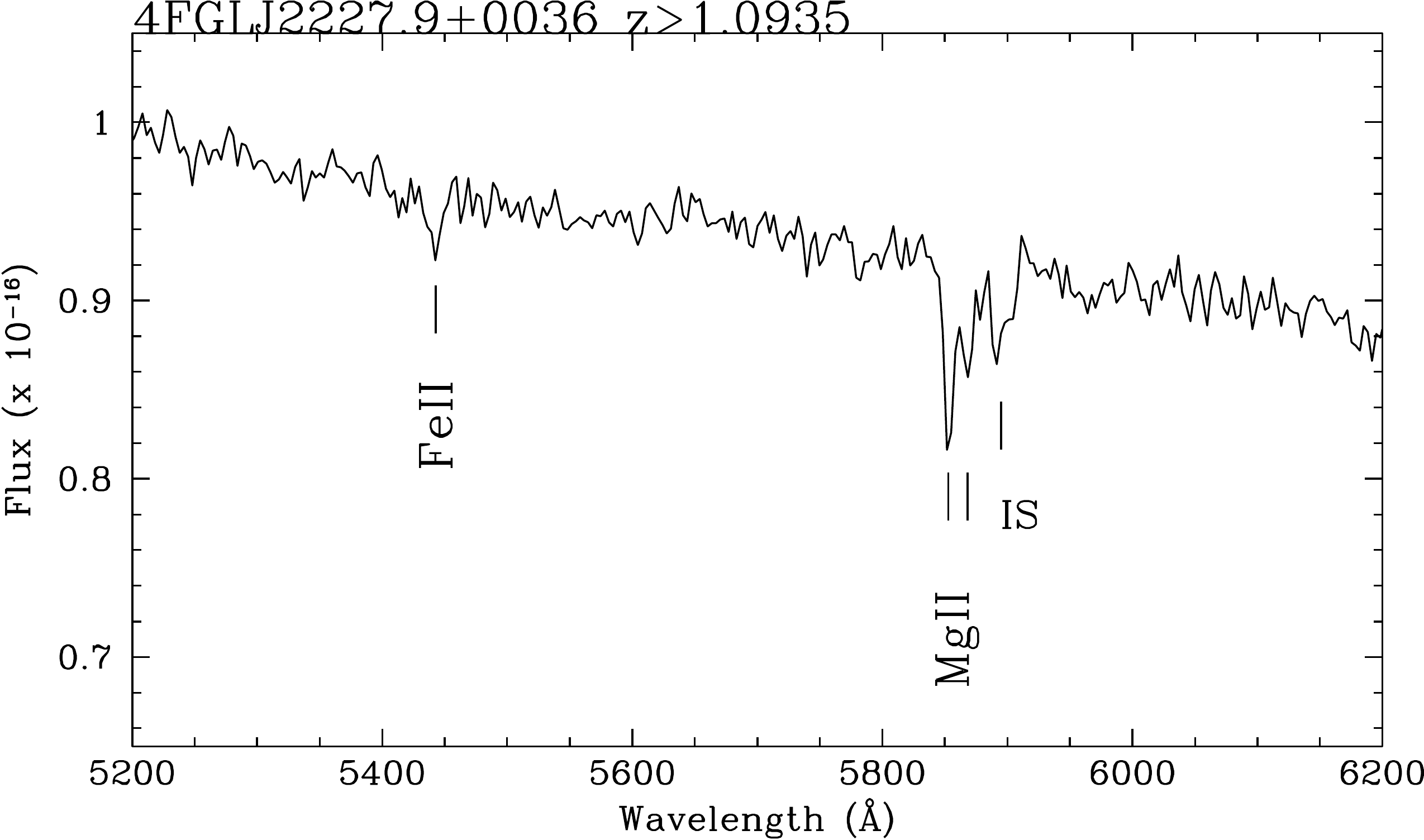}
\includegraphics[width=0.29\textwidth, angle=-90]{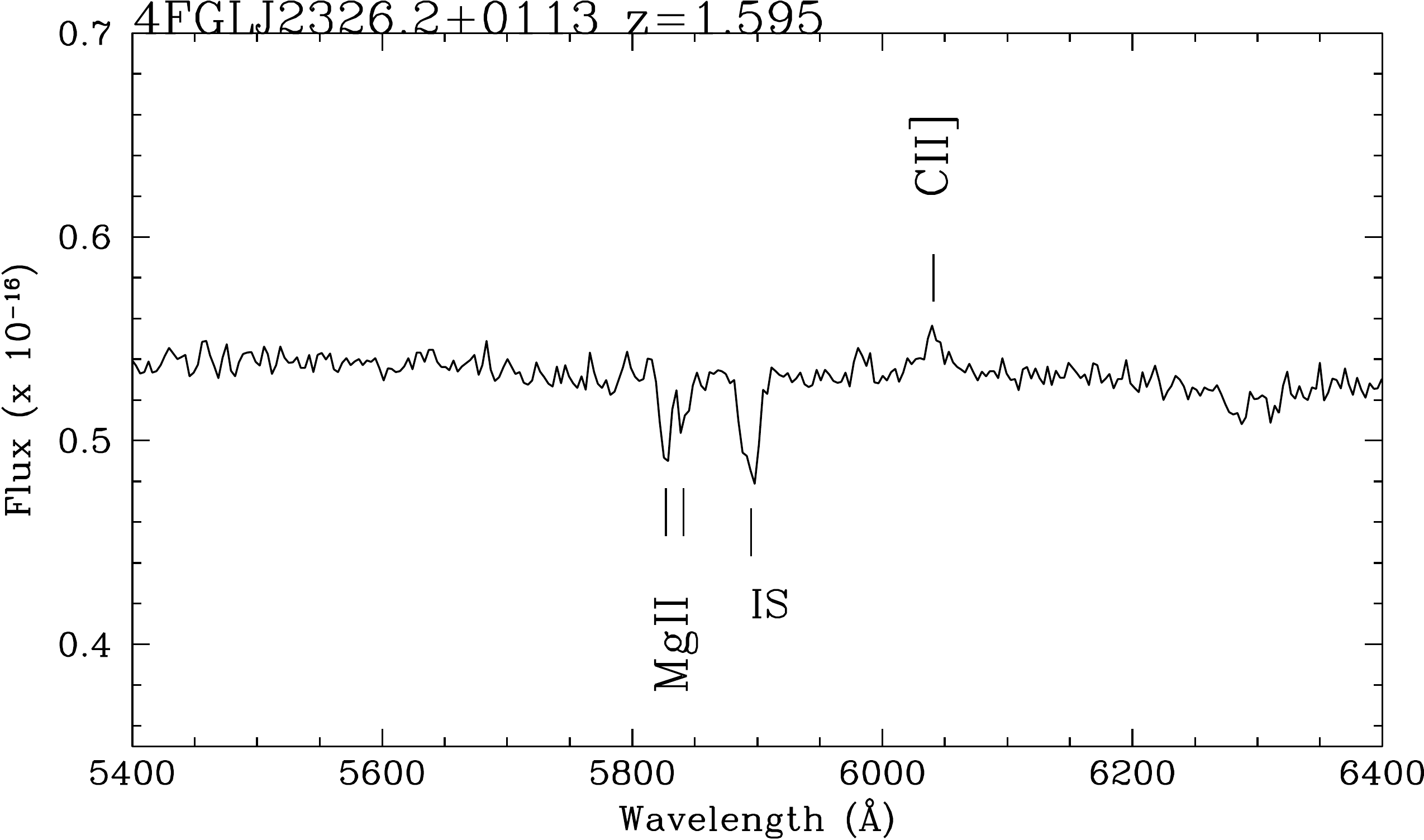}
\caption{Some examples of close-ups around the detected spectral lines of the spectra of the neutrino candidate blazars obtained at GTC and VLT. 
Main telluric bands are indicated by $\oplus$, spectral lines are marked by line identification. The red solid line for the source 4FGL~J0525.6-2008 is the fit of the spectrum as combination of an elliptical galaxy spectrum and a power law continuum (see Fig.~\ref{fig:decomposition} ): this allows us to highlight the detection of the [OIII]~5007 emission line.}
\label{fig:closeup}
\end{figure*}%[htbp]

\setcounter{figure}{2}
\begin{figure*}%[htbp]
\includegraphics[width=1.1\textwidth, angle=0]{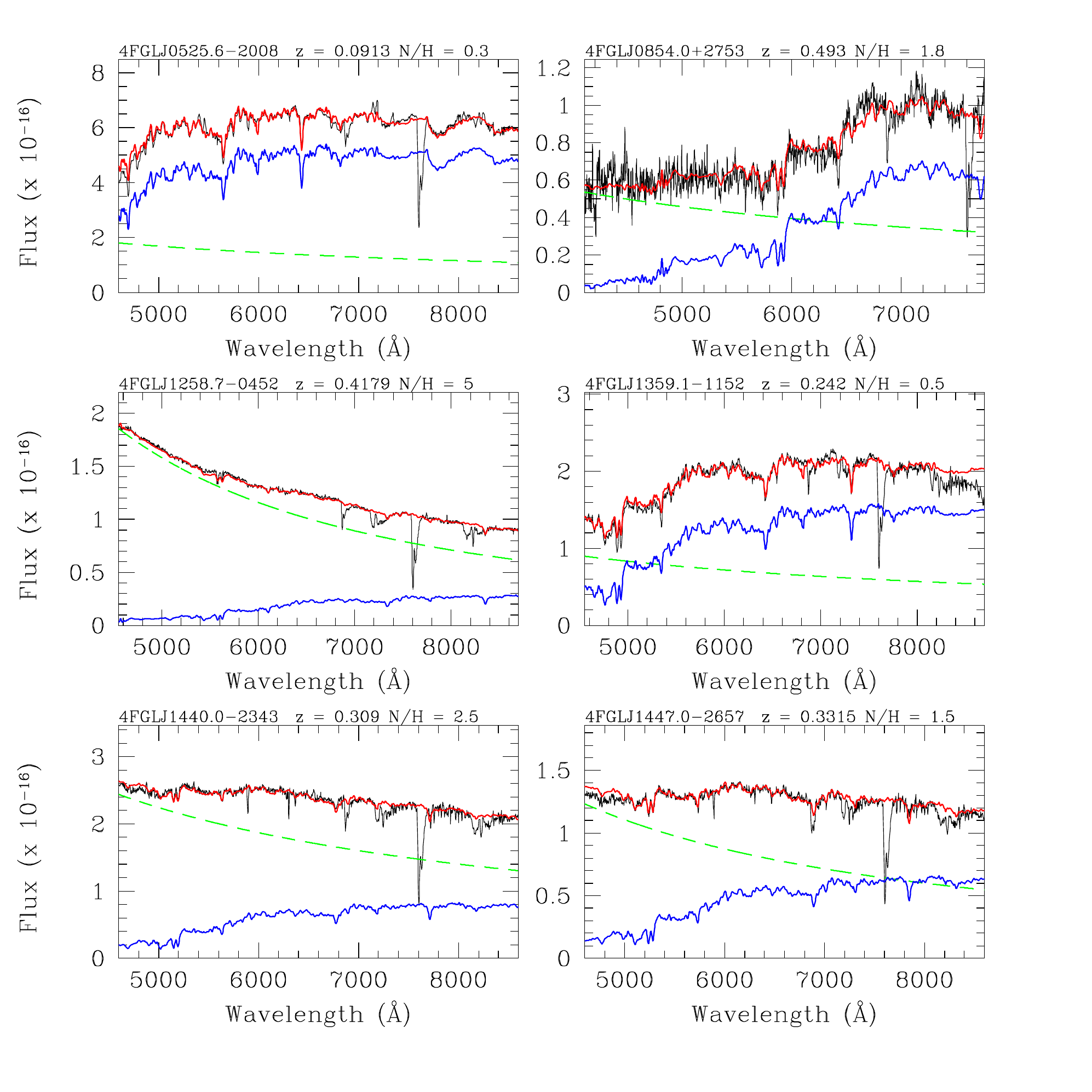}
\caption{Spectral decomposition of the observed optical spectrum (black line) of some target of our sample into a power law ( green dashed line ) and a template of elliptical for the host galaxy (blue line). The fit is given by the red solid line (see text in Section~4 for details). On each panel the nucleus to host ratio is given.}
\label{fig:decomposition}
\end{figure*}%[htbp]

\section*{Acknowledgments}
 We acknowledge helpful discussions and suggestions by Maria Petropoulou and Foteini Oikonomou. This work is based on observations collected at the European Southern Observatory under ESO programme 0104.B-0032(A) and Gran Telescopio Canarias under the programme GTC24-19B.
 Funding for the Sloan Digital Sky 
Survey IV has been provided by the 
Alfred P. Sloan Foundation, the U.S. 
Department of Energy Office of 
Science, and the Participating 
Institutions. 
SDSS-IV acknowledges support and 
resources from the Center for High 
Performance Computing  at the 
University of Utah. The SDSS 
website is www.sdss.org.
SDSS-IV is managed by the 
Astrophysical Research Consortium 
for the Participating Institutions 
of the SDSS Collaboration including 
the Brazilian Participation Group, 
the Carnegie Institution for Science, 
Carnegie Mellon University, Center for 
Astrophysics | Harvard \& 
Smithsonian, the Chilean Participation 
Group, the French Participation Group, 
Instituto de Astrof\'isica de 
Canarias, The Johns Hopkins 
University, Kavli Institute for the 
Physics and Mathematics of the 
Universe (IPMU) / University of 
Tokyo, the Korean Participation Group, 
Lawrence Berkeley National Laboratory, 
Leibniz Institut f\"ur Astrophysik 
Potsdam (AIP),  Max-Planck-Institut 
f\"ur Astronomie (MPIA Heidelberg), 
Max-Planck-Institut f\"ur 
Astrophysik (MPA Garching), 
Max-Planck-Institut f\"ur 
Extraterrestrische Physik (MPE), 
National Astronomical Observatories of 
China, New Mexico State University, 
New York University, University of 
Notre Dame, Observat\'ario 
Nacional / MCTI, The Ohio State 
University, Pennsylvania State 
University, Shanghai 
Astronomical Observatory, United 
Kingdom Participation Group, 
Universidad Nacional Aut\'onoma 
de M\'exico, University of Arizona, 
University of Colorado Boulder, 
University of Oxford, University of 
Portsmouth, University of Utah, 
University of Virginia, University 
of Washington, University of 
Wisconsin, Vanderbilt University, 
and Yale University.

\section*{Data Availability}
The flux-calibrated and dereddened spectra are available in our online data base ZBLLAC\footnote{http://web.oapd.inaf.it/zbllac/}.

%\newpage

%\end{document}

%%%%%%%%%%%%%%%%%%%%%%%%%%%%%%%%%%%%%%%%%%%%%%%%%%%%%%%%%%%%%%%%%%%%%%%%%%%%%%%%%%%%%
%\clearpage
\appendix
\section{SPECTROSCOPY OF ADDITIONAL SOURCES}

Here we focus on the seven sources that were considered in a preliminary list of possible counterparts of the neutrino events. 
They have been eliminated from the list of of 47 objects of Giommi et al 2020 (see Table~1) because either they do not belong to the 4FGL catalogue or the angular separations with respect to the neutrino events do not fulfill the criteria given in the Section~2. 
The list of these additional sources is given in Table~A1, the journal of observations in Table~A2 and the spectral properties and redshifts are reported in Table~A3.
Three objects have a quasar–like line spectrum (FL8YJ1302.5+2037, 4FGL~J2133.0+2610, 4FGL~J2311.7+2604) and are among the most distant ones
discussed in this paper. The other four appear bona fide BLL. 
The redshift of two derives from emission lines (4FGL~J1339.0-2400 and 4FGL~J2325.6+1644). 
Finally, for one object (4FGL~J2255.1+241) a robust lower limit to the redshift is derived from an Mg~II intervening absorption \citep[see also][]{paiano2019atel2255}.

\subsection{Notes on individual sources}

\begin{itemize}

\item[] \textbf{FL8YJ1302.5+2037}: % ---------------------------------------------
The object is reported in the FL8Y $\gamma$-ray catalogue, but not in the 4FGL.
In our spectrum, we clearly detect strong and broad (EW ranging from $\sim$33 to 127 $\textrm{\AA}$) emission lines attributed to C~IV, He~II, C~III] and Mg~II (see Tab. A.3) that lead to a redshift z~=~1.738.
The target can be classified as a quasar.

\item[] \textbf{4FGL~J1339.0-2400}: % ---------------------------------------------
We obtained a high quality (S/N$\sim$220) spectrum of the target (g~=~18.9) which exhibits faint emission lines attributed to [O~II], [Ne~III] and [O~III].
This yields a redshift of z~=~0.655, confirming the proposed uncertain redshift by \citet{hook2003}.

\item[] \textbf{4FGL~J1455.0+0247}: % ---------------------------------------------
This source was observed by \cite{Sandrinelli2013} and by the SDSS  and no emission/absorption lines were detected. 
Our (S/N$\sim$65) spectrum of the source (g~=~20.2) is still featureless and from the estimated minimum EW$\sim$0.50~$\textrm{\AA}$ we can set a redshift lower limit z~$>$~0.65.

\item[] \textbf{4FGL~J2133.0+2610}: % ---------------------------------------------
Our spectrum exhibits a prominent broad emission line (see Tab.~A3) that if attributed to Mg~II yields a redshift z$=$1.139.
In addition we detected two absorption doublets at 5340,5369~$\textrm{\AA}$ and 5774,5788~$\textrm{\AA}$ due to a Fe~II and Mg~II intervening system respectively, corresponding to a  z$=$1.065.

\item[] \textbf{4FGL~J2255.1+2411}: % ---------------------------------------------
The source has been also recently proposed as a possible counterpart of five ANTARES neutrino events with energy ranging from 3 to 40 TeV \citep[see][]{aublin2019,paiano2019atel2255,giommi2020a}.
In our very high quality (S/N~=~550) spectrum, we detect a Mg~II intervening system at 5209,5222~$\textrm{\AA}$ allowing us to set the spectroscopic redshift lower limit z~$\geq$~0.863.
 In previous spectra \cite[][and the SDSS survey]{shaw2013,massaro2015} no emission or absorption lines are identified but the feature at 5209,5222~$\textrm{\AA}$ is visible. 

\item[] \textbf{4FGL~J2311.7+2604}: % ---------------------------------------------
In our spectrum we clearly detected broad emission lines attributed to C~IV, He~II, CIII], and Mg~II (see Tab.~A3). This implies a redshift of z~=~1.7425, confirming the value of the SDSS spectrum.

\item[] \textbf{4FGL~J2325.6+1644}: % ---------------------------------------------
In the literature, \citet{massaro2015} report an optical spectrum of modest S/N without evident features.
In our spectrum, we detect two narrow emission lines due to forbidden transitions attributed to [O~II] and [O~III] (see Tab.~A3), yielding z~=~0.4817.

\end{itemize}

\newpage
\setcounter{table}{0}
\begin{table*}
\begin{center}
\caption{Additional Neutrino candidate blazars.}\label{tab:tablea1}
%\centering
%\begin{tabular}{llcccccccll}
\begin{tabular}{lllllllll}
\hline 
Object Name  &    Counterpart     &  IC event &  RA    &  DEC  &  Magnitude  &  $\nu_{\textit{peak}}^{\textit{syn}}$  & Redshift & Reference\\   
           &                    &           & (J2000)  &(J2000) &           &     &   \\
\hline
FL8Y~J1302.5+2037  & CRATES~J130215+203911 & IC-151017A & 13:02:14.6 & +20:39:18.6  & 20.1 & 12.5 & ?  & NED  \\
\hline
4FGL~J1339.0-2400  & 5BZU~J1339-2401      & IC-131202A & 13:39:01.7 & $-$24:01:14.0  & 18.2 & 14.1  & 0.665? & NED \\ 
\hline
4FGL~J1455.0+0247  & 5BZB~J1455+0250      & IC-111201A & 14:55:07.4 &  +02:50:40.1  & 19.7 & 14.3 & ? &  * \\ 
\hline
4FGL~J2133.0+2610  & NVSS~J213252+261143  & IC-150714A & 21:32:52.9 &  +26:11:43.8  & 20.5 & 13.0  & ? & * \\
\hline
4FGL~J2255.1+2411  & MG3~J225517+2409     & IC-100608A & 22:55:15.3 &  +24:10:11.3  & 18.6 & 14.7  & ? & SDSS \\ 
4FGL~J2311.7+2604  & MG3~J231144+2604	  & IC-100608A & 23:11:45.8 &  +26:04:47.8  & 21.2 & 13.0  & 1.74 & SDSS \\
\hline
4FGL~J2325.6+1644  & NVSS~J232538+164641  & IC-140522A & 23:25:38.1 &  +16:46:42.2  & 18.2 & 15.3 & ?  & ZBLLAC \\ 
\hline
\end{tabular}
\end{center}
\raggedright
\footnotesize \textit{Notes.} Column~1: Fermi name of the target in the 4FGL catalogue; Column~2: Counterpart name of the $\gamma$-ray source; Column~3: IceCube track;  Column~4: Angular separation (degrees) between the target and the centroid of the IceCube track; Column~5~-~6: Right ascension and declination of the optical counterpart; Column~7: g magnitude from SDSS survey and PANSTARRs; Column~8: Frequency of the synchrotron peak ($\nu_{\textit{peak}}^{\textit{syn}}$); Column~9: Redshift from the literature.\\ 
%\raggedright
% } 
\end{table*}

\newpage
\setcounter{table}{1}
\begin{table*}
\begin{center}
\caption{Journal of the observations of additional neutrino candidate blazars.}\label{tab:tablea2}
%\centering
%\begin{tabular}{llcccccccll}
\begin{tabular}{lllllll}
\hline 
Object Name  & E(B-V) & Telescope  & Date & t$_{Exp}$ & seeing & Air Mass \\    
             &        &            &      & (s)       &   (")  &   \\
\hline
FL8Y~J1302.5+2037  &  0.02 & GTC & 02 February 2020  & 6600  & 2.0  & 1.02\\
4FGL~J1339.0-2400  &  0.07 & VLT & 02 February 2020  & 5400  & 0.8  & 1.09 \\
4FGL~J1455.0+0247  &  0.04 & VLT & 21 March    2020  & 5400  & 1.4  & 1.31\\
4FGL~J2133.0+2610  &  0.11 & GTC & 12 October  2019  & 6600  & 1.7  & 1.32\\
4FGL~J2255.1+2411  &  0.06 & GTC & 04 October  2019  & 21600 & 0.9  & 1.10\\
4FGL~J2311.7+2604  &  0.05 & GTC & 21 October  2019  & 4500  & 2.0  & 1.14\\
4FGL~J2325.6+1644  &  0.03 & GTC & 09 October  2019  & 3000  & 1.0  & 1.13\\
\hline
\end{tabular}
\end{center}
\raggedright
\footnotesize \textit{Notes}. Column~1: Name of the target ; Column~2: E(B-V) taken from the NASA/IPAC Infrared Science Archive (https://irsa.ipac.caltech.edu/applications/DUST/);  Column~3: Telescope used for the observation; Column~4: Date of the observation; Column~5: Total integration time (sec); Column~6: Average seeing during the observation (arcsec). \\
%\tablenotetext{}{
%\raggedright
% } 
\end{table*}

\setcounter{table}{2}
\begin{table*}
\begin{center}
\caption{Spectral properties of the additional neutrino candidate blazars.}\label{tab:tablea3}
%\centering
\begin{tabular}{lllllll}
\hline
Object Name      &  g  & S/N  &   EW$_{min}$  & z &  Line type & $\alpha$  \\  
                 &     &      &   ($\textrm{\AA}$) &  &   & \\  
\hline
FL8Y~J1302.5+2037   &  20.1  & 35  &    -    & 1.738    & e,i &  1.2 $\pm$ 0.2\\
4FGL~J1339.0-2400   &  18.9  & 220 &    -    & 0.655    & e   & 0.3 $\pm$ 0.1\\ %18.3
4FGL~J1455.0+0247   &  20.2  & 65  &   0.50  & $>$0.65  & h   &  0.0 $\pm$ 0.1\\ %19.6
4FGL~J2133.0+2610   &  20.3  & 20  &    -    & 1.139    & e,i & 1.2 $\pm$ 0.3 \\
4FGL~J2255.1+2411   &  17.5  & 550 &   0.10  & $>$0.863 & i   &  0.95 $\pm$ 0.1\\
4FGL~J2311.7+2604   &  21.6  & 10  &    -    & 1.7425   & e   & 1.1 $\pm$ 0.1 \\
4FGL~J2325.6+1644   &  18.7  & 100 &    -    & 0.4817   & e   & 0.9 $\pm$ 0.1 \\
\hline
\end{tabular}
\end{center}
\raggedright
\footnotesize \textit{Notes}. Column~1: Name of the target;  Column~2: Magnitude (g) measured from the acquisition image;  Column~3: Median S/N of the spectrum;  Column~4: Minimum equivalent width (EW$_{min}$) derived in the 5500 - 6500 $\textrm{\AA}$ range (the measure is provided only in case of featureless spectrum);  Column~5: Redshift;  Column~6: Type of detected line to estimate the redshift: \textit{e} = emission line, \textit{g} = galaxy absorption line, \textit{i}= intervening absorption assuming Mg~II 2800~$\textrm{\AA}$ identification, \textit{h}= lower limit derived on the lack of detection of host galaxy absorption lines assuming a BLL elliptical host galaxy with M(R) = -22.9 \citep[see details in][]{paiano2017tev}. Column-7: Spectral index $\alpha$of the continuum described by a power law F$_\lambda \sim \lambda^{-\alpha}$ \\
%\tablenotetext{}{
%\raggedright
% } 
\end{table*}

\setcounter{table}{3}
\begin{table*}
\begin{center}
\caption{Properties of the emission lines.}\label{tab:tablea4}
%\centering
%\begin{tabular}{llcccccccll}
\begin{tabular}{llllll}
\hline
Object Name  &   z  & $\lambda$  &   EW  &   Line ID &  \textit{L} (line)   \\ 
             &             &  ($\textrm{\AA}$) &     ($\textrm{\AA}$) &    & (erg s$^{-1}$)  \\
\hline
FL8Y~J1302.5+2037 & 1.738    & 4242   & 127 $\pm$ 5      & C~IV~1549     & 5.6$\times$10$^{44}$ \\
                  &          & 4491   & 33  $\pm$ 2     & He~II~1640    & 4.7$\times$10$^{43}$ \\
                  &          & 5226   & 95 $\pm$  3      & C~III]~1909   & 1.5$\times$10$^{44}$ \\
                  &          & 7666   & 19: $\pm$ 5       & Mg~II~2800    & 1.2$\times$10$^{43}$*\\
4FGL~J1339.0-2400 & 0.655    & 6168   & 1.3 $\pm$ 0.3     & [O~II]~3727   & 4.0$\times$10$^{41}$ \\
                  &          & 6403   & 0.50 $\pm$ 0.2       & [Ne~III]~3869 & 1.5$\times$10$^{41}$ \\
                 %&          & 8208*  & $\sim$0.70* & [O~III]~4959  & 1.8$\times$10$^{41}$ \\nellatellurica
                  &          & 8287   & 2.2 $\pm$ 0.2        & [O~III]~5007  & 6.2$\times$10$^{41}$ \\
4FGL~J2133.0+2610 & 1.139    & 5989   & 50 $\pm$ 3       & Mg~II~2800    & 3.5$\times$10$^{43}$ \\
4FGL~J2311.7+2604 & 1.7425   & 4246   & 155 $\pm$ 6     & C~IV~1549     & 2.0$\times$10$^{44}$ \\
                  &          & 4499   & 66  $\pm$ 4    & He~II~1640    & 2.5$\times$10$^{43}$ \\
                  &          & 5235   & 16 $\pm$ 2     & C~III]~1909   & 5.6$\times$10$^{42}$ \\
                  &          & 7679   & 8:  $\pm$ 3       & Mg~II~2800    & 1.9$\times$10$^{42}$*\\
4FGL~J2325.6+1644 & 0.4817   & 5522   & 0.70  $\pm$ 0.2      & [O~II]~3727   & 1.0$\times$10$^{41}$ \\
                  &          & 7419   & 0.50 $\pm$ 0.2    & [O~III]~5007  & 4.8$\times$10$^{40}$ \\
\hline
\end{tabular}
\end{center}
\raggedright
\footnotesize \textit{Notes}. Column~1: Name of the target; Column~2: Redshift; Column~3: Barycenter of the detected line; Column~4: Measured equivalent width of the line; Column~5: Line identification; Column~6: Line luminosity. \\
(*) These lines are partially blended with the telluric band. \\
%\tablenotetext{}{
%\raggedright
% } 
\end{table*}

\setcounter{figure}{0}
\begin{figure*}%[htbp]
\includegraphics[width=0.25\textwidth, angle=-90]{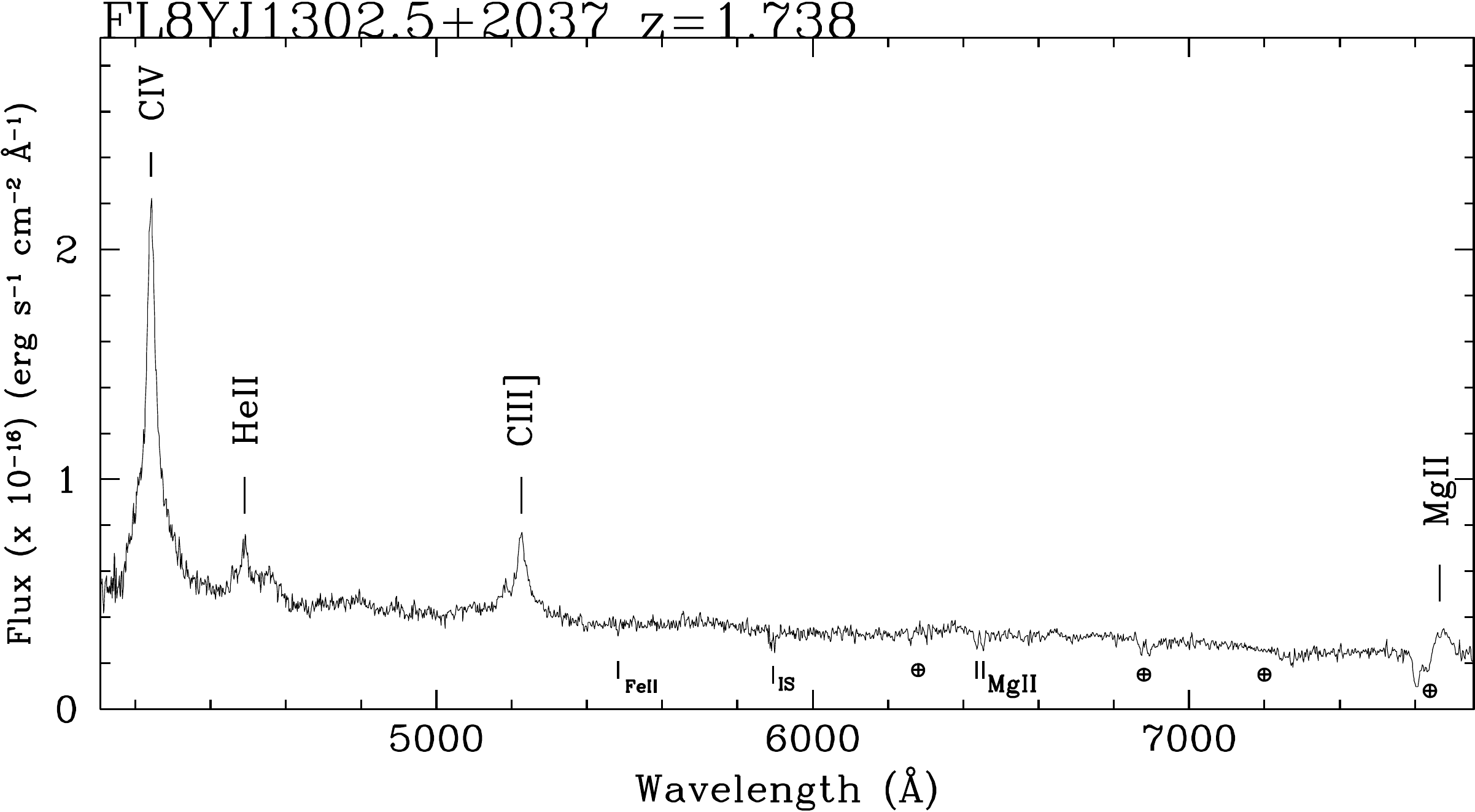}
\includegraphics[width=0.25\textwidth, angle=-90]{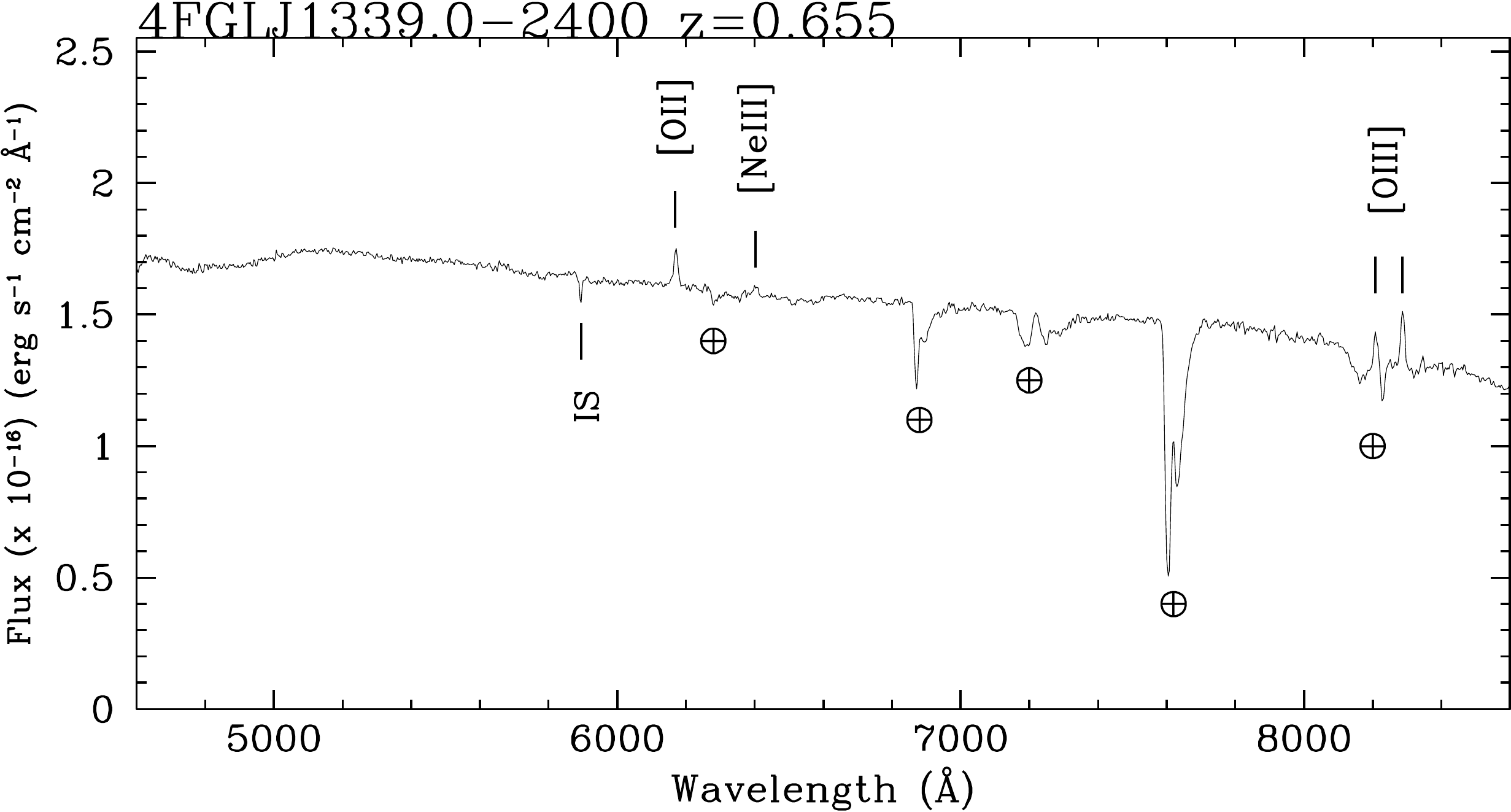}
\includegraphics[width=0.25\textwidth, angle=-90]{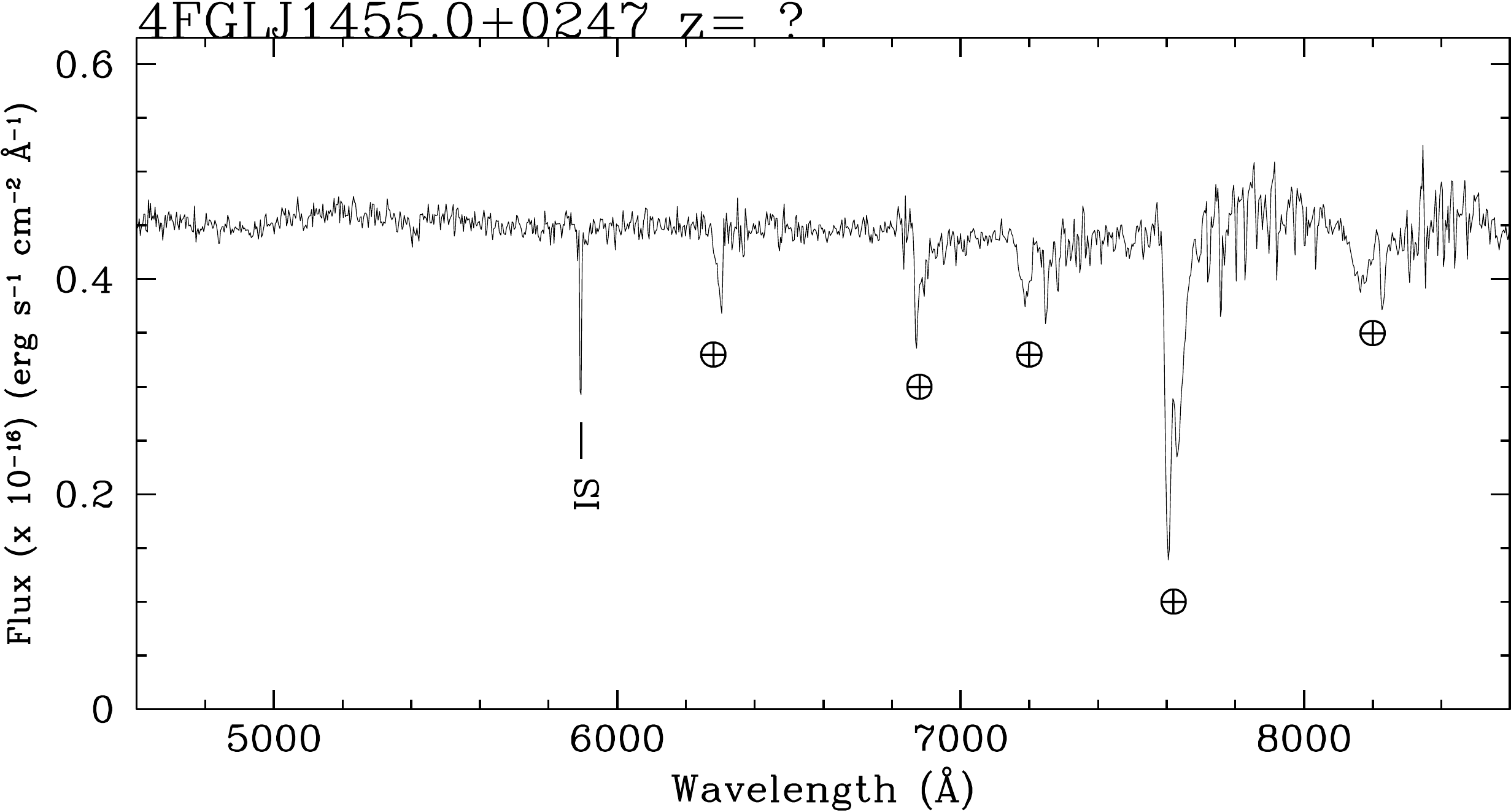}
\includegraphics[width=0.25\textwidth, angle=-90]{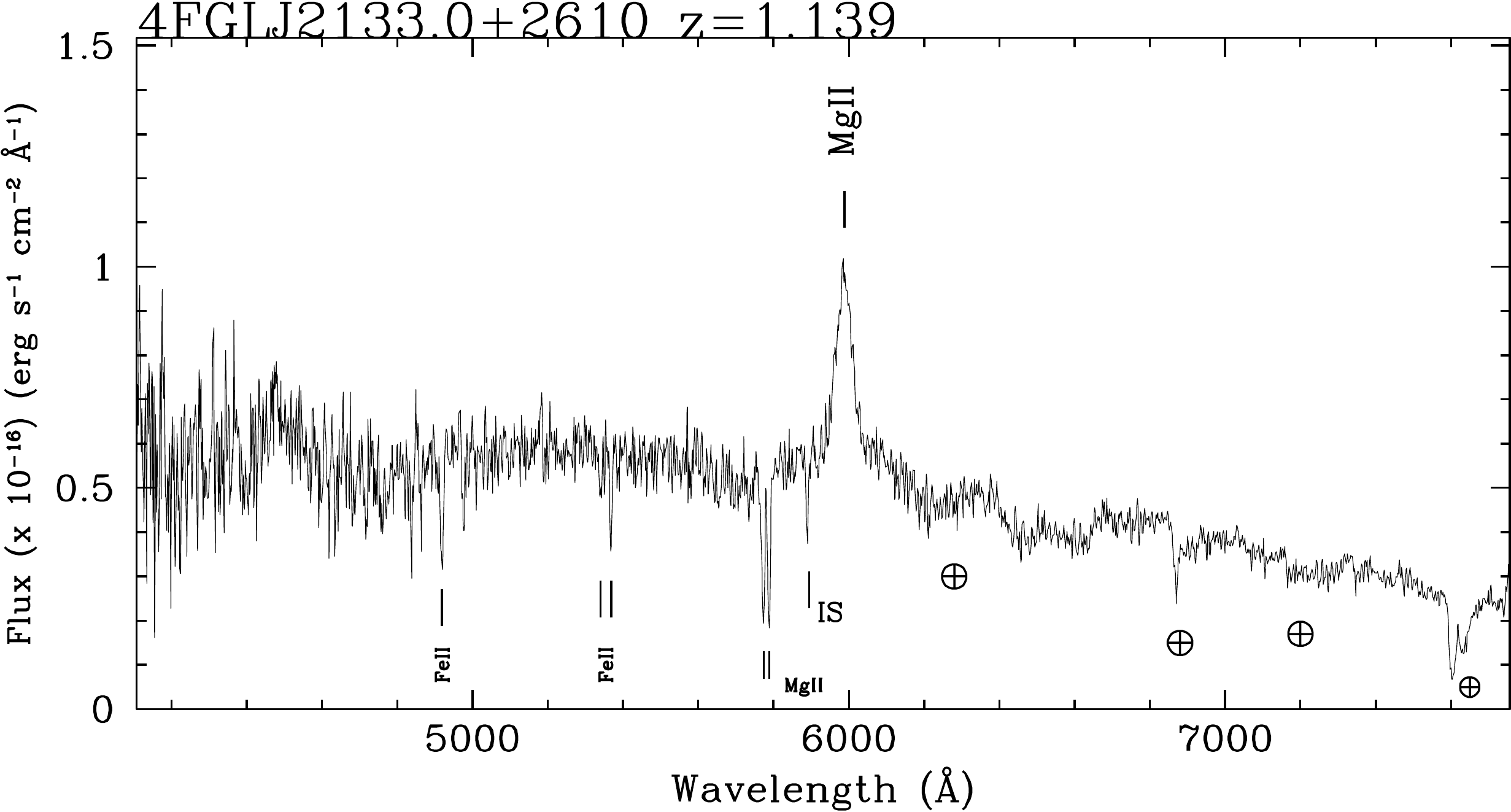}
\includegraphics[width=0.25\textwidth, angle=-90]{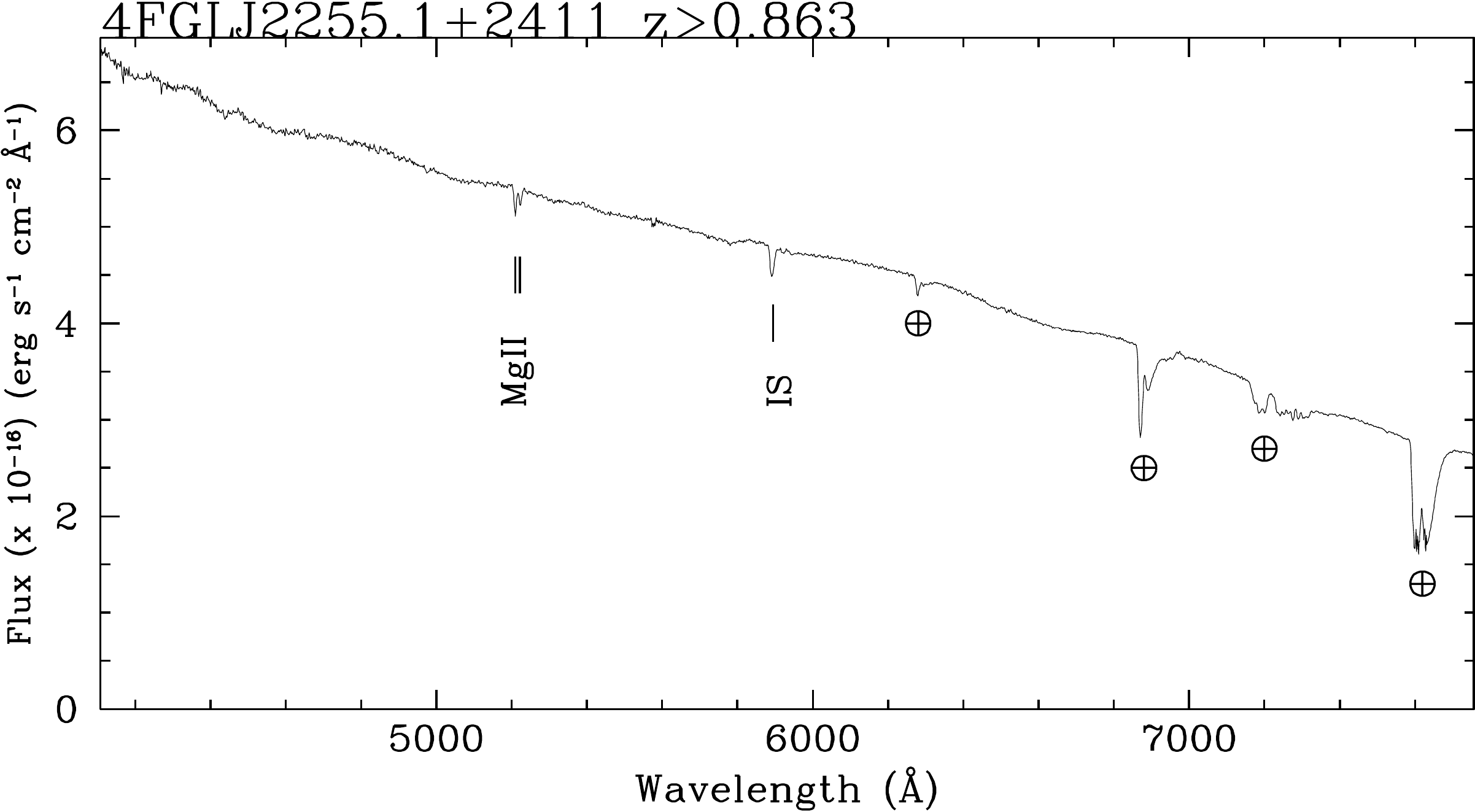}
\includegraphics[width=0.25\textwidth, angle=-90]{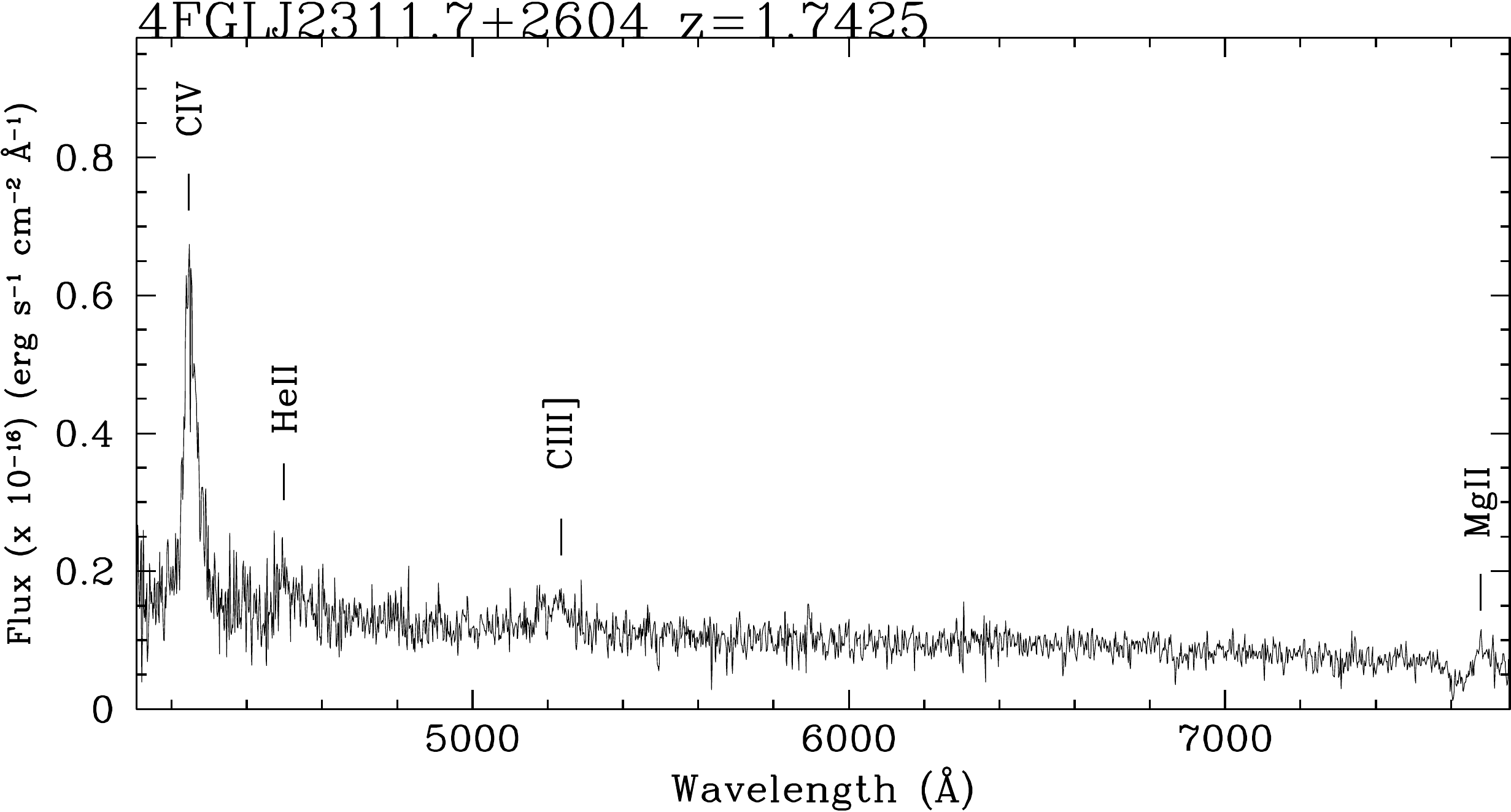}
\includegraphics[width=0.25\textwidth, angle=-90]{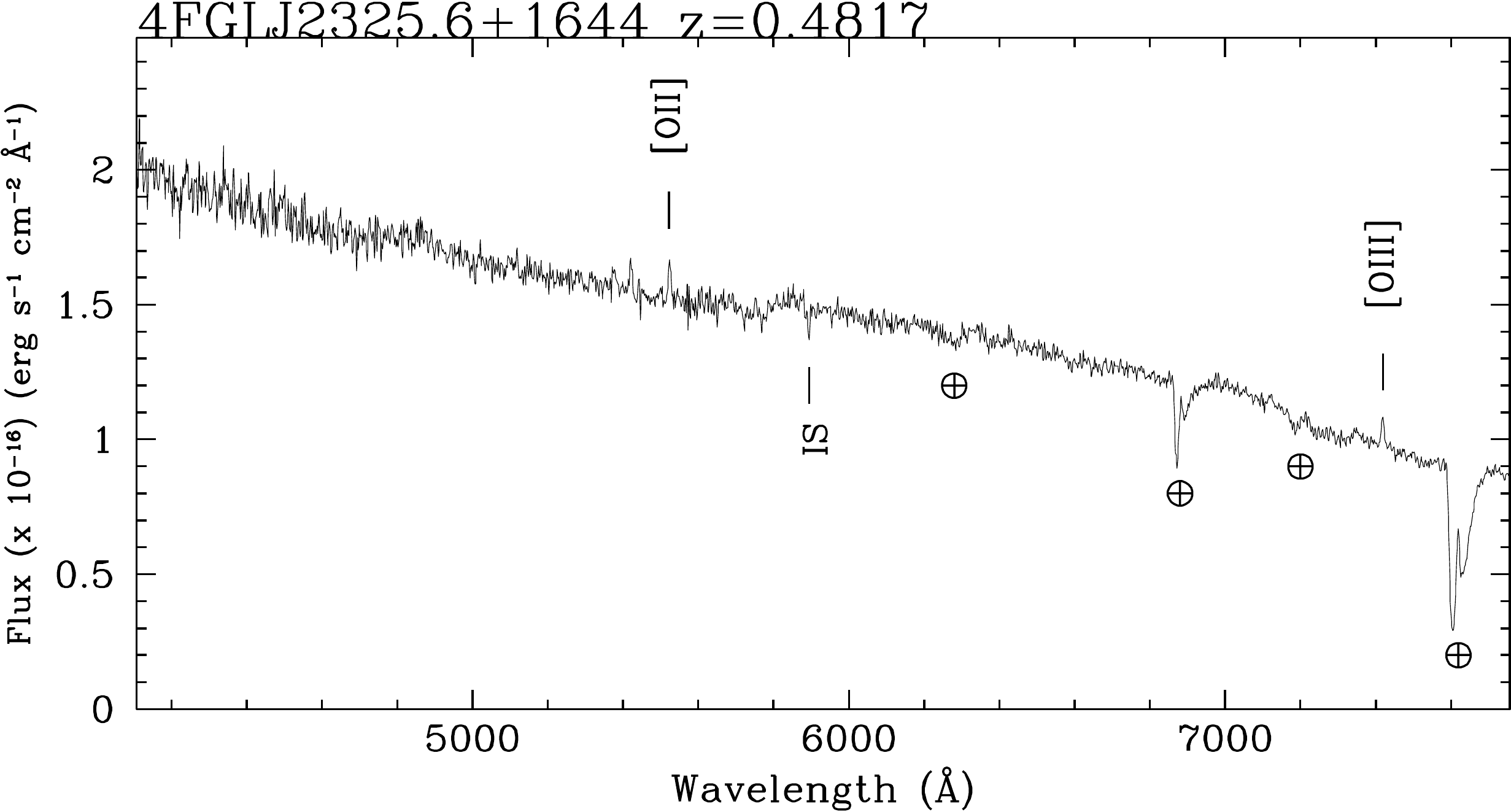}
\caption{Flux calibrated and dereddened spectra of the neutrino candidate blazars obtained at GTC and VLT. The main telluric bands are indicated by $\oplus$, the absorption features from interstellar medium of our galaxies are labelled as IS (Inter-Stellar).}
\end{figure*}%[htbp]

\label{lastpage}

\begin{thebibliography}{}
 \bibitem[\protect\citeauthoryear{Aartsen et al.}{2020}]{Aartsen2020} Aartsen M.~G., Ackermann M., Adams J., Aguilar J.~A., Ahlers M., Ahrens M., Alispach C., et al., 2020, PhRvL, 124, 051103
\bibitem[\protect\citeauthoryear{Abdo et al.}{2010}]{Abdo_2010} Abdo A. A.,
  Ackermann M., Ajello M., et al., 2010, ApJ, 716, 30 
\bibitem[\protect\citeauthoryear{Ahumada et al.}{2020}]{Ahumada2020} Ahumada R., Prieto C.~A., Almeida A., Anders F., Anderson S.~F., Andrews B.~H., Anguiano B., et al., 2020, ApJS, 249, 3
  
\bibitem[\protect\citeauthoryear{Appenzeller et al.}{1998}]{appenzeller1998} Appenzeller I., Fricke K., F{\"u}rtig W., G{\"a}ssler W., H{\"a}fner R., Harke R., Hess H.-J., et al., 1998, 
ESO Messenger, 94, 1  
\bibitem[\protect\citeauthoryear{Aublin}{2019}]{aublin2019} Aublin J., 2019, ICRC, 36, 840
\bibitem[\protect\citeauthoryear{Biteau et al.}{2020}]{biteau2020} Biteau J., Prandini E., Costamante L., Lemoine M., Padovani P., Pueschel E., Resconi E., et al., 2020, NatAs, 4, 124
\bibitem[\protect\citeauthoryear{Bressan, et al.}{2006}]{bressan2006} Bressan A., Falomo R., Vald{\'e}s J.~R., Rampazzo R., 2006, ApJL, 645, L101
\bibitem[\protect\citeauthoryear{Cardelli, Clayton \& Mathis}{1989}]{cardelli1989} Cardelli J.~A., Clayton G.~C., Mathis J.~S., 1989, ApJ, 345, 245
\bibitem[\protect\citeauthoryear{Cepa et al.}{2003}]{cepa2003} Cepa J., et al., 2003, SPIE, 4841, 1739, SPIE.4841
\bibitem[\protect\citeauthoryear{Cherenkov Telescope Array Consortium}{2020}]{CTA2020} Cherenkov Telescope Array Consortium T., 2020, arXiv, arXiv:2010.01349

\bibitem[\protect\citeauthoryear{de Menezes et al.}{2020}]{deMenezes2020} de Menezes R., Amaya-Almaz{\'a}n R.~A., Marchesini E.~J., Pe{\~n}a-Herazo H.~A., Massaro F., Chavushyan V., Paggi A., et al., 2020, Ap\&SS, 365, 12
\bibitem[\protect\citeauthoryear{Falomo, Pian \& Treves}{2014}]{falomo2014} Falomo R., Pian E., Treves A., 2014, A\&ARv, 22, 73
\bibitem[\protect\citeauthoryear{Freudling et al.}{2013}]{freudling2013} Freudling W., Romaniello M., Bramich D.~M., Ballester P., Forchi V., Garc{\'\i}a-Dabl{\'o} C.~E., Moehler S., et al., 2013, A\&A, 559, A964
\bibitem[\protect\citeauthoryear{Giommi et al.}{2020a}]{giommi2020a} Giommi P., Glauch T., Padovani P., Resconi E., Turcati A., Chang Y.~L., 2020a, MNRAS, 497, 865
\bibitem[\protect\citeauthoryear{Giommi et al.}{2020b}]{giommi2020b} Giommi P., Padovani P., Oikonomou F., Glauch T., Paiano S., Resconi E., 2020b, A\&A, 640, L4
\bibitem[\protect\citeauthoryear{Hook et al.}{2003}]{hook2003} Hook I.~M., Shaver P.~A., Jackson C.~A., Wall J.~V., Kellermann K.~I., 2003, A\&A, 399, 469
%\bibitem[\protect\citeauthoryear{IceCube Collaboration et al.}{2017}]{ICECube17_2} IceCube Collaboration, Aartsen M.~G., Ackermann M., Adams J., Aguilar J.~A., Ahlers M., Ahrens M., et al., 2017, arXiv, arXiv:1710.01179
\bibitem[\protect\citeauthoryear{IceCube Collaboration et al.}{2018a}]{icfermi}
 IceCube Collaboration, 2018, Science, 361, 147
\bibitem[\protect\citeauthoryear{IceCube Collaboration}{2018b}]{iconly}
 IceCube Collaboration et al., 2018, Science, 361, eaat1378
 
 \bibitem[\protect\citeauthoryear{Jones et al.}{2009}]{Jones2009} Jones D.~H., Read M.~A., Saunders W., Colless M., Jarrett T., Parker Q.~A., Fairall A.~P., et al., 2009, MNRAS, 399, 683
 
 \bibitem[\protect\citeauthoryear{Landoni, et al.}{2018}]{landoni2018} Landoni M., Paiano S., Falomo R., Scarpa R., Treves A., 2018, ApJ, 861, 130
\bibitem[\protect\citeauthoryear{Landoni et al.}{2020}]{landoni2020} Landoni M., Falomo R., Paiano S., Treves A., 2020, ApJS, 250, 37
\bibitem[\protect\citeauthoryear{Lucarelli et al.}{2019}]{Lucarelli_2019}
Lucarelli F., et al., 2019, ApJ, 870, 136 
\bibitem[\protect\citeauthoryear{Mannucci et al.}{2001}]{mannucci2001} Mannucci F., Basile F., Poggianti B.~M., Cimatti A., Daddi E., Pozzetti L., Vanzi L., 2001, MNRAS, 326, 745
\bibitem[\protect\citeauthoryear{Maselli et al.}{2015}]{maselli2015} Maselli A., Massaro F., D'Abrusco R., Cusumano G., La Parola V., Segreto A., Tosti G., 2015, Ap\&SS, 357, 141
\bibitem[\protect\citeauthoryear{Massaro et al.}{2014}]{massaro2014} Massaro F., Masetti N., D'Abrusco R., Paggi A., Funk S., 2014, AJ, 148, 66
\bibitem[\protect\citeauthoryear{Massaro et al.}{2015}]{massaro2015} Massaro F., Landoni M., D'Abrusco R., Milisavljevic D., Paggi A., Masetti N., Smith H.~A., et al., 2015, A\&A, 575, A124

\bibitem[\protect\citeauthoryear{Massaro et al.}{2015}]{BZCAT} Massaro E., Maselli A., Leto C., Marchegiani P., Perri M., Giommi P., Piranomonte S., 2015, Ap\&SS, 357, 75

\bibitem[\protect\citeauthoryear{Padovani \& Giommi}{1995}]{padgio95}
  Padovani P., Giommi P., 1995, ApJ, 444, 567
\bibitem[\protect\citeauthoryear{Padovani et al.}{2016}]{Padovani_2016}
  Padovani P., Resconi E., Giommi P., Arsioli B., Chang Y.~L., 2016, MNRAS,
  457, 3582 
\bibitem[\protect\citeauthoryear{Padovani et al.}{2017}]{Padovani_2017}
  Padovani P., et al., 2017, A\&ARv, 25, 2
 \bibitem[\protect\citeauthoryear{Padovani et  al.}{2018}]{padovani2018} Padovani P., Giommi P., Resconi E., Glauch T., Arsioli B.,  
Sahakyan N., Huber M., 2018, MNRAS, 480, 192
\bibitem[\protect\citeauthoryear{Padovani et al.}{2019}]{Padovani_2019} Padovani P., Oikonomou F., Petropoulou M., Giommi 
P., Resconi E., 2019, MNRAS, 484, L104
\bibitem[\protect\citeauthoryear{Paiano et al.}{2017a}]{paiano2017tev} Paiano S., Landoni M., Falomo R., Treves A., Scarpa R., Righi C., 2017, ApJ, 837, 144
\bibitem[\protect\citeauthoryear{Paiano et al.}{2017b}]{paiano20173fgl} Paiano S., Landoni M., Falomo R., Treves A., Scarpa R., 2017, ApJ, 844, 120
\bibitem[\protect\citeauthoryear{Paiano et al.}{2017c}]{paiano2017ufo1} Paiano S., Falomo R., Franceschini A., Treves A., Scarpa R., 2017, ApJ, 851, 135
\bibitem[\protect\citeauthoryear{Paiano et al.}{2018}]{paiano2018} Paiano S., Falomo R., Treves A., Scarpa R., 2018, ApJ, 854, L32
\bibitem[\protect\citeauthoryear{Paiano et al.}{2019a}]{paiano2019ufo2} Paiano S., Falomo R., Treves A., Franceschini A., Scarpa R., 2019, ApJ, 871, 162
\bibitem[\protect\citeauthoryear{Paiano et al.}{2019b}]{paiano2019atel2255} Paiano S., Padovani P., Falomo R., Giommi P., Scarpa R., Treves A., 2019, ATel, 13202
\bibitem[\protect\citeauthoryear{Paiano et al.}{2020a}]{paiano2020} Paiano S., Falomo R., Padovani P., Giommi P., Gargiulo A., Uslenghi M., Rossi A., et al., 2020, MNRAS, 495, L108
\bibitem[\protect\citeauthoryear{Paiano et al.}{2020b}]{paiano2020b} Paiano S., Falomo R., Treves A., Scarpa R., 2020, MNRAS, 497, 94
\bibitem[\protect\citeauthoryear{Patrignani \& Particle Data Group}{2016}]{Patrignani_2016} 
Patrignani C., Particle Data Group, 2016, ChPhC, 40, 100001, p. 528   
\bibitem[\protect\citeauthoryear{Pe{\~n}a-Herazo et al.}{2017}]{pena2017} Pe{\~n}a-Herazo H.~A., Marchesini E.~J., {\'A}lvarez Crespo N., Ricci F., Massaro F., Chavushyan V., Landoni M., et al., 2017, Ap\&SS, 362, 228
\bibitem[\protect\citeauthoryear{Petropoulou et al.}{2015}]{Petropoulou_2015} Petropoulou M., Dimitrakoudis S., Padovani P., Mastichiadis A., Resconi E., 2015, MNRAS, 448, 2412
\bibitem[\protect\citeauthoryear{Petropoulou et al.}{2020}]{Petropoulou_2020} Petropoulou M., Oikonomou F., Mastichiadis A., Murase K., Padovani P., Vasilopoulos G., Giommi P., 2020, ApJ, 899, 113  
\bibitem[\protect\citeauthoryear{Righi, Tavecchio, \& Inoue}{2019}]{righi2019} Righi C., Tavecchio F., Inoue S., 2019, MNRAS, 483, L127

\bibitem[\protect\citeauthoryear{Sandrinelli et al.}{2013}]{Sandrinelli2013} Sandrinelli A., Treves A., Falomo R., Farina E.~P., Foschini L., Landoni M., Sbarufatti B., 2013, AJ, 146, 163

\bibitem[\protect\citeauthoryear{Sbarufatti et al.}{2005}]{sbarufatti2005imaging} Sbarufatti B., Treves A., Falomo R., 2005, ApJ, 635, 173
\bibitem[\protect\citeauthoryear{Sbarufatti et al.}{2006}]{sbarufatti2006a} Sbarufatti B., Treves A., Falomo R., Heidt J., Kotilainen J., Scarpa R., 2006, AJ, 132, 1
\bibitem[\protect\citeauthoryear{Shaw et al.}{2013}]{shaw2013} Shaw M.~S. et al., 2013, ApJ, 764, 135
\bibitem[\protect\citeauthoryear{Schlafly \& Finkbeiner}{2011}]{schlafly2011} Schlafly E.~F., Finkbeiner D.~P., 2011, ApJ, 737, 103
 \bibitem[\protect\citeauthoryear{Schneider}{2019}]{schneider2019} Schneider, A., 2019, PoS, ICRC2019, 1004
 \bibitem[\protect\citeauthoryear{Stettner}{2019}]{stettner2019} Stettner, J., 2019, PoS, ICRC2019, 1017
\bibitem[\protect\citeauthoryear{Tody}{1986}]{tody1986} Tody D., 1986, SPIE, 627, 733, SPIE..627
\bibitem[\protect\citeauthoryear{Tody}{1993}]{tody1993} Tody D., 1993, ASPC, 52, 173, ASPC...52
\bibitem[\protect\citeauthoryear{Urry \& Padovani}{1995}]{UP95} Urry C.~M.,
  Padovani P., 1995, PASP, 107, 803
%\bibitem[\protect\citeauthoryear{van Dokkum}{2001}]{vandokkum2001} van Dokkum P.~G., 2001, PASP, 113, 1420


\end{thebibliography}
\end{document}